\documentclass{pinchcr}

\usepackage{amsmath}
\usepackage{graphicx}
\usepackage{amssymb}

\title{Dynamical mean-field theories of correlation and disorder}

\author{E. Miranda}

\affiliation{Instituto de F\'{\i}sica Gleb Wataghin, Unicamp, R. S\'{e}rgio Buarque de Holanda, 777,
 Campinas,
SP 13083-859, Brazil}

\author{V. Dobrosavljevi\'{c}}

\affiliation{Department of Physics and National High Magnetic Field Laboratory,
Florida State University, Tallahassee, FL 32306}
\authors{2}

\begin{document}

\maketitle

\tableofcontents

\maintext

\chapter{Dynamical mean-field theories of correlation and disorder}
\label{mirandavlad}

\section{Mott transitions in clean and disordered systems}

\label{sec:DMFT}

\vspace{12pt}
\emph{It is this fascination with the local and with the failures, not successes, of band theory, which...\hspace{6pt}contradicted the assumptions of the time...}\vspace{6pt}\\
\mbox{}\hfill\emph{P. W. Anderson, Nobel Prize Lecture, 1979} 

\subsection{Introduction}
\label{sub:DMFTIntro}

The Dynamical Mean Field Theory (DMFT) of interacting lattice fermions
is perhaps best described as the optimal description of these systems 
which takes into account only \emph{local} (on-site) correlation effects
\shortcite{Pruschke1995,georgesrmp}. Technically, this is implemented
through the assumption of a single-electron self-energy which, in
the lattice translationally invariant case, is independent of the 
site in real space or of wave vector in reciprocal space and therefore
is a function of frequency only \[
\Sigma\left(\boldsymbol{R_{i}},\omega\right)\to\Sigma\left(\omega\right)\qquad\mathrm{or}\qquad\Sigma\left(\boldsymbol{k},\omega\right)\to\Sigma\left(\omega\right).\]
This kind of approximation surely leaves out inter-site correlations.
However, local approaches to strong correlations have a long history
and several semi-phenomenological descriptions of classes of compounds
within this framework have been shown to be consistent with their
physical properties. A particularly well-studied example are heavy
fermion systems \shortcite{Stewart1984,Grewe1991}. Many properties
of heavy fermion materials suggest that local correlations are sufficient
for a good description, the large carrier effective mass (as derived
from the magnetic susceptibility or the Sommerfeld specific heat linear
coefficient) being the best known but by no means the only ones. Transport
properties such as DC and optical conductivity and ultrasound attenuation
can also be understood with the same assumptions \shortcite{varma85}.
Other examples of systems well described by a local approach include
systems close to a Mott metal-insulator transition, in particular,
the vicinity of the finite-temperature critical end-point \shortcite{Rozenberg1999a,Kotliar2000,Kotliar2002,Limelette03,Limelette2003,kagawaetal05}.
A particularly striking consequence of a local self-energy is the
cancellation of many-body renormalizations in the ratio of the coefficient
of the $T^{2}$-term of the resistivity to the square of the specific
heat coefficient, usually called the Kadowaki-Woods ratio \shortcite{KadowakiWoods,Miyakeetal89}.
Recently, it has been shown that, when materials-specific effects
(such as carrier density, density of states and Fermi velocity values)
are properly taken into account, then the Kadowaki-Woods ratio appears
to be universal across a much wider range of compounds, including,
besides heavy fermion systems, organic charge-transfer salts, transition-metal
oxides, and transition metals \shortcite{jackoetal09}. Thus, a local approach
to strong correlations seems to be much more generally valid than
initially thought.

Several starting points lead to theories that ultimately predict a
self-energy of this form, most notably descriptions based on the Gutzwiller
wave-function \shortcite{Gutzwiller1963,Gutzwiller1965,Vollhardt1984}
or the large-N limit \shortcite{colemanlong,millislee87}. However,
these theories usually end up imposing further restrictions, beyond
a local self-energy, e. g., inelastic scattering effects are not included,
higher-energy incoherent features are absent, among others. A description 
which incorporates \emph{all} possible local effects in a fully self-consistent
fashion is provided by DMFT.

Historically, DMFT was proposed by a recourse to the infinite-dimensional
limit of lattice systems \shortcite{MetznerVollhardt89}. Indeed, when
appropriate rescaling of parameters is done (as is usual when considering
this limit), the theory remains meaningful and non-trivial as $d\to\infty$
and the self-energy becomes completely local. The reader can find
many alternative derivations of DMFT in this limit in the review \shortcite{georgesrmp}.
Alternatively, DMFT can also be viewed as the best local description
of three-dimensional systems. The focus here will \emph{not} be to
derive the theory by resorting to the infinite-dimensional limit but
rather to highlight the physical content of a local description of
correlation effects. This is specially important since we will later
explore other, more general local theories which do not become exact
in any particular limit but which inherit the insights gained from
DMFT. We will therefore focus mostly on a Bethe lattice, which is
most transparent and lends itself particularly well to generalizations
to the disordered case. Furthermore, we will highlight the key physical
assumptions involved, which are kept in the other approximations.

\subsection{The clean case}

\label{sub:cleanDMFT}

Consider for concreteness the Hubbard model with only nearest-neighbor
hopping on a lattice with finite coordination $z$, in usual notation,\begin{eqnarray}
H & = & -\sum_{\left\langle ij\right\rangle ,\sigma}t_{ij}c_{i\sigma}^{\dagger}c_{j\sigma}+U\sum_{i}n_{i\uparrow}n_{i\downarrow}.\label{eq:HubbardHam}\end{eqnarray}
We focus on a particular site, call it $j$, which in the clean case
can be any site. The effective dynamics of this site alone can be
obtained by integrating out all the other sites. This is no longer
a Hamiltonian dynamics and the procedure requires an action description, 
which we will write in imaginary time

\begin{eqnarray}
S_{eff}\left(j\right) & = & \sum_{\sigma}\int_{0}^{\beta}d\tau c_{j\sigma}^{\dagger}\left(\tau\right)\left(\partial_{\tau}-\mu\right)c_{j\sigma}\left(\tau\right)\nonumber \\
 & + & \sum_{\sigma}\int_{0}^{\beta}d\tau\int_{0}^{\beta}d\tau^{\prime}c_{j\sigma}^{\dagger}\left(\tau\right)\Delta\left(\tau-\tau^{\prime}\right)c_{j\sigma}\left(\tau^{\prime}\right)\nonumber \\
 & + & U\int_{0}^{\beta}d\tau n_{j\uparrow}\left(\tau\right)n_{j\downarrow}\left(\tau\right).\label{eq:effaction}
\end{eqnarray}
The second term above comes from integrating out the other sites,
the first and third ones being the local contributions, already present
before the integration. The thing to note here is the fact that, in
general, the integration over \emph{interacting} sites generates other
higher-order terms, involving four and more fermionic fields. In the
high-dimensional limit or in DMFT in general, \emph{these higher-order
terms are absent or neglected}. It is clear that this means that only
single-particle inter-site correlations are kept in this limit/approximation
and this is precisely what is encoded in the second, retarded term 
in Eq.~(\ref{eq:effaction}). The {}``hybridization function'' 
$\Delta\left(\tau\right)$ describes the {}``leaking'' of electrons  
in and out of site $j$. It can be written as

\begin{equation}
\Delta\left(\tau\right)=t^{2}\sum_{l,m=1}^{z}G_{lm}^{\left(j\right)}\left(\tau\right),\label{eq:hyb}\end{equation}
where $G_{lm}^{\left(j\right)}\left(\tau\right)$ is the Green's function
for propagation from site $m$ to site $l$ \emph{in a lattice from
which site $j$ has been removed} (hence the superscript $\left(j\right)$) 

\begin{equation}
G_{lm}^{\left(j\right)}\left(\tau\right)=-\left\langle T\left[c_{l\sigma}\left(\tau\right)c_{m\sigma}^{\dagger}\left(0\right)\right]\right\rangle ^{\left(j\right)},\label{eq:greenremov}\end{equation}
and the sums extend over the $z$ nearest-neighbors of site $j$.
Let us not dwell on how this is calculated for now.

The action (\ref{eq:effaction}) is equivalent to the one of an Anderson
single-impurity problem \shortcite{Anderson1961}, whose Hamiltonian
is\begin{eqnarray}
H & = & \sum_{\boldsymbol{k},\sigma}E_{\boldsymbol{k}}a_{\boldsymbol{k}\sigma}^{\dagger}a_{\boldsymbol{k}\sigma}-\sum_{\sigma}\mu c_{j\sigma}^{\dagger}c_{j\sigma} \nonumber \\
&+&\sum_{\boldsymbol{k},\sigma}\left(\frac{V_{\boldsymbol{k}}}{\sqrt{N_{s}}}a_{\boldsymbol{k}\sigma}^{\dagger}c_{j\sigma}+\mathrm{H.c.}\right)+Uc_{j\uparrow}^{\dagger}c_{j\uparrow}c_{j\downarrow}^{\dagger}c_{j\downarrow},\label{eq:AndseronHam}\end{eqnarray}
\emph{provided} we choose $E_{\boldsymbol{k}}$ and $V_{\boldsymbol{k}}$
above in such a way that the Fourier transform, in Matsubara frequency
space, of the hybridization function in Eq.~(\ref{eq:hyb}) is such
that\begin{equation}
\Delta\left(i\omega_{n}\right)=\frac{1}{N_{s}}\sum_{\boldsymbol{k}}\frac{\left|V_{\boldsymbol{k}}\right|^{2}}{i\omega_{n}-E_{\boldsymbol{k}}}.\label{eq:AndersonBath}\end{equation}
This equivalence proves to be extremely useful since the well-studied
behavior of the Anderson single-impurity problem serves as a guide
to physical insight \shortcite{georgeskotliar92}.

Suppose now that we can somehow find the full interacting Green's
function of the system described by the action of Eq.~(\ref{eq:effaction})
\begin{equation}
G_{jj}\left(\tau\right)=-\left\langle T\left[c_{j\sigma}\left(\tau\right)c_{j\sigma}^{\dagger}\left(0\right)\right]\right\rangle _{eff},\label{eq:greenloc}\end{equation}
where the subscript $eff$ emphasizes that it is to be calculated
under the dynamics dictated by (\ref{eq:effaction}). We can repackage
our ignorance about this function by defining a self-energy $\Sigma\left(i\omega_{n}\right)$
such that

\begin{equation}
G_{jj}\left(i\omega_{n}\right)=\frac{1}{i\omega_{n}+\mu-\Delta\left(i\omega_{n}\right)-\Sigma\left(i\omega_{n}\right)}.\label{eq:selfen}\end{equation}

Having quantified the local dynamics, we now need to bring in information
from the rest of the lattice. In principle, this is quite straightforward:
\emph{since only local correlations are included, }\textbf{$\Sigma\left(i\omega_{n}\right)$
}\emph{is also the self-energy for generic lattice propagation}\begin{equation}
G\left(\boldsymbol{k},i\omega_{n}\right)=\frac{1}{i\omega_{n}-\varepsilon_{\boldsymbol{k}}+\mu-\Sigma\left(i\omega_{n}\right)},\label{eq:latticeselfen}\end{equation}
where $\varepsilon_{\boldsymbol{k}}$ is the non-interacting dispersion.
From this expression, we can obtain the Green's function with one
site removed from Eq.~(\ref{eq:greenremov}) and from that the hybridization
function (\ref{eq:hyb}), which closes the self-consistency loop.
This procedure to get $\Delta\left(i\omega_{n}\right)$ from $G\left(\boldsymbol{k},i\omega_{n}\right)$
is described, e. g., in the review \shortcite{georgesrmp}. However,
we will proceed in the simpler and more illuminating case of the Bethe
lattice with coordination $z$, for which

\begin{equation}
G_{lm}^{\left(j\right)}\left(\tau\right)=\delta_{lm}G_{ll}^{\left(j\right)},\label{eq:greenbethe}\end{equation}
since the removal of site $j$ completely disconnects the two branches
that start at the nearest-neighbors $l$ and $m$, if $l\neq m$.
Thus, \begin{eqnarray}
\Delta\left(\tau\right) & = & t^{2}\sum_{l=1}^{z}G_{ll}^{\left(j\right)}\left(\tau\right)\label{eq:hybbethe1a}\\
 & = & zt^{2}G_{ll}^{\left(j\right)}\left(\tau\right),\label{eq:hybbethe1b}\end{eqnarray}
since all sites are equivalent. We can now take the limit $z\to\infty$,
noting that, in this limit, the removal of one nearest-neighbor site
($j$) is irrelevant for the local propagation at $l$ and that an
appropriate rescaling, namely $zt^{2}\to\tilde{t}^{2}$, is necessary
\begin{equation}
\Delta\left(\tau\right)=\tilde{t}^{2}G_{ll}\left(\tau\right)=\tilde{t}^{2}G_{jj}\left(\tau\right),\label{eq:hybbethe2}\end{equation}
since all sites are equivalent. This is the self-consistency condition
in this case: the solution of the problem is the one for which, if
we plug in the local Green's function $G_{jj}\left(\tau\right)$ in
Eq.~(\ref{eq:hybbethe2}), then insert it into the action (\ref{eq:effaction})
and find the expectation value in Eq.~(\ref{eq:greenloc}), we get
back $G_{jj}\left(\tau\right)$.

A more physical alternative route and one which is not restricted
to the Bethe lattice is to note \emph{that the local Green's function
obtained from} (\ref{eq:latticeselfen}) \emph{must coincide with
the one defined in} (\ref{eq:selfen})
\begin{eqnarray}
G_{jj}\left(i\omega_{n}\right)&=&\frac{1}{N_{s}}\sum_{\boldsymbol{k}}G\left(\boldsymbol{k},i\omega_{n}\right)\Rightarrow \nonumber \\
\frac{1}{i\omega_{n}+\mu-\Delta\left(i\omega_{n}\right)-\Sigma\left(i\omega_{n}\right)}&=&\int d\varepsilon\frac{\rho_{0}\left(\varepsilon\right)}{i\omega_{n}-\varepsilon+\mu-\Sigma\left(i\omega_{n}\right)}.\label{eq:self-cons}
\end{eqnarray}
Here, $\rho_{0}\left(\varepsilon\right)$ is the bare density of states
generated from $\varepsilon_{\boldsymbol{k}}$. The two procedures
can be shown to be equivalent and establish the necessary self-consistency
condition.

It is possible to extend the above analysis to two-particle correlation
functions \shortcite{georgesrmp} but we will not delve into it. It is,
however, worthwhile to notice that the current-current correlation
function, which provides the conductivity through the Kubo formula
acquires no vertex correction within DMFT and is given by a simple
bubble of renormalized single-particle Green's functions.

\begin{figure}

\begin{centering}
\includegraphics[width=4in]{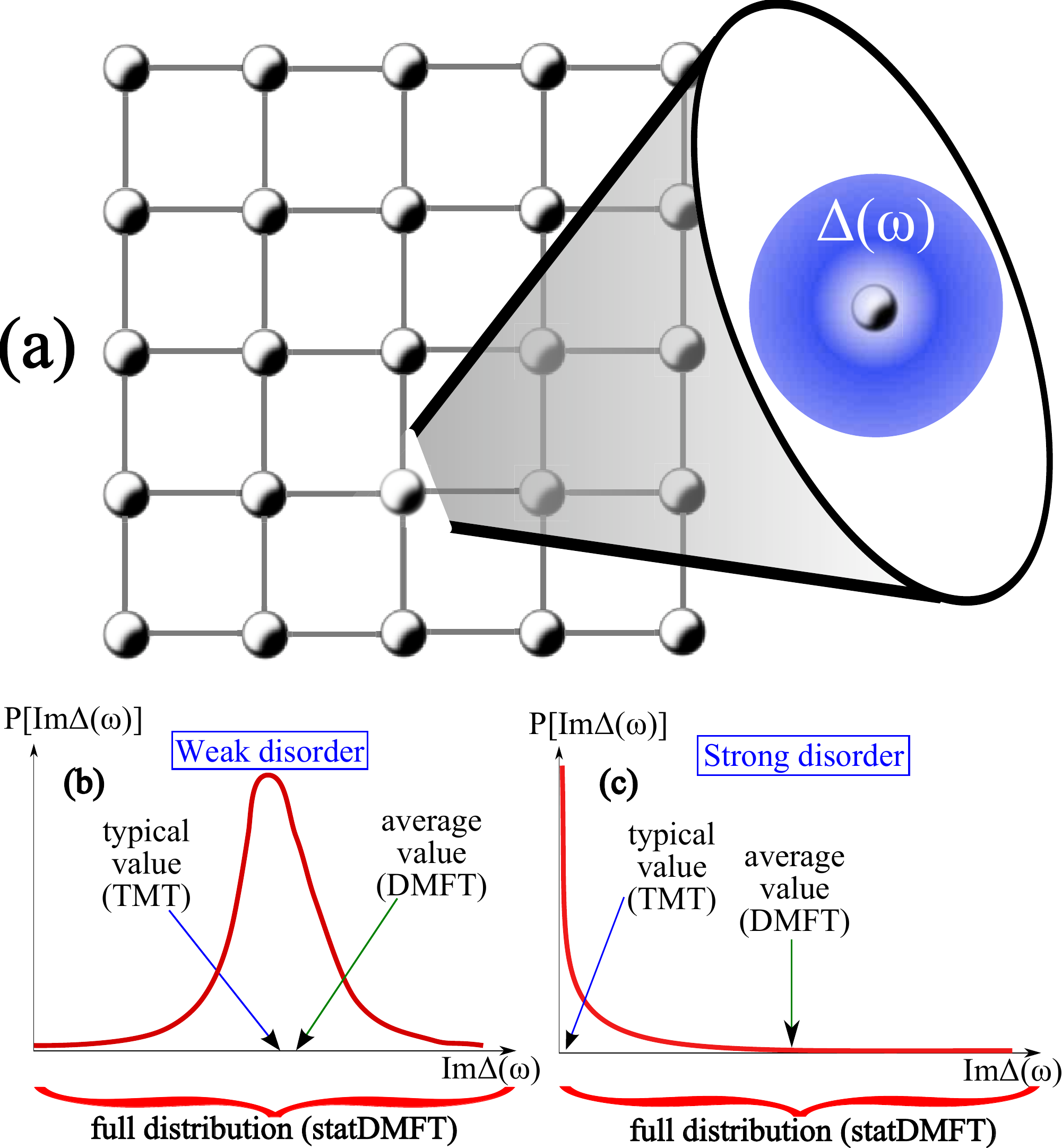}

\caption{\label{fig:DMFTscheme} Schematic view of all DMFT-inspired theories
of disordered strongly correlated systems. (a) All these theories
share the same ingredient: the local site {}``knows'' about the
other sites through the self-consistently determined hybridization
function $\Delta\left(\omega\right)$ (see Eq.~(\ref{eq:effaction})),
which acts as an order parameter. They differ in the level of description
of this order parameter, which is in general a random quantity (panels
b and c). Both DMFT and the Typical Medium Theory (TMT) replace the
actual local realization of this random variable by a fixed, non-random
function, namely, the average and the typical values of $\Delta\left(\omega\right)$,
respectively. These two functions are very similar at weak disorder
(panel b), but become increasingly different as the disorder grows
(panel c). The statistical dynamical mean field theory (statDMFT),
by contrast, retains the full spectrum of spatial fluctuations, since
each site {}``sees'' the actual local realization $\Delta_{i}\left(\omega\right)$.}

\end{centering}

\end{figure}

It is perhaps wise to highlight yet again what the key assumptions
of this approach are:
\begin{enumerate}
\item The effects of interactions that are included are on-site only, or
equivalently, the self-energy is purely local.
\item Different sites {}``know'' about each other through single-particle
processes only, see Fig.~\ref{fig:DMFTscheme}(a).
\item The local dynamics, as dictated by the local effective action and
usually encoded in the local Green's function, must coincide with
the local dynamics as derived from the lattice propagation.
\end{enumerate}
The DMFT approach, like the original Bragg-Williams (BW) mean-field
theory of magnetism \shortcite{goldenfeldbook}, focuses on a single lattice
site, but replaces its environment by a self-consistently determined
{}``effective medium'' \shortcite{georgesrmp}. Unlike the BW theory,
the effect of the environment cannot be captured by a static external
field, but must be encoded in a full complex \emph{function} $\Delta\left(i\omega_{n}\right)$,
which contains information about the dynamics of an electron moving
in and out of the given site. The calculation then reduces to solving
an appropriate quantum impurity problem, Eq.~(\ref{eq:effaction}),
supplemented by an additional self-consistency condition, Eq.~(\ref{eq:self-cons}),
that ultimately determines this hybridization function $\Delta(i\omega_{n})$. 

The approach has been very successful in examining the vicinity of
the Mott transition in clean systems, in which it has met spectacular
success in elucidating various properties of several transition metal
oxides \shortcite{georgesrmp}, heavy fermion systems, and even Kondo
insulators \shortcite{re:Rozenberg96}.

\subsection{The clean Mott transition}

The Mott transition in a single-band Hubbard model can be regarded as a prototype for 
a interaction-driven metal-insulator transition, a phenomenon with plausible relevance
to many physical systems of current interest. Its basic mechanism has been correctly 
understood for more than fifty years \shortcite{mott1949}, yet the precise nature of this 
phase transition has long remained controversial and ill-understood. Part of the 
confusion stems from the fact that at low temperatures the Mott insulator is 
typically unstable to antiferromagnetic ordering, leading many authors \shortcite{slater51}
to focus on magnetism as a proposed driving force. The shortcoming of this view was
most lucidly emphasized by Anderson \shortcite{andersonlocrev} who stressed that the 
Mott insulating state persists well above the Néel temperature. It is thus transmutation of conduction electrons 
into local magnetic moments - not the long range magnetic ordering - that should
be regarded as the fundamental physical process behind the Mott transition. The two phenomena 
can be most clearly separated in systems where the tendency for magnetic ordering
can be appreciably weakened due to frustration effects, such as often found in orbitally degenerate 
transition metal oxides. Here, the competition between antiferromagnetic superexchange and 
ferromagnetic tendencies due to Hund's rule couplings typically lead to large cancellations, 
resulting in very weak magnetic correlations in the paramagnetic phase. 

From the theoretical point of view, this situation can be most clearly formulated by focusing on the
``maximally frustrated'' Hubbard model, with infinite-range hopping of random sign \shortcite{georgesrmp}. 
In this model the magnetic frustration is so strong as to completely suppress any magnetic ordering,
while the DMFT approximation becomes exact, allowing  precise and detailed characterization of 
such an ``ideal" Mott transition within the paramagnetic phase. In the following we briefly describe 
the main features of the resulting DMFT picture of the bandwidth-driven Mott transition.

\subsubsection{Critical behavior and mass divergence at $T=0$}

Within DMFT, the critical regime between the Fermi liquid metal and the Mott insulator features 
\shortcite{georgesrmp} a finite temperature coexistence region and a first-order transition line ending 
at the critical end-point at $T=T_c$. At $T=0$, however, the metallic solution is the stable (lower energy) one
throughout the coexistence regime. It is characterized by heavy quasiparticles (QP) with an effective mass
that diverges as the transition is approached. 

In general, the effective mass is evaluated from the 
single-particle self-energy using the expression
\begin{equation}
{\frac{m^{\ast}}{m}}=\left.  {\frac{1-{\frac{\partial}{\partial\omega}}
\Sigma^{\prime}(\mathbf{k},\omega)}{1+{\frac{m}{k}}{\frac{\partial}{\partial
k}}\Sigma^{\prime}(\mathbf{k},\omega)}}\right\vert _{k=k_{F},\omega=0},
\label{eq:offshellm}
\end{equation}
where $k_{F}$ denotes the Fermi momentum, and $\Sigma^{\prime}$ is the real part of the 
self-energy $\Sigma(\mathbf{k},\omega )$.  The QP weight, on
the other hand, is defined by
\begin{equation}
Z^{-1}\equiv\left.  1-{\frac{\partial}{\partial\omega}}\Sigma^{\prime
}(\mathbf{k},\omega)\right\vert _{k=k_{F},\omega=0}.
\end{equation}
Within DMFT, $\Sigma(\mathbf{k},\omega )=\Sigma(\omega)$ is momentum independent, and $m^{\ast}/m=Z^{-1}$.
Note that, since generally $\left.  {\frac{\partial}{\partial\omega}}%
\Sigma^{\prime}(\omega)\right\vert _{\omega=0}<0$, the interactions increase
the effective mass. 
The actual divergence is obtained only if the quantity
$A\equiv\left.  -{\frac{\partial}{\partial\omega}}\Sigma^{\prime}
(\omega)\right\vert _{\omega=0}$ itself diverges. This scenario is realized,
for example, in the Brinkmann-Rice theory of the Mott transition, as well as
in the more recent DMFT solution. Since the QP weight is
simply $Z^{-1}=m^{\ast}/m$, it must diverge at the same place as $m^{\ast}$ does.

We should emphasize that this result is exact within the DMFT approach and is an excellent
approximation for many Mott compounds where magnetic frustration is sufficiently strong. To put this
result in perspective, we contrast it with a popular but uncontrolled weak-coupling approach, 
based on the so-called ``on-shell approximation'' \shortcite{quinn75prl} for the effective 
mass of the correlated electron gas. Here, an approximate expression for the effective mass is proposed
\begin{equation}
{\frac{m^{\ast}}{m}}\approx\left.  {\frac{1}{1+{\frac{m}{k}}{\frac{d}{dk}%
}\Sigma^{\prime}(\mathbf{k},\xi_\mathbf{k})}}\right\vert _{k=k_{F}},
\label{eq:onshellm}
\end{equation}
where $\xi_\mathbf{k}$ is the unrenormalized band dispersion. When this approximation 
is applied to the low-density electron gas within the Random Phase Approximation 
(RPA) scheme \shortcite{dassarma05prb}, 
one finds that the effective mass diverges {\em before} the QP weight $Z$ vanishes. 
This result seems quite pathological, since the natural interpretation of the effective mass 
divergence is the localization of itinerant electrons, where one also expects the breakdown 
of the quasiparticle picture. 

To benchmark the 
validity of the proposed ``on-shell approximation'', we apply it to the maximally frustrated 
Hubbard model, where DMFT provides us with an exact result for both the effective mass 
and the full self-energy. In this case $\Sigma^{\prime}(\mathbf{k},\omega=\xi_{\mathbf{k}}) =
\Sigma^{\prime}(\omega\mathbf{=}\xi_{\mathbf{k}})$. Noting that
\[
\left.  {\frac{m}{k}}{\frac{d}{dk}}\Sigma^{\prime}(\omega=\xi_{\mathbf{k}%
})\right\vert _{k=k_{F}}=\left.  {\frac{\partial}{\partial\omega}}%
\Sigma^{\prime}(\omega)\right\vert _{\omega=0},
\]
we get the ``on-shell" result
\[
{\frac{m^{\ast}}{m}}\approx\left.  {\frac{1}{1+{\frac{\partial}{\partial
\omega}}\Sigma^{\prime}(\omega)}}\right\vert _{\omega=0} ={\frac{1}{1-A}}.
\]

As we can see, this expression is equivalent to the exact expression $m^{\ast
}/m=1+A$, only to leading order, i.e. for $(1-Z)\ll1$. On
the other hand, the positive quantity $A$ is expected to grow with the
interaction. As long as it is finite, neither will the properly defined
effective mass $m^{\ast}/m,$ nor the inverse QP weight $Z^{-1}$ ever
diverge. In contrast, if one uses the ``on-shell" expression, then the
effective mass will blow up as soon as $A=1$, and this will happen at some
point in any approximation where $A$ grows with the interaction. However, as
we can see, this will not lead to the divergence of the inverse QP weight
$Z^{-1}$. What we can see from these expressions is that the essence of the
``on-shell" approximation is simply to linearize the expression for $(m^{\ast
}/m)^{-1}$, by expanding it in the quantity $A\equiv\left.  -{\frac{\partial
}{\partial\omega}}\Sigma^{\prime}(\omega)\right\vert _{\omega=0}.$ Instead of
appearing in the numerator of the effective mass expression, it now enters the
denominator, leading to an unphysical effective mass divergence. This example 
provides a perfect illustration of how dangerous it is to indiscriminately apply
weak-coupling results to non-perturbative phenomena near the Mott transition. 
It also shows how DMFT not only correctly captures the essence of strong 
correlations, but also provides a simple and transparent insight into their
physical content.

\subsubsection{Quantum-critical behavior at $T > T_c$}

Many systems close to the metal-insulator transition (MIT) often
display surprisingly similar transport features in the high
temperature regime
Here, the family of resistivity curves typically assumes a
characteristic ``fan-shaped'' form, reflecting a
gradual crossover from metallic to insulating transport. At the
highest temperatures the resistivity depends only weakly on the
control parameter (concentration of charge carriers or pressure) 
while as $T$ is lowered, the system seems to ``make up its
mind'' and rapidly converges towards either a metallic or an
insulating state. Since temperature acts as a natural cutoff scale
for the metal-insulator transition, such behavior is precisely
what one expects for quantum criticality. In some cases \shortcite{abrahams-rmp01}, the
entire family of curves displays beautiful scaling behavior, with
a remarkable ``mirror symmetry'' of the relevant scaling
functions \shortcite{gang4me}. But under which microscopic
conditions should one expect such scaling phenomenology?
Should one expect similar or very different transport phenomenology in the Mott
picture? Is the paradigm of quantum criticality even a useful
language to describe high temperature transport around the Mott
point?

\begin{figure}[t]
\includegraphics[width=4in]{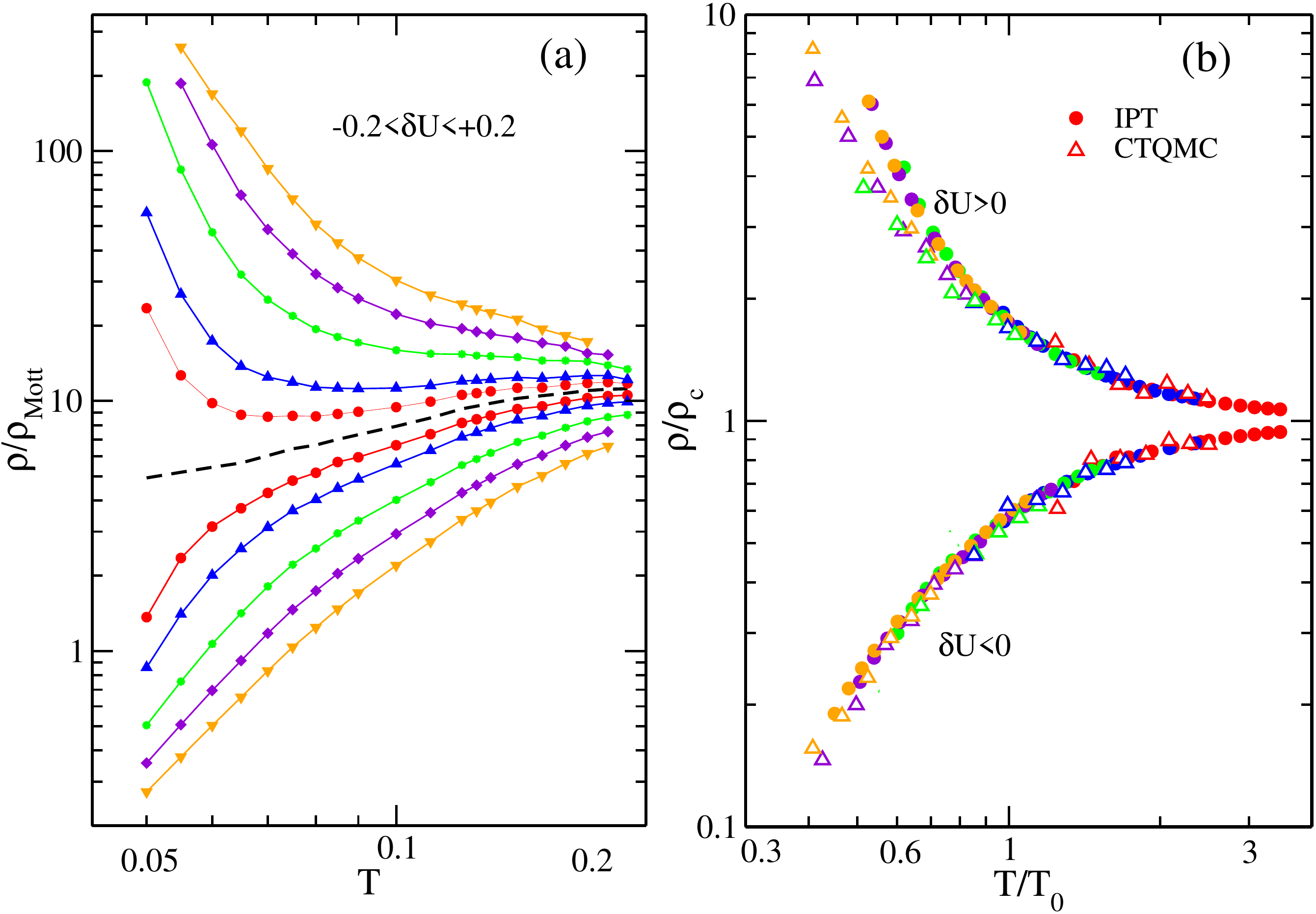}

\caption{(a) DMFT resistivity curves as functions of the temperature
along different trajectories across the Mott transition in a half-filled Hubbard model \protect\shortcite{terletska-mott11prl}. (b) Resistivity
scaling displaying remarkable ``mirror symmetry'' \protect\shortcite{gang4me}.}

\end{figure}

Somewhat surprisingly, most DMFT studies of the Mott transition focused on the lowest 
temperature regime, paying little attention to the high temperature crossover regime
relevant to many experiments. On the other hand, it is well known that at very low temperatures $T<T_{c}\sim0.03 T_F$,
this model features a first order metal-insulator transition terminating
at the critical end-point $T_{c}$ (Fig. 2), very similar to the familiar liquid-gas
transition. For $T>T_{c}$, however, different crossover
regimes have been tentatively identified \shortcite{georgesrmp} but they
have not been studied in any appreciable detail. The fact that the
first order coexistence region is restricted to such very low temperatures
provides strong motivation to examine the high temperature crossover
region from the perspective of ``hidden quantum criticality''.
In other words, it is very plausible that the presence of a coexistence dome at $T<T_{c}\ll T_F$,
an effect with very small energy scale, is not likely to influence
the behavior at much higher temperatures $T\gg T_{c}$.
In this high temperature regime smooth crossover is found, which may
display behavior consistent with the presence of a ``hidden''
quantum critical (QC) point at $T=0$. To test this idea, very recent work 
\shortcite{terletska-mott11prl} utilized standard
scaling methods appropriate for quantum criticality and computed the
resistivity curves along judiciously chosen trajectories respecting
the symmetries of the problem. Characteristic scaling behavior for the entire
family of resistivity curves has been identified, with the corresponding 
beta-function displaying striking ``mirror symmetry'' consistent with 
experiments. These findings provide compelling arguments in support 
of the suggestion that finite temperature behavior in many Mott systems 
should be interpreted from the perspective of quantum criticality \shortcite{christos-vlad05}.
It should stressed, however, that the DMFT solution in question does not contain 
any physical processes associated with approach to magnetic or charge ordering. 
If quantum criticality is indeed at play here, it has a fundamentally different nature, 
one that is associated with the destruction of a Fermi liquid without the aid of
any static symmetry breaking pattern - in dramatic contrast to most known 
critical phenomena.

\subsection{The disordered case}

\label{sub:disDMFT}

Let us now proceed to write down the equations for the dynamical mean
field theory description of a disordered system \shortcite{janisvoll92,janisetal93,dk-prl93,dk-prb94}.
We will focus on the case of diagonal site disorder, which for the
Hubbard model reads\begin{eqnarray}
H & = & -\sum_{\left\langle ij\right\rangle ,\sigma}t_{ij}c_{i\sigma}^{\dagger}c_{j\sigma}+\sum_{j,\sigma}\varepsilon_{j}c_{j\sigma}^{\dagger}c_{j\sigma}+U\sum_{i}n_{i\uparrow}n_{i\downarrow},\label{eq:DisordHubbardHam}\end{eqnarray}
 where $\varepsilon_{j}$ are assumed to be independent random variables
drawn from a given distribution $P\left(\varepsilon\right)$ whose
strength is $W$ (say, a uniform distribution from $-W/2$ to $W/2$).
Focusing once more on the local dynamics, it is clear it must now
be dictated by\begin{eqnarray}
S_{eff}\left(j\right) & = & \sum_{\sigma}\int_{0}^{\beta}d\tau c_{j\sigma}^{\dagger}\left(\tau\right)\left(\partial_{\tau}+\varepsilon_{j}-\mu\right)c_{j\sigma}\left(\tau\right)\nonumber \\
 & + & \sum_{\sigma}\int_{0}^{\beta}d\tau\int_{0}^{\beta}d\tau^{\prime}c_{j\sigma}^{\dagger}\left(\tau\right)\Delta\left(\tau-\tau^{\prime}\right)c_{j\sigma}\left(\tau^{\prime}\right)\nonumber \\
 & + & U\int_{0}^{\beta}d\tau n_{j\uparrow}\left(\tau\right)n_{j\downarrow}\left(\tau\right).\label{eq:disordeffaction}\end{eqnarray}
Notice that the effective action is now different for different sites
because of the $\varepsilon_{j}$ term. However, the hybridization
function $\Delta\left(\tau\right)$ is \emph{not} site-dependent.
This is motivated again by the infinite-dimensional limit. Indeed,
if the number of nearest-neighbors is infinite, the sum in Eq.~(\ref{eq:hyb})
is effectively an \emph{averaging procedure over all possible realizations
of the Green's function}. In fact, in the infinite-dimensional Bethe
lattice, Eq.~(\ref{eq:hybbethe1a}) becomes \begin{eqnarray}
\Delta\left(\tau\right) & = & zt^{2}\left[\frac{1}{z}\sum_{l=1}^{z}G_{ll}^{\left(j\right)}\left(\tau\right)\right]\label{eq:disordhybbethe1a}\\
 & \xrightarrow[z\to\infty]{} & \tilde{t}^{2}\overline{G_{ll}\left(\tau\right)},\label{eq:disordhybbethe1b}\end{eqnarray}
where the overbar denotes average over quenched disorder. Again, the
effect of removing one nearest-neighbor is negligible in this case.
Thus, the hybridization function is proportional to the \textbf{average
local Green's function} (see Fig.~\ref{fig:DMFTscheme}b and c).
Solving the DMFT equations in the disordered case then entails solving
an \emph{ensemble} of single-impurity problems as in Eq.~(\ref{eq:disordeffaction}),
one for each value of $\varepsilon_{j}$, and finding for each of
them the local Green's function and self-energy (which are now also
site-dependent)

\begin{eqnarray}
G_{jj}\left(\tau\right) & = & -\left\langle T\left[c_{j\sigma}\left(\tau\right)c_{j\sigma}^{\dagger}\left(0\right)\right]\right\rangle _{eff},\label{eq:disordgreenloc}\\
G_{jj}\left(i\omega_{n}\right) & = & \frac{1}{i\omega_{n}-\varepsilon_{j}+\mu-\Delta\left(i\omega_{n}\right)-\Sigma_{j}\left(i\omega_{n}\right)}.\label{eq:disordselfen}\end{eqnarray}
The self-consistency can then be written, for the Bethe lattice, as
(cf. Eqs.~(\ref{eq:hybbethe1a}-\ref{eq:hybbethe2}))\begin{equation}
\Delta\left(i\omega_{n}\right)=\tilde{t}^{2}\overline{G_{jj}\left(i\omega_{n}\right)}=\tilde{t}^{2}\int d\varepsilon\frac{P\left(\varepsilon\right)}{i\omega_{n}-\varepsilon+\mu-\Delta\left(i\omega_{n}\right)-\Sigma\left[\varepsilon,i\omega_{n}\right]},\label{eq:disordselfconsist}\end{equation}
where we have slightly modified the notation in order to show that
the denominator on the right-hand side depends on the site-energy
$\varepsilon$ both explicitly and implicitly through the self-energy
$\Sigma\left[\varepsilon,i\omega_{n}\right]$. It is obvious that
the above procedure reduces to the original DMFT in the clean case
(cf. Eq.~(\ref{eq:hybbethe2})) but what does it reduce to in the
non-interacting, disordered case? 

It turns out that the treatment of disordered non-interacting systems
obtained from these equations is equivalent to the so-called Coherent
Potential Approximations (CPA) \shortcite{elliotetal74,EconomouCPA},
which is known to become exact in infinite dimensions \shortcite{vlamingvollhardt92}.
The CPA equations are usually obtained through a strategy that consists
in replacing the effects of scattering off the exact disorder potential
by an effective \emph{average medium.} Formally, one writes the average
Green's function in terms of an average medium self-energy $\Sigma_{AM}\left(i\omega_{n}\right)$
(a frequency-dependent complex quantity) \begin{equation}
\overline{G\left(\boldsymbol{k},\boldsymbol{k}^{\prime},i\omega_{n}\right)}=\frac{\delta_{\boldsymbol{k},\boldsymbol{k}^{\prime}}}{i\omega_{n}-\varepsilon_{\boldsymbol{k}}+\mu-\Sigma_{AM}\left(i\omega_{n}\right)},\label{eq:cpaselfen1}\end{equation}
where again $\varepsilon_{\boldsymbol{k}}$ is the clean non-interacting
dispersion. $\Sigma_{AM}\left(i\omega_{n}\right)$ is calculated by
replacing the average medium by the exact potential (as defined by
the actual values of $\varepsilon_{j}$) at a \emph{single} generic
site (while keeping it at the other sites) and imposing that the difference
between the exact and the average scattering t-matrices vanishes on
the average \shortcite{elliotetal74,EconomouCPA}. A similar, effective
medium approach, incidentally, can be used to derive the DMFT of clean
interacting systems \shortcite{georgesrmp}, so it is no surprise that
one recovers CPA in this case. In the generic case of disordered interacting
systems, $\Sigma_{AM}\left(i\omega_{n}\right)$ is obtained from the
local part of the average Green's function
\begin{eqnarray}
\overline{G_{jj}\left(i\omega_{n}\right)}
&=&\int d\varepsilon\frac{P\left(\varepsilon\right)}{i\omega_{n}-\varepsilon+\mu-\Delta\left(i\omega_{n}\right)-\Sigma\left[\varepsilon,i\omega_{n}\right]}
\nonumber \\
&=&\frac{1}{N_{s}}\sum_{\boldsymbol{k}}\frac{1}{i\omega_{n}-\varepsilon_{\boldsymbol{k}}+\mu-\Sigma_{AM}\left(i\omega_{n}\right)}
.\label{eq:cpaselfen2}
\end{eqnarray}

One may wonder what is the form of the DMFT self-consistency for generic
disordered interacting systems, beyond the Bethe lattice case, in
other words, the analogue of Eq.~(\ref{eq:self-cons}). This is most
easily done through the analogy with CPA. Once we have the local self-energy
for every value of $\varepsilon_{j}$, $\Sigma_{j}\left(i\omega_{n}\right)$,
for a given $\Delta\left(i\omega_{n}\right)$, Eqs.~(\ref{eq:disordgreenloc},\ref{eq:disordselfen}),
we first find the average medium self-energy $\Sigma_{AM}\left(i\omega_{n}\right)$
through Eq.~(\ref{eq:cpaselfen2}). We then note that the average
local Green's function within CPA is also given by\begin{equation}
\overline{G_{jj}\left(i\omega_{n}\right)}=\frac{1}{i\omega_{n}+\mu-\Delta\left(i\omega_{n}\right)-\Sigma_{AM}\left(i\omega_{n}\right)},\label{eq:cpalocalgreen}\end{equation}
since this is what you get if you replace the actual scattering potential
$\varepsilon_{j}+\Sigma_{j}\left(i\omega_{n}\right)$ in Eq.~(\ref{eq:disordselfen})
by the effective medium self-energy $\Sigma_{AM}\left(i\omega_{n}\right)$.
Finally, from comparing Eqs.~(\ref{eq:cpaselfen2}) and (\ref{eq:cpalocalgreen})
we arrive at the desired self-consistency condition\begin{equation}
\frac{1}{N_{s}}\sum_{\boldsymbol{k}}\frac{1}{i\omega_{n}-\varepsilon_{\boldsymbol{k}}+\mu-\Sigma_{AM}\left(i\omega_{n}\right)}=\frac{1}{i\omega_{n}+\mu-\Delta\left(i\omega_{n}\right)-\Sigma_{AM}\left(i\omega_{n}\right)},\label{eq:cpa}\end{equation}
 which would give an improved hybridization function $\Delta\left(i\omega_{n}\right)$
in an iterative procedure. We should note in passing that the average
medium self-energy and the average Green's function (\ref{eq:cpaselfen1})
are the key ingredients in the calculation of the conductivity, which
as mentioned before involves no vertex corrections within DMFT \shortcite{dk-prb94}.

It is important to know what the limitations of this approach are.
The main one is its inability to describe the disorder-induced Anderson
metal-insulator transition \shortcite{anderson58}. As the self-consistency
condition makes quite apparent, the central order parameter of this
mean field theory is the average local Green's function, see Eqs.~(\ref{eq:disordhybbethe1b}),
(\ref{eq:cpaselfen2}) or (\ref{eq:cpa}). However, as explained by
Anderson in the original 1958 paper \shortcite{anderson58}, the average
local Green's function, which is non-critical and finite at the mobility
edge, is unable to signal the phase transition between extended and
localized states. Indeed, the spatial fluctuations of the local Green's
function are so large that its \emph{typical} value is far removed
from the \emph{average} one (see Fig.~\ref{fig:DMFTscheme}b and
c). Thus, DMFT cannot describe the Anderson localization transition
and one needs to go beyond this approximation if Anderson localization
effects are to be incorporated. It had long been known that CPA has
no Anderson transition, so this should not come as a big surprise.
We will show below, however, that one can in fact address the effects
of localization, while at the same time retaining the local description of all correlation
effects.

\subsection{Applications of the disordered DMFT}

\label{sub:dmftappl}

Early applications of the disordered DMFT scheme focused on the phase
diagram of the disordered Hubbard model. In particular, the fate of
the antiferromagnetic phase of the clean model when disorder is introduced
was investigated \shortcite{ulmkeetal95,singhetal98}. Appropriate incorporation
of broken symmetry phases, like antiferromagnetism, requires generalizing
the procedure of Section~\ref{sub:disDMFT} through the introduction
of sub-lattice structure and spin-dependent single particle quantities,
which is quite straightforward and will not be discussed here \shortcite{georgesrmp}.
The main findings are a surprising enhancement of the ordering tendencies
at weak disorder and strong interactions, which was attributed to
a peculiar disorder-induced delocalization effect \shortcite{ulmkeetal95,singhetal98}.
More recently, the phase diagram of the paramagnetic Hubbard model
(suitable for systems with a high degree of magnetic frustration)
has been determined, showing a gradual suppression of the region of
coexistence of metallic and insulating phases found in the clean case
\shortcite{aguiaretal05}.

\subsubsection{Kondo disorder}

\label{sub:Kondodisorder}

\begin{figure}

\begin{centering}
\includegraphics[bb=36bp 36bp 542bp 449bp,scale=0.5]{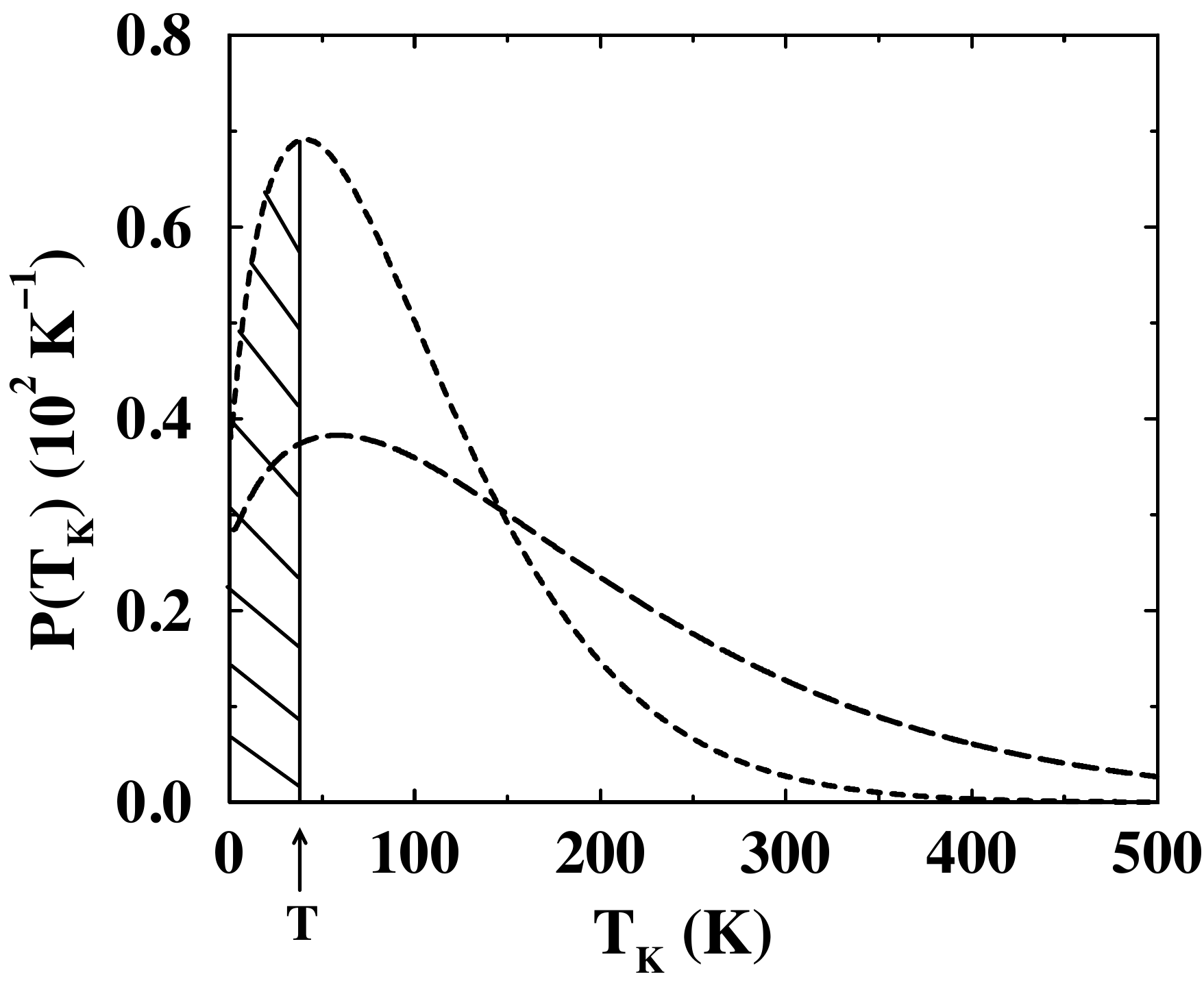}\medskip{}

\par\end{centering}

\centering{}\caption{\label{fig:PTKKondoDisModel} Distribution of Kondo temperatures for
the alloys $\mathrm{UCu_{4}Pd}$ (long dashed line) and $\mathrm{UCu_{3.5}Pd_{1.5}}$
(short dashed line). Spins with $T_{K}$ in the hatched area ($T>T_{K}$)
behave as effectively free and lead to a singular thermodynamic response.
From reference \protect\shortcite{mirandavladgabi1}.}

\end{figure}

A very attractive feature of the disordered DMFT approach is its ability
to provide \emph{full distributions of local quantities}, which in
turn may have profound effects on the low temperature behavior of
physical systems. Consider for example the \emph{ensemble} of effective
actions in Eq.~(\ref{eq:disordeffaction}), which, as explained before,
can be viewed as an \emph{ensemble} of Anderson single-impurity problems,
with the same conduction electron bath (Eq.~(\ref{eq:AndersonBath})),
but different impurity-site energies $\varepsilon_{j}$. As is well
known, at sufficiently strong coupling $U$, a local magnetic moment
can be stabilized at these impurity sites at high temperatures \shortcite{Anderson1961}.
However, below an energy scale set by the Kondo temperature $T_{Kj}$,
the moments are ``quenched'' by the conduction electrons and form
a singlet bound state (or Kondo resonance) \shortcite{Anderson1961,Kondo1964,Yuval-Anderson2,Yuval-Anderson1,Nozi`eres1974,K.G.Wilson1975,Hewson1993}.
The dependence of $T_{Kj}$ on $\varepsilon_{j}$ is given by\begin{equation}
T_{Kj}\approx D\exp\left(-1/\rho_{F}J_{j}\right),\label{eq:kondotemp}\end{equation}
where $D$ and $\rho_{F}$ are the conduction electron half band-width
and density of states at the Fermi level, respectively, and $J_{j}$
is the local Kondo exchange coupling constant \shortcite{Schrieffer1966}\begin{equation}
J_{j}=2\left|V_{k_{F}}\right|^{2}\left[\frac{1}{\left|\varepsilon_{j}\right|}+\frac{1}{\left|\varepsilon_{j}+U\right|}\right].\label{eq:kondocoupling}\end{equation}
Therefore, because of the strong exponential dependence of the Kondo
temperature on the local parameters, a distribution of site energies
can give rise to a wide distribution of Kondo temperatures \shortcite{Dobrosavljevi'c1992b}.
As a consequence, depending on whether a specific site has $T_{Kj}<T$
or $T_{Kj}>T$, it will behave as a free spin in the former case or
as a quenched inert impurity in the latter one, with significant effects
on thermodynamic and transport properties, see Fig.~(\ref{fig:PTKKondoDisModel}).

This scenario has received strong experimental support in the context
of disordered heavy fermion systems. This was initially sparked by
NMR experiments done on the Kondo alloy $\mathrm{UCu_{4-x}Pd_{x}}$,
whose broad temperature-dependent line-widths were analyzed in terms
of a distribution of Kondo temperatures \shortcite{Bernal1995}. The same
distribution was then used to calculate the magnetic susceptibility
and the specific heat, with very good agreement with the observed
behavior \shortcite{Bernal1995}. In this context, this phenomenology has
been dubbed the \emph{Kondo disorder model}. This was particularly
striking because this system was among many intensively studied heavy
fermion compounds \shortcite{stewartNFL,Stewart2006} whose properties
are in apparent contradiction with Landau's theory of Fermi liquids
\shortcite{Landau1957,Landau1957a,Landau1959}. In particular, the magnetic
susceptibility showed an approximately logarithmic divergence with
lowering temperatures, in contrast with the usual saturation to a
constant found in weakly, or even some strongly correlated Fermi liquid
metals. The reason for the observed anomalous behavior was quite clear
within the Kondo disorder model. Indeed, the distribution of Kondo
temperatures needed to explain the NMR line-widths was so broad that
$P\left(T_{K}\right)\approx P_{0}=\mathrm{const.}$ when $T_{K}<\Lambda$,
where $\Lambda$ is some low energy scale of the distribution. In
this case, no matter how low the temperature is, there are always
a few unquenched spins left over with $T_{K}<T$ whose contribution
to the susceptibility is Curie-like and large (Fig.~(\ref{fig:PTKKondoDisModel}))\begin{equation}
\chi\left(T\right)\sim\frac{1}{T}.\label{eq:curie}\end{equation}
Thus, by using a fairly accurate parametrization of the Kondo susceptibility
\shortcite{K.G.Wilson1975}\begin{equation}
\chi_{Kondo}\left(T\right)\sim\frac{1}{T+\alpha T_{K}},\label{eq:kondosusc}\end{equation}
one can immediately find that the bulk susceptibility obtained from
an average over the local contributions calculated with the empirical
$P\left(T_{K}\right)$ distribution is dominated by the low-$T_{K}$
spins (with $T_{K}\ll T\sim\Lambda$) and is logarithmically divergent
\begin{equation}
\overline{\chi}\left(T\right)\sim\int\frac{P\left(T_{K}\right)}{T+\alpha T_{K}}dT_{K}\sim\int_{0}^{\Lambda}\frac{P_{0}}{T+\alpha T_{K}}dT_{K}\sim\ln\left(\frac{T_{0}}{T}\right),\label{eq:Kondodisordsusc}\end{equation}
where $T_{0}$ is a distribution-dependent constant.

Clearly, the Kondo disorder model found a natural setting within the
DMFT approach to disordered systems, which put the phenomenology obtained
from NMR on a firmer basis \shortcite{mirandavladgabi1,mirandavladgabi2,mirandavladgabi3}.
Because heavy fermion systems are characterized by a lattice of ions
with incomplete f-shells, the most appropriate model Hamiltonian is
a disordered Anderson lattice Hamiltonian
\begin{eqnarray}
H_{And}&=&-\sum_{\left\langle ij\right\rangle ,\sigma}t_{ij}c_{i\sigma}^{\dagger}c_{j\sigma}+\sum_{j,\sigma}\left(\varepsilon_{j}-\mu\right)c_{j\sigma}^{\dagger}c_{j\sigma}+\sum_{j,\sigma}\left(E_{fj}-\mu\right)f_{j\sigma}^{\dagger}f_{j\sigma}\nonumber \\
&+&\sum_{j,\sigma}\left(V_{j}c_{j\sigma}^{\dagger}f_{j\sigma}+\mathrm{H.c.}\right)+U\sum_{j}f_{j\uparrow}^{\dagger}f_{j\uparrow}^{\phantom{\dagger}}f_{j\downarrow}^{\dagger}f_{j\downarrow}^{\phantom{\dagger}},\label{eq:disordandersonlattice}
\end{eqnarray}
in usual notation, and in which we have in general assumed that both
f- and c-site energies, $E_{fj}$ and $\varepsilon_{j}$, as well
as the local hybridizations $V_{j}$ between them are random quantities,
each with its own independent distribution. The effective local action
in this case reads\begin{eqnarray}
S_{eff}\left(j\right) & = & \int_{0}^{\beta}d\tau\sum_{\sigma}\left[c_{j\sigma}^{\dagger}\left(\tau\right)\left(\partial_{\tau}+\varepsilon_{j}-\mu\right)c_{j\sigma}\left(\tau\right)+f_{j\sigma}^{\dagger}\left(\tau\right)\left(\partial_{\tau}+E_{fj}-\mu\right)f_{j\sigma}\left(\tau\right)\right]\nonumber \\
 & + & \int_{0}^{\beta}d\tau\int_{0}^{\beta}d\tau^{\prime}\sum_{\sigma}\left[c_{j\sigma}^{\dagger}\left(\tau\right)\Delta\left(\tau-\tau^{\prime}\right)c_{j\sigma}\left(\tau^{\prime}\right)\right]\nonumber \\
 & + & \int_{0}^{\beta}d\tau\left[V_{j}\sum_{\sigma}c_{j\sigma}^{\dagger}\left(\tau\right)f_{j\sigma}\left(\tau\right)+Uf_{j\uparrow}^{\dagger}\left(\tau\right)f_{j\uparrow}\left(\tau\right)f_{j\downarrow}^{\dagger}\left(\tau\right)f_{j\downarrow}\left(\tau\right)\right],\label{eq:disordandeffaction}\end{eqnarray}
and the self-consistency condition is analogous to the one in Eq.~(\ref{eq:disordselfconsist})\begin{equation}
\Delta\left(i\omega_{n}\right)=\tilde{t}^{2}\overline{G_{jj}^{c}\left(i\omega_{n}\right)},\label{eq:disorderselfconsistAndLattModel}\end{equation}
where the local $c$-electron Green's function $G_{jj}^{c}\left(i\omega_{n}\right)$
is\begin{equation}
\left[G_{jj}^{c}\left(i\omega_{n}\right)\right]^{-1}=i\omega_{n}-\varepsilon_{j}+\mu-\frac{V_{j}^{2}}{i\omega_{n}-E_{fj}-\Sigma_{j}\left(i\omega_{n}\right)}-\Delta\left(i\omega_{n}\right),\label{eq:AndLattModelCGreen}\end{equation}
and the averaging procedure is performed over the random quantities
$\varepsilon_{j}$, $V_{j}$ and $E_{fj}$. Thus, $T_{K}$ fluctuations
can have several origins in general, as the local Kondo temperature
is affected by $\varepsilon_{j}$, $E_{fj}$ and $V_{j}$. Within
DMFT, it was possible to better justify the \emph{ad hoc} assumptions
of the Kondo disorder model. In particular, one could quantify the
validity of and thus justify the approximation of calculating the
bulk susceptibility as an average over single-site contributions \shortcite{mirandavladgabi1}.
In addition, good agreement was also found with the dynamic magnetic
susceptibility obtained through neutron scattering experiments \shortcite{Aronson1995}.
Furthermore, going well beyond the simple Kondo disorder phenomenology,
the DMFT approach is able to give direct information about transport
properties. It was found that \shortcite{mirandavladgabi1,mirandavladgabi2,mirandavladgabi3}
\begin{itemize}
\item There is a strong interaction-induced renormalization of the disorder
seen by the conduction electrons.
\item This, in turn, leads to a rapid suppression of the low-temperature
Fermi liquid coherence characteristic of clean heavy fermion materials,
as a function of increasing disorder.
\item Finally, when the quasi-particle coherence is completely destroyed
and the distribution of Kondo temperatures develops a finite intercept
in the limit of $T_{K}\to0$, the non-Fermi liquid thermodynamics
described above is accompanied by a non-Fermi liquid \emph{linear
in $T$ resistivity}\begin{equation}
\rho\left(T\right)=\rho_{0}-AT,\label{eq:Kondodisorresist}\end{equation}
where $A>0$. Like in the case of the thermodynamic properties, the
anomalous resistivity is also due to left-over low-$T_{K}$ free spins
off which the conduction electrons scatter incoherently.
\end{itemize}
It was possible to verify that the self-consistency does not lead
to a large disorder dependence of the hybridization function $\Delta\left(\tau\right)$.
As a result, the distribution of Kondo temperatures is fairly sensitive
to the \emph{bare} distribution of random parameters $\varepsilon_{j}$,
$E_{fj}$ and $V_{j}$. Going beyond DMFT, as we will discuss later,
one finds that this is an artifact of the approximations and, in general,
self-consistency leads to a much more robust dependence on disorder.

\subsubsection{Elastic and inelastic scattering in the disordered Hubbard model}

\label{sub:elastinelastDisHubModel}

The interplay between local correlation effects and transport is a
striking feature which is made almost obvious by the DMFT scheme.
This has been demonstrated in studies of the disordered Hubbard model,
Eq.~(\ref{eq:DisordHubbardHam}), in references \shortcite{Tanaskovi'c2003}
and \shortcite{Aguiar2004}, as we now describe. 

The first study \shortcite{Tanaskovi'c2003} was confined to $T=0$ and
thus addresses only the effects of elastic scattering. This was done
using the Kotliar-Ruckenstein slave boson mean field theory as the
impurity solver \shortcite{kotliarruckenstein} (see Section~\ref{sub:TMTMottAnderson}
for further details). The relevant question is how interactions renormalize
the scattering of quasiparticles by the disorder potential at the
Fermi level (hence at $T=0$). In Hartree-Fock theory, the renormalized
disorder potential is determined by the self-consistently determined
distribution of the electronic charge. This, in turn, is governed
by the charge compressibility if the charge can be assumed to readjust
itself to the disorder potential in a fashion dictated by linear response
theory. For small $U$, the response is that of a good metal and allows
for a flexible adjustment of the charge to the bare random potential,
leading to a weakened renormalized disorder ({}``disorder screening'')
\shortcite{Herbut2001}. For strong interactions close to Mott localization,
however, the charge compressibility is significantly reduced (the
metal becomes increasingly less compressible) and Hartree-Fock theory
predicts poor disorder screening.

It was shown in \shortcite{Tanaskovi'c2003} that indeed the efficient
disorder screening predicted by Hartree-Fock theory is recovered by
DMFT at weak interactions. However, it was found that another phenomenon
intervenes and strong disorder screening does occur even as the system
approaches the Mott transition. The reason why this happens is once
again related to the peculiarities of the Kondo effect discussed above.
Indeed, as several DMFT studies have shown \shortcite{georgesrmp}, the
Mott transition is signaled by the disappearance of the metallic quasiparticles
\shortcite{Brinkman1970}. The coherent nature of these quasiparticles
exists only within a narrow energy range around the Fermi level whose
width is set by the Kondo temperature of the associated single-impurity
problem: as $U\to U_{c}$, $T_{K}\to0$. In the disordered case, there
is a distribution of $T_{K}$'s, but all of them vanish at the transition.
Now, the value assumed by the renormalized disorder potential on a
given site is set by \emph{the position of the local Kondo resonances}
(within DMFT), which are known to be \emph{strongly pinned} to the
Fermi level \shortcite{Hewson1993}. Therefore, Kondo resonance pinning
strongly reduces the bare disorder fluctuations and screens the disorder
rather effectively, leading to a correlation-induced suppression of
the renormalized disorder, in sharp contrast to the simple Hartree-Fock
prediction.

Kondo physics again comes in when one looks at inelastic scattering
\shortcite{Aguiar2004}. This was done by means of iterative perturbation
theory \shortcite{georgeskotliar92,X.Y.Zhang1993,kajuetergabi} (see Section~\ref{sub:statDMFTIPT}
for more details). Indeed, $T_{K}$ governs also the temperature above
which inelastic scattering dominates over elastic scattering. It is
found that a gradual and mild temperature dependence of the resistivity
is observed in the weakly correlated regime $U\approx W\ll D$, which
is reasonably captured by Hartree-Fock theory. As interactions become
of the order of (or larger than) the Fermi energy $U\approx W\gg D$,
however, a sharper temperature dependence sets in. This is due to
the strong suppression of the low-temperature scales of the associated
Kondo impurity problems, which is not well described within Hartree-Fock
theory. 

\begin{figure}
\centering{}\includegraphics[bb=0bp 0bp 687bp 534bp,clip,scale=0.5]{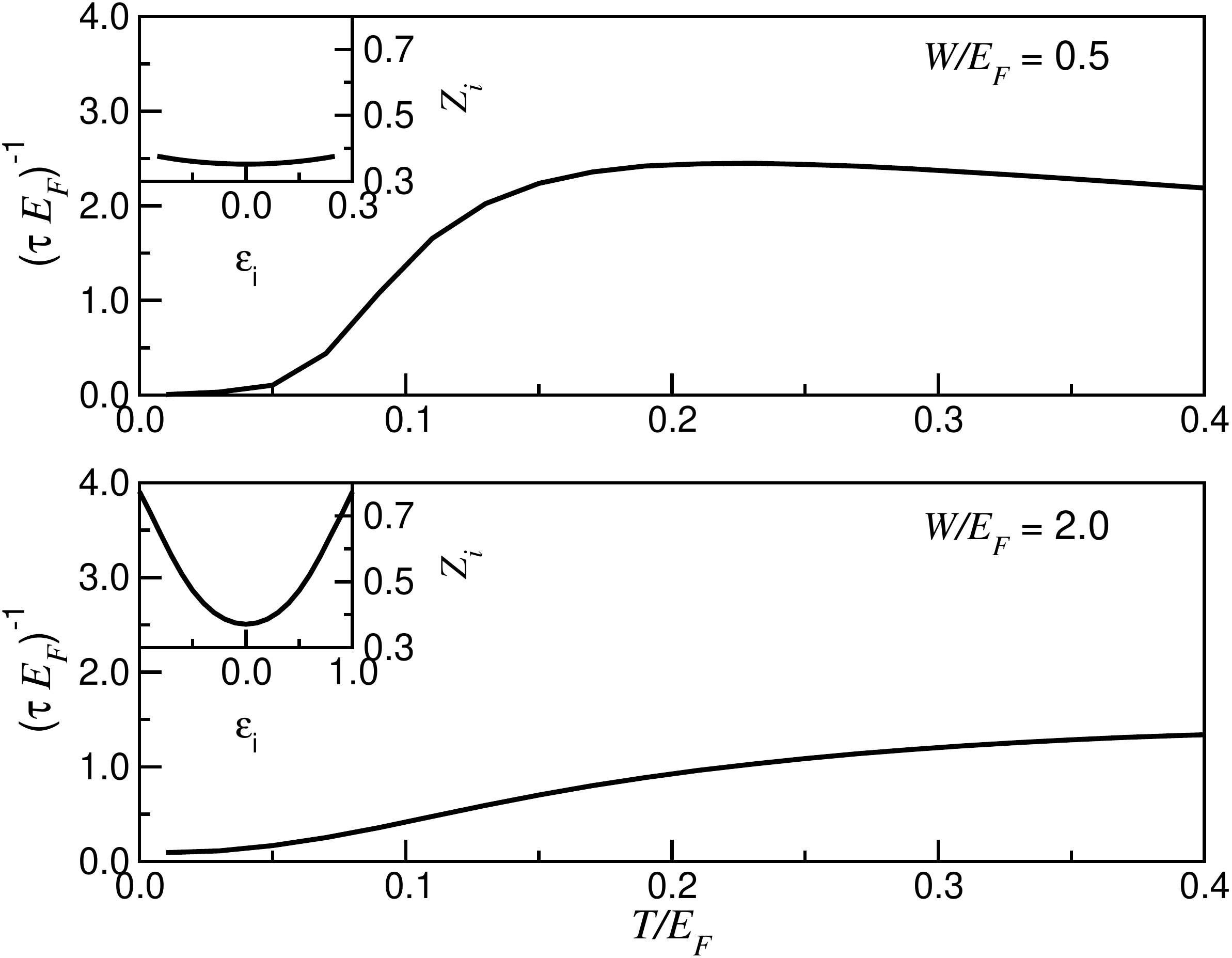}\caption{\label{fig:InelasticCPA} Scattering rate $1/\tau$ as a function
of temperature for the disordered Hubbard model within DMFT at $U=2D\equiv2E_{F}$
for weak ($W=0.5D$, top panel) and strong disorder ($W=2D$, bottom
panel). Insets show the dependence of the local Kondo temperature
$\sim Z_{i}$ on the site energy $\varepsilon_{i}$. At weak disorder,
a sharp distribution of Kondo temperatures gives rise to a steep rise,
whereas stronger disorder leads to a broad distribution and a slow
temperature dependence. From reference \protect\shortcite{Aguiar2004}.}

\end{figure}

Furthermore, varying the disorder strength $W$ at strong interactions
($U\gg D$) leads to vastly different temperature dependences of the
resistivity. Here, the distribution of Kondo temperatures defines
the range over which inelastic processes become progressively more
dominant. When $U\approx W$, a wide distribution of Kondo temperatures
is generated and a rather slow growth of the resistivity with temperature
is found. In contrast, in the cleaner case $W\ll U,$ there is a much
narrower distribution of $T_{K}$'s resulting in a sharp temperature
dependence of the resistivity, typical of the onset of coherence in
heavy fermion materials \shortcite{Stewart1984} (see Fig.~(\ref{fig:InelasticCPA})).

\section{Mott-Anderson transitions: Typical Medium Theory}

\label{sec:beyondDMFT}

Although the dynamical mean field theory described above is able to
capture many features which are expected to be quite independent of
its underlying assumptions (e. g., a distribution of local energy
scales governing both thermodynamic and transport properties), it
is clear its limitations call for improvements at several points.
In particular, it would be highly desirable to incorporate Anderson
localization effects. As we have seen, these are conspicuously absent
in the original DMFT, as the latter is essentially a mean field theory
whose order parameter is $\Delta\left(\omega\right)$, which in turn
is determined by the average local Green's function, see Eq.~(\ref{eq:disordselfconsist})
or (\ref{eq:cpalocalgreen}-\ref{eq:cpa}). As Anderson localization
originates precisely in the spatial fluctuations of this quantity,
this is not enough. Physically, $\Delta\left(\omega\right)$ represents
the \emph{available electronic states} to which an electron can {}``jump''
on its way out of a given lattice site. From Fermi's golden rule,
the transition ({}``escape'') rate to a neighboring site is proportional
to the imaginary part of $\Delta\left(\omega\right)$. If this is
zero at the Fermi energy, the electrons cannot hop out and are effectively
localized. In a clean system, in which this quantity is the same at
every site, this is a good order parameter for localization, as in
the case of the clean Mott transition.
In a highly disordered system, $\Delta\left(\omega\right)$ shows
strong spatial fluctuations from site to site and its average value
is not a good measure of the conducting properties. A {}``typical''
site in an Anderson insulator will have a hybridization function $\Delta_{i}\left(\omega\right)$
with large gaps and a few isolated peaks, reflecting the nearby localized
wave functions that have an overlap with it.
The vanishing of its imaginary part signals the electron's inability
to leave the site and is a good indicator of localized behavior. However,
averaging over the whole sample washes out these gaps hiding the true
insulating behavior.
The discrepancy between the typical and average values of $\Delta_{i}\left(\omega\right)$
persists even on the metallic side, where $\Delta_{typ}\left(\omega\right)$
can be much smaller than $\overline{\Delta\left(\omega\right)}$. 

%



Two alternative routes can be taken at this point. The ideal solution
is to track the actual local hybridization or escape rate at each
site. We will focus on this possibility in Section~\ref{sub:statDMFT},
where we analyze the so-called Statistical Dynamical Mean Field Theory.
The other option is to focus on a simpler, yet meaningful measure
of the escape rate which, though incapable of incorporating the richness
of the actual local realizations, does not ``throw away the (localization)
baby with the bath water''. Here we should take as guidance the remark
by Anderson that ``no real atom is an average atom'' \shortcite{andersonlocrev}
and seek a more apt description of a ``real atom''. Indeed a good
measure of the typical escape rate $\mathrm{Im}\Delta_{typ}\left(\omega\right)$
is the \emph{geometric average} $\exp\left\langle \ln\left[\mathrm{Im}\Delta_{i}\left(\omega\right)\right]\right\rangle $.
This has the great advantage of serving as an order parameter for
the localization transition: indeed, in contrast to the algebraic
average, the geometric average vanishes at the mobility edge \shortcite{anderson58}
(see also \shortcite{M.Janssen1998,A.D.Mirlin2000,schubertetal10}). We
can then reason by analogy with the regular Dynamical Mean Field Theory
approach explained above and construct a self-consistent extension
centered around this \emph{typical} escape rate function. This theory
has been dubbed the Typical Medium Theory (TMT) \shortcite{tmt}.

\subsection{Formulation of the theory}

\label{sub:TMT}

We can proceed by analogy with the DMFT equations as explained in
Section~\ref{sub:disDMFT}, see Eqs.~(\ref{eq:cpaselfen1}-\ref{eq:cpa}),
to obtain the Typical Medium Theory. We again focus on a disordered
Hubbard model, Eq.~(\ref{eq:DisordHubbardHam}), and imagine replacing
the disordered medium by an effective \emph{typical} medium described
by a self-energy function $\Sigma_{TMT}\left(\omega\right)$. How
do we determine this self-energy? Focusing on a generic site $j$,
it is described by an effective action which has the same form as
Eq.~(\ref{eq:disordeffaction}). The hybridization function $\Delta\left(\omega\right)$
is still left unspecified at this point, but we envisage that it will
reflect a ``typical'' site as opposed to an ``average'' one,
so we set $\Delta\left(\omega\right)=\Delta_{typ}\left(\omega\right)$
in Eq.~(\ref{eq:disordeffaction}). The local Green's function is
still defined as in Eqs.~(\ref{eq:disordgreenloc}) and (\ref{eq:disordselfen}).
The local density of states is given by the imaginary part of the
local Green's function\begin{equation}
\rho_{j}\left(\omega\right)=\frac{1}{\pi}\mathrm{Im}G_{jj}\left(\omega-i\delta\right).\label{eq:localDOS}\end{equation}
The \emph{typical} local density of states can be defined through
its geometric average\begin{equation}
\rho_{typ}\left(\omega\right)=\exp\left[\int d\varepsilon_{j}P\left(\varepsilon_{j}\right)\ln\rho_{j}\left(\omega\right)\right].\label{eq:typicalDOS}\end{equation}
Note that we have reverted to the real frequency axis because we need
a positive-definite quantity in order to be able to define a geometric
average. To preserve causality, the typical local Green's function
is obtained through the usual Hilbert transform \begin{equation}
G_{typ}\left(\omega\right)=\int_{-\infty}^{\infty}d\omega^{\prime}\frac{\rho{}_{typ}\left(\omega^{\prime}\right)}{\omega-\omega^{\prime}}.\label{eq:typicallocalgreen}\end{equation}
Note the analogous \emph{average} quantity in the second equality
of Eq.~(\ref{eq:cpaselfen2}), which appears in DMFT. The typical
medium self-energy $\Sigma_{TMT}\left(\omega\right)$ 
is then defined through the inversion of the following
equation\begin{equation}
G_{typ}\left(\omega\right)=\frac{1}{N_{s}}\sum_{\boldsymbol{k}}\frac{1}{\omega-\varepsilon_{\boldsymbol{k}}+\mu-\Sigma_{TMT}\left(\omega\right)},\label{eq:TMTselfen}\end{equation}
which is the analogue of the first equality in Eq.~(\ref{eq:cpaselfen2}).
Finally, the loop is closed by setting\begin{equation}
G_{typ}\left(\omega\right)=\frac{1}{\omega+\mu-\Delta_{typ}\left(\omega\right)-\Sigma_{TMT}\left(\omega\right)},\label{eq:TMTlocalgreen}\end{equation}
which can be used in an iterative scheme to generate an updated hybridization
function $\Delta_{typ}\left(\omega\right)$ and is the analogue of
Eq.~(\ref{eq:cpa}). It becomes clear that the \emph{crucial difference
between TMT and DMFT is the replacement of the average local Green's
function by the typical one} (see Fig.~\ref{fig:DMFTscheme}b and
c). This has been shown to capture even quantitative features of the
Anderson localization transition, as we will discuss below. For a
more comprehensive review, see \shortcite{TMTbook}.

\subsection{Applications of TMT}

\label{sub:applicationsTMT}

We will now describe the most important results obtained from the
TMT theory of disordered systems. We will focus on the non-interacting
case in Section~\ref{sub:criticalnonintTMT} and on the disordered
Hubbard model in Section~\ref{sub:TMTMottAnderson}. We should also
mention a study of the Falicov-Kimball model within TMT in reference
\shortcite{Byczuk2005a}.

\subsubsection{Critical behavior in the non-interacting case}

\label{sub:criticalnonintTMT}

\begin{figure}
\begin{centering}
\includegraphics[clip,scale=0.65]{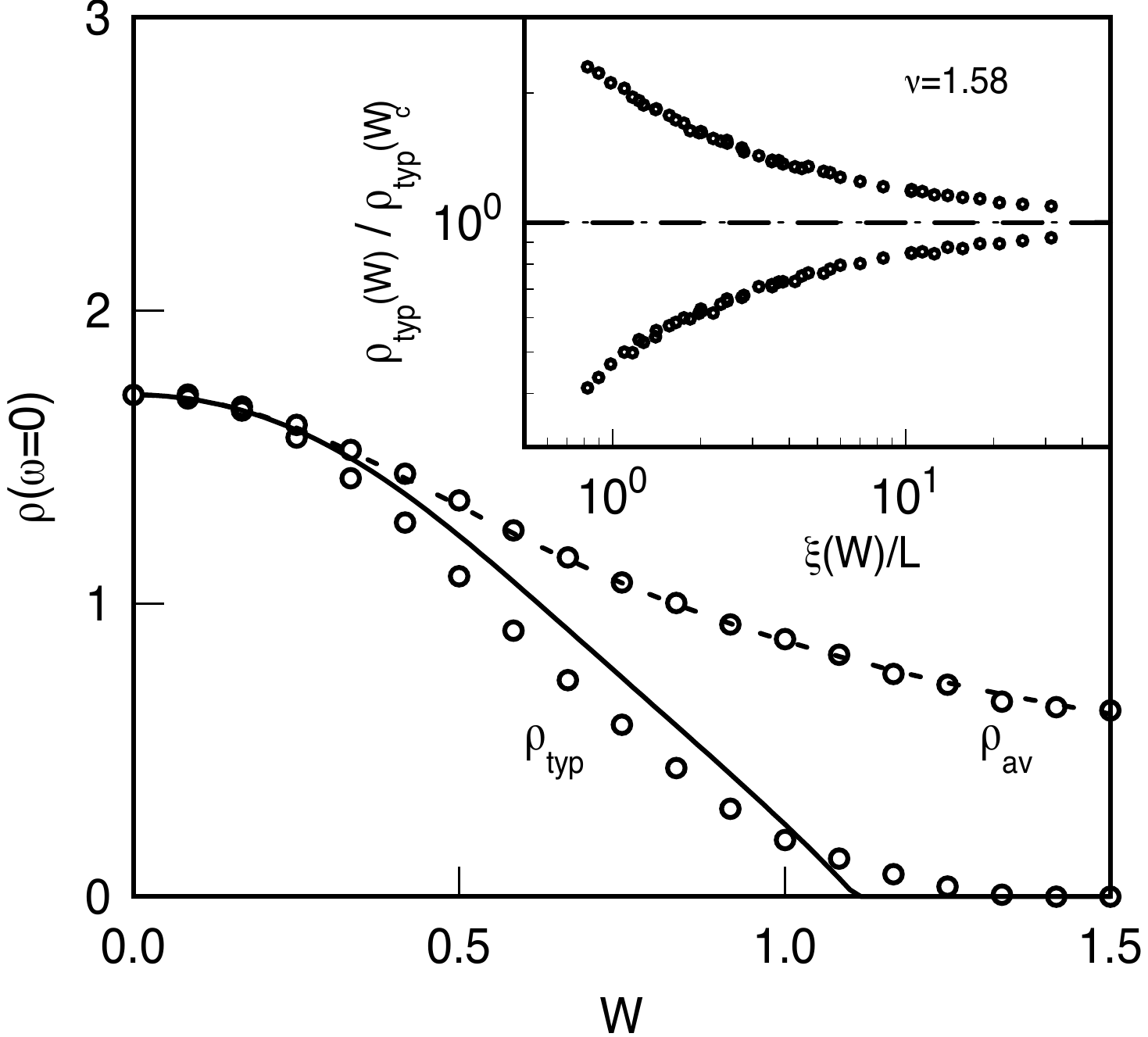}
\par\end{centering}

\caption{\label{fig:TMTXNumerics}Comparison between the TMT and CPA with numerical
results (circles) on the non-interacting 3D tight-binding model with
diagonal disorder. The average local density of states at the band
center $\rho_{av}$ is remarkably well captured by the Coherent Potential
Approximation (dashed line) but is \emph{not critical} at the Anderson
localization transition at $W_{c}\approx1.38$ (in units of the clean
bandwidth). The geometrically averaged local density of states $\rho_{typ}$,
on the other hand, vanishes at the transition and is well described
by the Typical Medium Theory (full line), except for the detailed
critical behavior. The inset shows the numerics for the critical behavior
of $\rho_{typ}$ as a function of the correlation length $\xi\left(W\right)$,
where $\xi\left(W\right)\sim\left|W-W_{c}\right|^{\nu}$, $\nu\approx1.58$.
From reference \protect\shortcite{tmt}.}

\end{figure}

As a first test of the usefulness of the TMT approach, it was first
applied to the non-interacting three-dimensional case, Eq. (\ref{eq:DisordHubbardHam})
with $U=0$ \shortcite{tmt}. In fact, the results of
applying the TMT equations to a cubic lattice were directly compared
to a numerical diagonalization of the Hamiltonian. In particular,
the numerically determined arithmetic and geometric averages of the
local density of states at the Fermi level, $\rho_{av}\left(\omega=0\right)$
and $\rho_{geo}\left(\omega=0\right)$, were compared to the results
of CPA \shortcite{elliotetal74,EconomouCPA} and of TMT (see Fig.~\ref{fig:TMTXNumerics}).
A remarkably accurate agreement between $\rho_{av}\left(\omega=0\right)$
and CPA was observed. As is known, this quantity is not critical at
the Anderson transition. On the other hand, the geometric average
of the local density of states does vanish at a critical disorder
strength $W_{c}$. A reasonably good agreement between the numerical
$\rho_{geo}\left(\omega=0\right)$ and TMT is found for most values
of the disorder strength, even though TMT misses the correct critical
behavior. This is not too surprising as TMT has the flavor of a mean
field theory. Physically, it is clear the $\rho_{geo}\left(\omega\right)$
should be viewed as the density of extended states of system, decreasing
with increasing disorder and eventually vanishing altogether for sufficiently
large randomness. Thus, the spectral weight described by $\rho_{geo}\left(\omega\right)$
is not conserved. 

In fact, further insight into the critical behavior of TMT can be
achieved analytically \shortcite{tmt}. By assuming an
elliptic density of states for the clean lattice, $\rho_{0}\left(\omega\right)=\frac{2}{\pi D}\sqrt{1-\left(\frac{\omega}{D}\right)^{2}}$,
it can be proved that in the critical region $W\to W_{c}=eD$, the
typical density of states (given, as usual, by the geometric average)
assumes a universal form\begin{equation}
\rho_{typ}\left(\omega,W\right)\approx\rho\left(\omega=0,W\right)f\left[\omega/\omega_{0}\left(W\right)\right],\label{eq:TMTrhotypscalingform}\end{equation}
where \begin{equation}
\rho\left(\omega=0,W\right)=\left(\frac{4}{\pi}\right)^{2}\left(W_{c}-W\right),\label{eq:TMTcriticalrhotyp}\end{equation}
the frequency scale \begin{equation}
\omega_{0}\left(W\right)=\sqrt{\frac{e}{4}\left(W_{c}-W\right)},\label{eq:TMTcriticalfreqscale}\end{equation}
 and the scaling function has a simple form $f\left(x\right)=1-x^{2}$.
Note that Eq.~(\ref{eq:TMTcriticalrhotyp}) gives an order parameter
critical exponent $\beta_{TMT}=1$, which should be compared to the
accepted value in three dimensions $\beta_{3D}\approx1.58$ \shortcite{slevinohtsuki99}.

\subsubsection{The disordered Hubbard model at half filling}

\label{sub:TMTMottAnderson}

We now direct our attention to disordered \emph{interacting} systems.
The TMT was applied to the disordered Hubbard model \shortcite{byczuketal05,aguiaretal09}.
The phase diagram of the paramagnetic half-filled case at $T=0$ was
obtained with two different impurity solvers: Wilson's numerical renormalization
group (NRG) \shortcite{K.G.Wilson1975} was used in \shortcite{byczuketal05}
and the Kotliar-Ruckenstein slave boson mean field theory (SB4) \shortcite{kotliarruckenstein}
was applied in \shortcite{aguiaretal09}. The results obtained largely
agree with each order but there are some small discrepancies in the
details. Essentially, three different phases are observed: a disordered
correlated metal phase, characterized by $\rho_{c}\equiv\rho_{typ}\left(\omega=0\right)\neq0$,
a Mott-like insulating phase (which we will call simply a Mott insulator)
and a Anderson-like insulating phase (for which we will use the name
Mott-Anderson insulator), the latter two phases having $\rho_{c}=0$. We will
discuss each set of results and then the discrepancies. For reviews,
see \shortcite{TMTbook} and \shortcite{byczuketal10}.

For small values of disorder and interaction, both approaches find
that the system is metallic, although this is not too surprising.
There is also agreement on the fact that, for a fixed small disorder
strength, the order parameter $\rho_{c}$ increases with increasing
interaction ($U>W$). This is a reflection of disorder screening by
interactions (see Section~\ref{sub:elastinelastDisHubModel}). Eventually,
at a critical value of the interaction strength $U_{c}\left(W\right)$,
the order parameter $\rho_{c}$ exhibits a finite jump and drops to
zero, signaling a metal-Mott insulator phase transition. Furthermore,
both methods agree that, for a fixed small value of $U$, as one increases
the disorder ($W>U$) $\rho_{c}$ decreases monotonically, indicating
that the spectral weight due to extended states is decreasing. At
the critical value $W_{c}\left(U\right)$, $\rho_{c}$ vanishes and
the system enters a Mott-Anderson insulating phase.

The differences in the results of the two approaches are the following.
The NRG-based TMT \shortcite{byczuketal05} predicts the metal-Mott insulator
transition for $U>W$ to be first order in character, with typical
hysteretic behavior: the metallic solution is locally stable for $U<U_{c2}\left(W\right)$
and the Mott-insulating solution is locally stable for $U>U_{c1}\left(W\right)$,
where $U_{c1}\left(W\right)<U_{c2}\left(W\right)$. In the coexistence
region $U_{c1}\left(W\right)<U<U_{c2}\left(W\right)$, both solutions
can be stabilized (although only one is a true global energy minimum
at each $U$). The SB4-based TMT \shortcite{aguiaretal09}, on the other
hand, predicts no hysteresis. To understand this, it should be mentioned
that SB4 is a good description of the low-energy properties of the
impurity spectral function, although it misses the higher-energy features
which give rise to the Hubbard bands in the lattice. An important
ingredient of this description is the quasiparticle weight $Z_{i}\equiv Z\left(\varepsilon_{i}\right)$,
well known from Fermi liquid theory, which in the impurity problem
\nobreakdash- Eq.~(\ref{eq:disordeffaction}) \nobreakdash- determines
the width of the Kondo resonance (essentially the Kondo temperature).
Formally, it appears in the local Green's function at site $i$\begin{equation}
G_{ii}\left(i\omega_{n}\right)=\frac{Z_{i}}{i\omega_{n}-\tilde{\varepsilon}_{i}+\mu-Z_{i}\Delta_{typ}\left(i\omega_{n}\right)},\label{eq:SB4localgreen}\end{equation}
where $\tilde{\varepsilon}_{i}$ is the renormalized local site energy,
which gives the position of the resonance. In the clean lattice case,
$Z$ vanishes continuously as the interaction strength is tuned to
its critical value $U\to U_{c}$, $Z\sim U_{c}-U\to0$, signaling
the Mott localization of the itinerant electrons, which become localized
magnetic moments. The continuous nature of the transition in the clean
case, with no accompanying hysteresis, survives the introduction of
disorder within the SB4-based TMT. In fact, in this approach all the
$Z_{i}$'s vanish at a unique $U_{c}\left(W\right)$. It should be
stressed that in both approaches $\rho_{c}$ is discontinuous at the
transition, a fact which can be ascribed (at least for small disorder)
to the observed perfect screening of disorder ($\rho_{av}\left(\omega=0\right)\to\rho_{geo}\left(\omega=0\right)$
as $U\to U_{c}\left(W\right)$) and the pinning of the clean density
of states at the Fermi level to its non-interacting value (see Section~\ref{sub:elastinelastDisHubModel}).

\begin{figure}
\begin{centering}
\includegraphics[clip,scale=0.45]{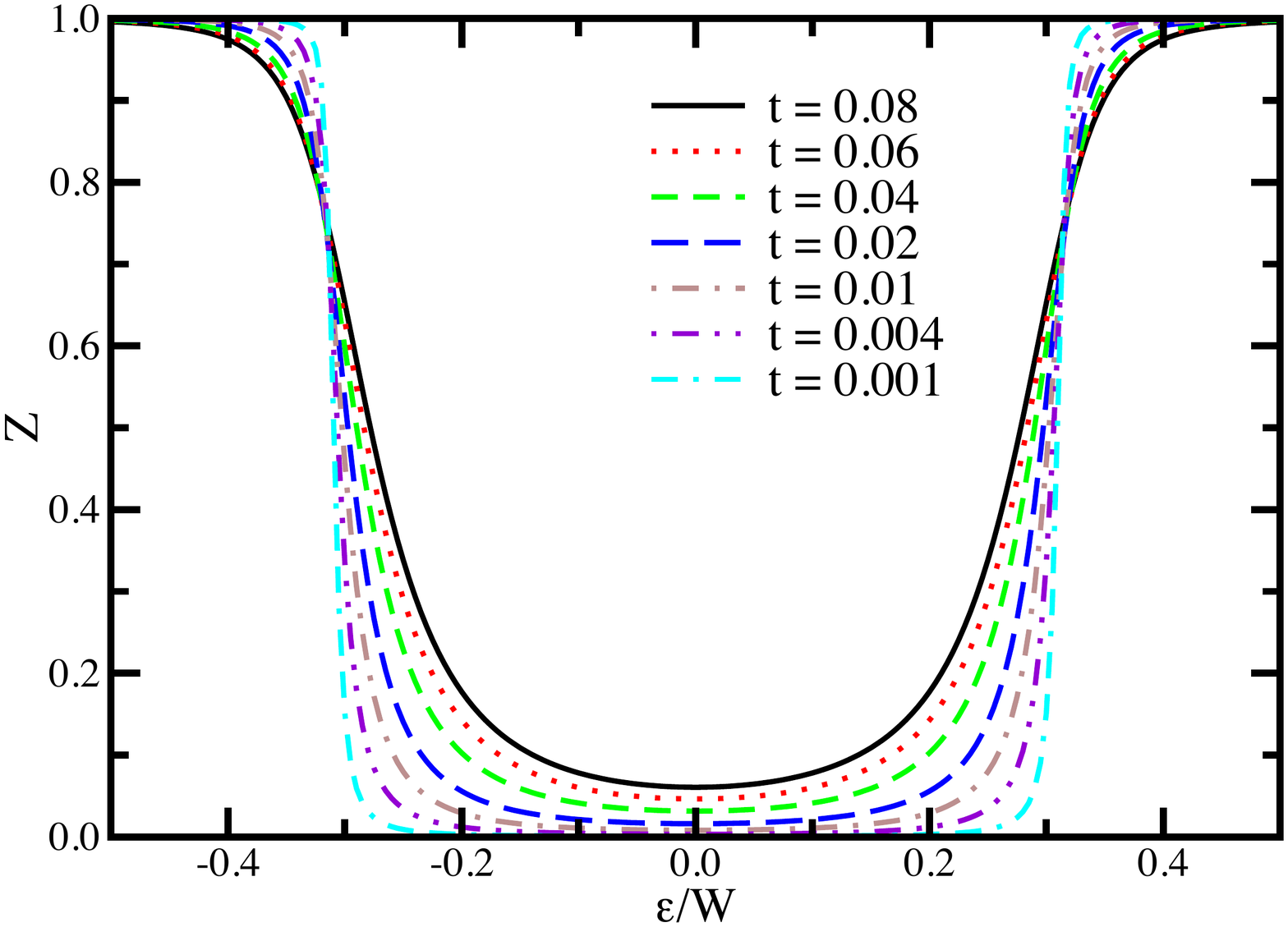}
\par\end{centering}

\caption{\label{fig:TMT2fluids}$Z\left(\varepsilon_{i}\right)$ showing the
two-fluid behavior of the Mott-Anderson transition within TMT. As
the transition is approached ($t\to0$) for $W=2.8>U=1.75$, sites
with $\left|\varepsilon_{i}\right|<U/2\approx0.31W$ represent Mott
insulating regions with $Z\left(\varepsilon_{i}\right)\to0$, whereas
sites with $\left|\varepsilon_{i}\right|>U/2$ are Anderson localized
with $Z\left(\varepsilon_{i}\right)\to1$. From reference \protect\shortcite{aguiaretal06}.}

\end{figure}

Moreover, the results obtained with the NRG impurity solver indicate
that the transition in the region $W>U$ is such that $\rho_{c}\to0$
continuously as $W\to W_{c}\left(U\right)$ \shortcite{byczuketal05}.
This is in contrast to the SB4-based approach, which finds that $\rho_{c}$
exhibits a discontinuous jump to zero at $W_{c}\left(U\right)$ \shortcite{aguiaretal09}.
In fact, the latter method brings out an important ingredient which
significantly enhances the physical understanding of the TMT approach
to this problem. This is achieved by tracking the behavior of the
quasiparticle weights $Z_{i}$. For a given fully-converged hybridization
function $\Delta_{typ}\left(i\omega_{n}\right)$, the ensemble of
impurity problems is characterized by the function $Z\left(\varepsilon_{i}\right)$
for $\left|\varepsilon_{i}\right|<W/2$ (a uniform disorder distribution
is assumed). This function has the property that, as the phase transition
is approached, $Z\left(\varepsilon_{i}\right)\to0$ for $\left|\varepsilon_{i}\right|<U/2$,
whereas $Z\left(\varepsilon_{i}\right)\to1$ if $\left|\varepsilon_{i}\right|>U/2$
(see Fig.~\ref{fig:TMT2fluids}) \shortcite{aguiaretal06}. In the SB4
language \shortcite{kotliarruckenstein}, $Z\to0$ implies a singly occupied
site with a localized magnetic moment, whereas $Z\to1$ means either
a doubly occupied or a singly occupied site, either of which is essentially
non-interacting. Thus, in the region $W>U$, a fraction of the sites,
those with $\left|\varepsilon_{i}\right|<U/2$, experience Mott localization,
while those sites with $\left|\varepsilon_{i}\right|>U/2$ undergo
Anderson localization. The picture that emerges is that of a \emph{spatially
inhomogeneous system}, composed of Mott-localized droplets intermingled
with Anderson insulating regions. This situation has been dubbed a
{}``site-selective Mott transition'' \shortcite{aguiaretal09}. Analytical
insight into the SB4-results can be brought to bear in order to show
that in that approach any finite $U$ renders the vanishing of $\rho_{c}$
\emph{discontinuous}, in sharp contrast to the non-interacting case
\shortcite{tmt}.

Finally, in the intermediate region of $W\approx U$, it was suggested,
based on the NRG results that there might be a crossover from a metal
to a disordered Mott insulator \shortcite{byczuketal05}. However, since
their results clearly show regions where $\rho_{c}\neq0$ and regions
where $\rho_{c}=0,$ we believe the correct interpretation is to identify
the former as metallic and the latter as insulating. Besides, this
is expected from the sharp distinction between an insulator and a
metal at zero temperature. Having said that, however, it is clear
that the nature of the transition in this problematic region certainly
deserves a further more detailed investigation.

The disordered Hubbard model was also studied within TMT by allowing
antiferromagnetic order on a bipartite lattice \shortcite{byczuketal09}.
The phase diagram at zero temperature as a function of disorder and
interactions was determined, with the identification of both paramagnetic
and antiferromagnetic metallic phases (for finite disorder only),
an antiferromagnetic Mott insulating phase and a paramagnetic Anderson
insulating phase at large disorder strength.

\section{Mott-Anderson transitions: Statistical DMFT}

\label{sub:statDMFT}

As outlined above, the most natural and accurate extension of the
DMFT philosophy which can incorporate Anderson localization effects
is one which replaces the algebraic average of the hybridization function
$\overline{\Delta\left(\omega\right)}$ (as in the original DMFT)
or its typical value/geometric average $\Delta_{typ}\left(\omega\right)$
(as in the TMT), by the actual realizations of $\Delta_{j}\left(\omega\right)$
at each site \shortcite{motand,london} (see
Fig.~\ref{fig:DMFTscheme}b and c). As is to be expected, the complexity
of the equations increases considerably and one has to rely heavily
on numerical computations. However, many insights have been obtained
regarding the systems analyzed. Besides, a much larger degree of universality
of the distributions is observed when compared with the much more
{}``rigid'' approaches of DMFT or TMT.

\subsection{Formulation of the theory}

\label{sub:statDMFTformulation}

Let us examine how the statDMFT works. We begin with the by now familiar
disordered Hubbard model of Eq.~(\ref{eq:DisordHubbardHam}) and
focus, as usual, on the dynamics of a given site $j$, dictated by
an effective action (we now revert back to imaginary time) \begin{eqnarray}
S_{eff}\left(j\right) & = & \sum_{\sigma}\int_{0}^{\beta}d\tau c_{j\sigma}^{\dagger}\left(\tau\right)\left(\partial_{\tau}+\varepsilon_{j}-\mu\right)c_{j\sigma}\left(\tau\right)\nonumber \\
 & + & \sum_{\sigma}\int_{0}^{\beta}d\tau\int_{0}^{\beta}d\tau^{\prime}c_{j\sigma}^{\dagger}\left(\tau\right)\Delta_{j}\left(\tau-\tau^{\prime}\right)c_{j\sigma}\left(\tau^{\prime}\right)\nonumber \\
 & + & U\int_{0}^{\beta}d\tau n_{j\uparrow}\left(\tau\right)n_{j\downarrow}\left(\tau\right).\label{eq:effactionstatDMFT}\end{eqnarray}
Notice the crucial difference now: the hybridization function $\Delta_{j}\left(i\omega_{n}\right)$
is now \emph{site-dependent.} Each site, besides having a different
energy $\varepsilon_{j}$, also {}``sees'' a different local environment
$\Delta_{j}\left(i\omega_{n}\right)$. The local dynamics is again
encoded in a site-dependent self-energy $\Sigma_{j}\left(i\omega_{n}\right)$,
obtained from the local Green's function as before, see Eqs.~(\ref{eq:disordgreenloc}) and \ref{eq:disordselfen}).
It is important to note that this description assumes a \emph{diagonal}
(albeit site-dependent) self-energy function, adhering to the generic
philosophy of incorporating only \emph{local} interaction effects.

Unlike the previous DMFT or TMT approaches, statDMFT does not try
to mimic this self-energy function $\Sigma_{j}\left(i\omega_{n}\right)$
through any type of effective medium. Instead, we choose to include
its full spatial fluctuations. We do so by appealing to the physical
picture of the self-energy as a \emph{shift of the local site energy},
albeit a complex, frequency-dependent one: $\varepsilon_{j}\to\varepsilon_{j}+\Sigma_{j}\left(i\omega_{n}\right)$.
Single-particle propagation can thus be viewed as described by the
effective resolvent\begin{equation}
\widehat{G}\left(i\omega_{n}\right)=\left[i\omega_{n}\widehat{1}-\widehat{t}-\widehat{\varepsilon}-\widehat{\Sigma}\left(i\omega_{n}\right)\right]^{-1},\label{eq:resolvent}\end{equation}
where $\widehat{1}$ is the identity operator, $\widehat{t}$ and
$\widehat{\varepsilon}$ are, respectively, the hopping and site-energy
terms (first and second terms of the Hamiltonian~(\ref{eq:DisordHubbardHam})),
and the matrix elements of the self-energy operator $\widehat{\Sigma}\left(i\omega_{n}\right)$
are given in the site basis as\begin{equation}
\left\langle i\left|\widehat{\Sigma}\left(i\omega_{n}\right)\right|j\right\rangle =\Sigma_{j}\left(i\omega_{n}\right)\delta_{ij}.\label{eq:selfenoperator}\end{equation}
Although any matrix element of the resolvent~(\ref{eq:resolvent}),
both intra- and inter-site, can in principle be calculated (which
is important, for example, for a Landauer-type calculation of the
conductivity), the self-consistency requires only the diagonal part,
related to the local Green's function\begin{equation}
\left\langle j\left|\widehat{G}\left(i\omega_{n}\right)\right|j\right\rangle =G_{jj}\left(i\omega_{n}\right)=\frac{1}{i\omega_{n}-\varepsilon_{j}-\Delta_{j}\left(i\omega_{n}\right)-\Sigma_{j}\left(i\omega_{n}\right)}.\label{eq:selfconsstatDMFT}\end{equation}
This last equation closes the self-consistency loop by providing,
in an iterative scheme, an updated hybridization function \emph{for
each site} $\Delta_{j}\left(i\omega_{n}\right)$. We summarize the
self-consistency loop for completeness:
\begin{enumerate}
\item For a given realization of disorder, $\varepsilon_{j}$, start from
a set of {}``initial trial'' hybridization functions $\Delta_{j}\left(i\omega_{n}\right)$,
one for each site. 
\item For the set of effective actions (\ref{eq:effactionstatDMFT}), calculate
the local Green's function $G_{jj}\left(i\omega_{n}\right)$ (Eq.~(\ref{eq:disordgreenloc}))
and the local self-energy $\Sigma_{j}\left(i\omega_{n}\right)$ (Eq.~(\ref{eq:disordselfen}))
for each site.
\item Invert the matrix resolvent (\ref{eq:resolvent}) and get its diagonal
elements $G_{jj}\left(i\omega_{n}\right)$.
\item Obtain an updated set of hybridization functions $\Delta_{j}\left(i\omega_{n}\right)$
by equating these diagonal elements to the expression of the local
Green's functions, Eq.~(\ref{eq:selfconsstatDMFT}).
\end{enumerate}
In general, the set of Eqs.~(\ref{eq:effactionstatDMFT}-\ref{eq:selfconsstatDMFT})
forms the so-called Statistical Dynamical Mean Field Theory of disordered
correlated electron systems. It should be noted that, besides the
challenge of solving the \emph{ensemble} of impurity problems represented
by Eq.~(\ref{eq:effactionstatDMFT}), Eq.~(\ref{eq:selfconsstatDMFT})
poses the numerical problem of inversion of a complex matrix for each
value of the frequency, which can be very time-consuming. The pay-off
is a description which incorporates all Anderson localization effects.
Indeed, when interactions are turned off, the theory becomes exact,
since, e. g., Eq.~(\ref{eq:resolvent}) becomes the exact single-particle
Green's function (from which transport properties can be obtained
with the Landauer formalism). In the absence of randomness, we recover,
of course, the DMFT equations. In the presence of both disorder and
interactions, this is the optimal theory of disordered interacting
lattice fermions which includes only \emph{local} correlation effects.

\subsection{Early implementations of statDMFT: the Bethe lattice}

\label{sub:statDMFTBethe}

It had long been known that the Bethe lattice (or {}``Cayley tree'')
leads to considerable simplifications of the treatment of non-interacting
disordered systems. For example, the so-called self-consistent theory
of localization of Abou-Chacra, Anderson and Thouless \shortcite{abouetal}
becomes exact on a Bethe lattice. This is so because the local Green's
function with a neighboring site removed, Eq.~(\ref{eq:greenremov}),
satisfies a single compact stochastic equation in that lattice, which
allows for quite an efficient analysis, both analytically and numerically.
It was shown, for example, that for a coordination number $z>2$,
there is always an Anderson transition at a non-zero critical disorder
strength $W_{c}$. This transition has been extensively studied \shortcite{mirlinfyodorov}
and is regarded as a large-dimensionality limit of the Anderson transition.
Although it has an anomalous critical behavior, with an exponential
rather than a power-law dependence, this description has been very
fruitful, specially if one is interested in the non-critical region.

Given the complexity of the full statDMFT equations, this suggested
that the preliminary investigations could be carried out on a Bethe
lattice. Here, it is appropriate to comment on what is, in our view,
a misunderstanding of the conceptual basis of the statDMFT approach.
It has been stated \shortcite{semmleretal10b} that there is somehow a
\emph{conceptual} difference between the statDMFT as applied to the
Bethe lattice and the statDMFT used to analyze realistic lattices,
like the square or cubic ones (for these, see the later Section~\ref{sub:statDMFTreallattices}).
The misunderstanding comes from assuming that whereas on a realistic
lattice one has a fixed disorder realization, which is then solved
by statDMFT, therefore defining a \emph{deterministic} problem, on
the infinite Bethe lattice one does not deal with fixed disorder realizations
but with distributions, and the approach becomes \emph{non-deterministic}
and {}``statistical'' (we realize the origin of the misunderstanding
may have been the use of this word in the name of the method \shortcite{london}).
This distinction is unfounded if one realizes that on \emph{any} infinite
lattice with a single fixed disorder realization (with random, spatially
uncorrelated, parameters), the infinite values of local quantities
(such as the local density of states) on each lattice site give rise
to a \emph{statistical distribution of local quantities} (assuming
the lattice-translational invariance of the distributions, see \shortcite{lifshitzbook}
for a careful discussion). Thus, when one solves a given Hamiltonian
with statDMFT on an infinite Bethe lattice, one is actually solving,
in practice as well as in principle, for a single fixed disorder realization.
Each iteration of the method outlined in \shortcite{abouetal} corresponds
to {}``going outwards'' on the branches of a fixed disorder realization
of an infinite Bethe lattice, while at the same time accumulating
random variables and building a histogram of local quantities. By
the same token, if one could solve the statDMFT equations for a single
disorder realization on an \emph{infinite} realistic (say, square)
lattice, each one of its sites would contribute one random variable
for a histogram of the same local quantities (which would be, obviously,
different from the ones obtained from the differently connected Bethe
lattice). In practice, of course, one solves many disorder realizations
of \emph{finite}, hopefully large, realistic lattices in order to
generate distributions with good statistics. However, fundamentally,
there is no \emph{conceptual} difference between the statDMFT solutions
on the two types of lattices.

\subsubsection{The Mott-Anderson transition}

\label{sub:statDMFTMottAnderson}

In \shortcite{motand,london}, the first implementation
of the statDMFT theory, as applied to the Mott-Anderson transition
described by the disordered Hubbard model was completed. This was
done by using a Fermi liquid parametrization of the associated zero-temperature
impurity problems, namely, the infinite-$U$ slave boson mean-field
theory \shortcite{N.Read1983,colemanlong}. Like its Kotliar-Ruckenstein
finite-$U$ counterpart \shortcite{kotliarruckenstein}, this theory captures
the low energy sector and is known to give a quantitatively good description
of this limit. This Fermi liquid description is encapsulated in just
two parameters: the quasiparticle weight $Z_{j}$ and the effective
level energy (or Kondo resonance location) $\tilde{\varepsilon}_{j}$.
The local self-energy is written as \begin{equation}
\Sigma_{j}\left(i\omega_{n}\right)=\left(1-Z_{j}^{-1}\right)i\omega_{n}+\frac{\tilde{\varepsilon}_{j}}{Z_{j}}-\varepsilon_{j}+\mu,\label{eq:slavebosonselfen}\end{equation}
 leading to a local Green's function\begin{equation}
G_{j}\left(i\omega_{n}\right)=\frac{Z_{j}}{i\omega_{n}-\tilde{\varepsilon}_{j}-Z_{j}\Delta_{j}\left(i\omega_{n}\right)}.\label{eq:slavebosongreenloc}\end{equation}
The results of \shortcite{motand,london} reveal
that, as in the non-interacting case, the Anderson-Mott transition
can be identified by the vanishing of the typical local density of
states (as described by the geometric average) at a certain critical
disorder strength. Interestingly, in contrast to non-interacting electrons,
the critical behavior is conventional (power-law) and the average
density of states is \emph{divergent}. There is at present no good
understanding of this divergence. 

Furthermore, a great opportunity afforded by statDMFT is the ability
to investigate \emph{distributions} of local quantities. By tracking
the distribution of quasiparticle weights, it was found that it broadens
considerably with increasing disorder, showing a characteristic\emph{
power-law form} at large randomness\begin{equation}
P\left(Z\right)\sim Z^{\alpha-1},\label{eq:powerlawZ}\end{equation}
with the exponent $\alpha=\alpha\left(W\right)$ a smooth function
of disorder. It should be remembered that $Z_{j}$ determines the
local Kondo temperature $T_{Kj}\sim Z_{j}$ and thus governs the local
contribution to thermodynamic quantities such as the magnetic susceptibility
and specific heat (see Section~\ref{sub:Kondodisorder}). In fact,
by averaging over this distribution of $Z$'s as in Eq.~(\ref{eq:Kondodisordsusc}),
one finds power-law dependences for these quantities as well\begin{equation}
\chi\left(T\right)\sim C\left(T\right)/T\sim T^{\alpha-1}.\label{eq:griffsuscandspecheat}\end{equation}
As disorder increases, $\alpha$ decreases and eventually becomes
smaller than 1 well before the Mott-Anderson transition. When this
happens, the thermodynamic response becomes singular and non-Fermi
liquid-like\begin{equation}
\chi\left(T\to0\right)\to\infty.\label{eq:NFLsusc}\end{equation}

\begin{figure}
\begin{centering}
\includegraphics[bb=0bp 0bp 534bp 438bp,clip,scale=0.6]{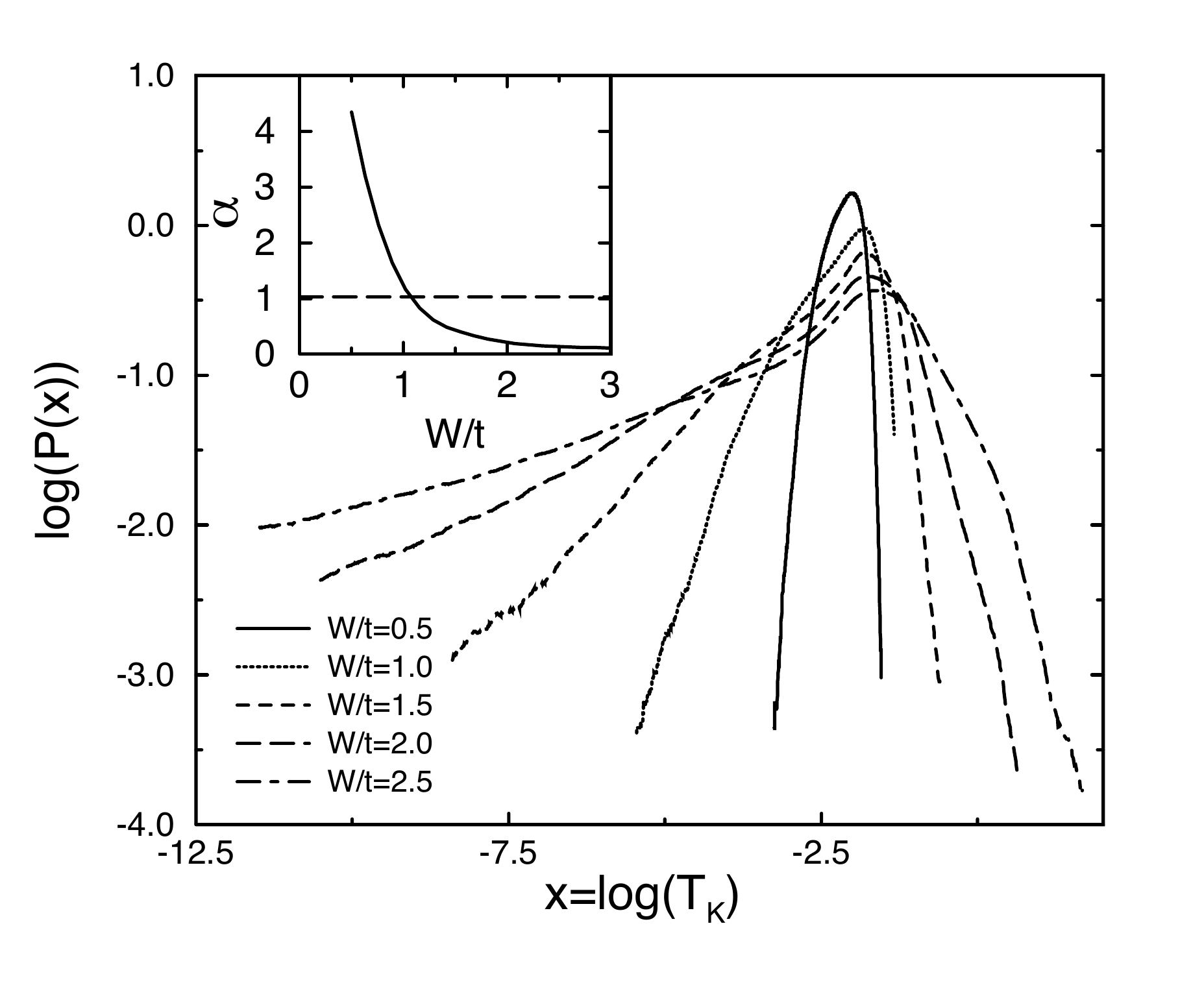}
\par\end{centering}

\caption{\label{fig:powerlawKondotemp}Power-law distributions of Kondo temperatures
in a disordered Anderson lattice model for different levels of disorder:
the linear behavior for small values of $\log T_{K}$ implies a power-law
$P\left(T_{K}\right)\sim T_{K}^{\alpha-1}$. The inset shows the exponent
$\alpha$ as a function of disoder $W$. From reference \protect\shortcite{Miranda2001}.}

\end{figure}

Many different correlated systems have indeed been shown to exhibit
this form of anomalous behavior \shortcite{stewartNFL}, with non-universal
exponents $\alpha$. The presence of non-universal, smoothly varying
exponents characterizing divergences in physical quantities is reminiscent
of a large class of disordered systems and is usually dubbed a quantum
Griffiths phase (for reviews, see \shortcite{Miranda2005,Vojta2006}),
by analogy with a similar situation in classical systems first analyzed
by Griffiths \shortcite{griffiths}. Most other known examples of quantum
Griffiths phases had been found in the vicinity of magnetic phase
transitions in the presence of disorder, most notably in insulating
magnets \shortcite{D.S.Fisher92,D.S.Fisher95,guo-bhatt-huse-prb96,pichetal98,motrunichetal01},
but also in metallic systems \shortcite{castronetoetal1,M.C.deAndrade1998,castroneto-jones-prb00,Millis2002}.
Here, however, the characteristic power laws are found in the vicinity
of the paramagnetic Anderson metal-insulator transition and hence
the name Electronic Griffiths phase was adopted. This anomalous behavior
is apparently not at all dependent on the particular details of the
disordered Hubbard model. Very similar power-law distributions of
Kondo temperatures were also found in Bethe lattice implementations
of statDMFT for the disordered Anderson lattice Hamiltonian (\ref{eq:disordandersonlattice})
(see Fig.~(\ref{fig:powerlawKondotemp})) \shortcite{Miranda1999,Miranda2001,M.C.O.Aguiar2003}.

It should be remembered that the forms of the distributions of Kondo
temperatures obtained within DMFT were strongly dependent on the shape
of the bare distributions of parameters. This is in sharp contrast
to the ubiquitous power-law distributions found in the statDMFT approach.
Thus, although the exponent $\alpha$ is disorder-dependent and non-universal,
the power-law \emph{shape} is quite independent of whether the bare
parameters are given by, say, uniform, Gaussian or binary distributions
\shortcite{M.C.O.Aguiar2003}. This is again easy to understand if we note
that the local Kondo temperature depends exponentially on the local
density of states at the Fermi level (and less strongly on its value
at higher energies). Now, due to the extended nature of the electronic
wave functions in metallic systems, the density of states at one site
is influenced by spatial fluctuations at very distant sites and thus
samples a great number of local environments. The resulting distributions
of local quantities thus reflect this long-distance sampling. 

In order to understand why this effect leads \emph{specifically} to
a power law, an effective model was proposed in \shortcite{tanaskovicetal04}.
The effective model consisted of a disordered Anderson lattice model
with Gaussian distributed conduction electron disorder ($\varepsilon_{j}$
in Eq.~(\ref{eq:disordandersonlattice})) \begin{equation}
P\left(\varepsilon_{j}\right)=\frac{1}{\sqrt{2\pi}W}\exp\left(-\varepsilon_{j}^{2}/2W^{2}\right),\label{eq:gaussiancondelecdisorder}\end{equation}
treated within DMFT. In DMFT, the hybridization function $\Delta\left(i\omega_{n}\right)$
is \emph{site-in\-de\-pen\-dent} and the Kondo temperature distribution
is solely determined by fluctuations of $\varepsilon_{j}$. It can
then be shown that \shortcite{tanaskovicetal04} \begin{equation}
T_{Kj}=T_{K}^{0}e^{-\lambda\varepsilon_{j}^{2}},\label{eq:kondotemptoymodel}\end{equation}
where $T_{K}^{0}$ is the Kondo temperature at $\varepsilon_{j}=0$
and $\lambda$ is determined by other model parameters but does not
depend on $\varepsilon_{j}$. It is easy to show from Eqs.~(\ref{eq:gaussiancondelecdisorder})
and (\ref{eq:kondotemptoymodel}) that\begin{equation}
P\left(T_{K}\right)=P\left[\varepsilon_{j}\left(T_{K}\right)\right]\left|\frac{d\varepsilon_{j}}{dT_{K}}\right|\sim T_{K}^{\alpha-1},\label{eq:powerlawdistkondotemptoy}\end{equation}
which is precisely the power-law distribution of Kondo temperatures
found generically \emph{within statDMFT treatments}. We note, in passing,
that this kind of argument is generic to all known quantum Griffiths
phases: the relevant energy scales are exponentially suppressed by
a certain random parameter (see Eq.~(\ref{eq:kondotemptoymodel})),
whose probability is in turn also exponentially small (see Eq.~(\ref{eq:gaussiancondelecdisorder}))
\shortcite{Miranda2005,Vojta2006}. 

What is the relation between these results, obtained within DMFT,
and the power laws observed in the applications of statDMFT? In statDMFT,
the single (average) hybridization function of DMFT gets replaced
by a strongly fluctuating distribution of local hybridizations $\Delta_{j}\left(i\omega_{n}\right)$.
The imaginary part of each of these functions describes the available
density of states for Kondo screening at site $j$ and enters the
expression for the local Kondo temperature much like $\rho_{F}$ does
in Eq.~(\ref{eq:kondotemp}). From the central-limit theorem, fluctuations
of the available densities of states around the mean value are generically
Gaussian for weak and intermediate disorder and lead to a power-law
distribution of $T_{K}$'s in a fashion quite similar to the effective
model. Indeed, detailed calculations showed that the effective model
is quite accurate when compared with full statDMFT results \shortcite{tanaskovicetal04}.
Crucially, these arguments can be used to show that the non-Fermi
liquid behavior occurs already at quite moderate values of disorder
and \emph{strictly precedes} the Anderson metal-insulator transition.
This elucidates then the microscopic origin of the electronic Griffiths
phase.

\subsubsection{Iterative perturbation theory as impurity solver}

\label{sub:statDMFTIPT}

Most of the early statDMFT results on the Bethe lattice were obtained
through the use of the infinite-U slave boson mean-field theory \shortcite{N.Read1983,colemanlong}
as impurity solver. These are good descriptions of the low-energy
coherent Fermi-liquid part of the impurity spectrum but fail to account
for inelastic scattering at low energies as well as higher-energy
incoherent features such as upper and lower Hubbard bands. A technique
which is able to incorporate there features is the so-called iterative
perturbation theory \shortcite{georgeskotliar92,X.Y.Zhang1993,kajuetergabi}.
The iterative perturbation theory approach lends itself also more
easily for an analysis of the temperature dependence of physical quantities.
It has been used to analyze the disordered Anderson lattice model
\shortcite{M.C.O.Aguiar2003} as well as the disordered Hubbard model \shortcite{semmleretal10}.
It suffers from the disadvantage of not being able to capture the
correct exponential dependence of the low-energy Kondo scale, see
Eqs.~(\ref{eq:kondotemp}) and (\ref{eq:kondotemptoymodel}), leaving
out, therefore, the possibility of characterizing Griffiths phase
behavior (Section~\ref{sub:statDMFTMottAnderson}).

\begin{figure}
\begin{centering}
\includegraphics[bb=0bp 0bp 491bp 403bp,clip,scale=0.6]{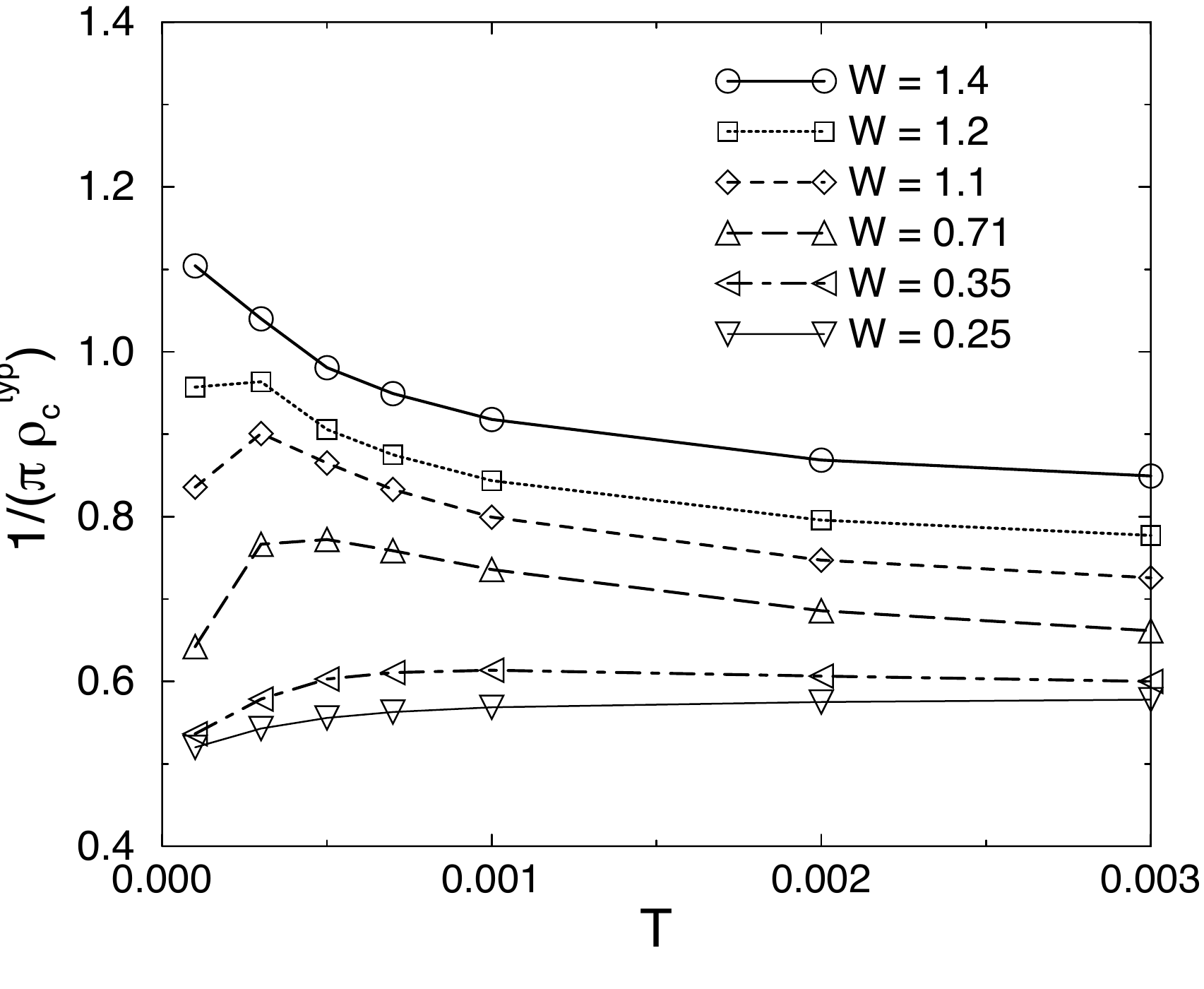}
\par\end{centering}

\caption{\label{fig:mooij}Temperature dependence of the inverse typical conduction
electron density of states of the disordered Anderson lattice model
for different levels of disorder. This {}``fan-like'' family of
curves is generic to many disordered, strongly correlated systems
({}``Mooij correlations''). From reference \protect\shortcite{M.C.O.Aguiar2003}.}

\end{figure}

In the case of the disordered Anderson lattice, one of the interesting
findings was the interplay between elastic scattering off the disorder
potential and inelastic electron-electron scattering \shortcite{M.C.O.Aguiar2003}.
If one uses the inverse of the typical density of states at the Fermi
level as a rough guide to the resistivity (a direct calculation of
the resistivity was numerically prohibitive at the time of those studies),
its \emph{temperature dependence} is found to be quite sensitive to
the amount of disorder. Indeed, low-disorder regimes are marked by
an increase of the resistive properties with rising temperatures,
signaling the onset of inelastic scattering processes, much like in
the clean case. On the other hand, strongly disordered samples exhibit
a decrease in the resistivity with increasing temperatures, because
a decrease in the \emph{effective} elastic scattering outweighs the
increase in the inelastic one. This type of fan-like family of resistivity
curves (see Fig.~\ref{fig:mooij}) has been seen in several disordered
strongly correlated materials, being known as Mooij correlations \shortcite{mooij}.
They are seen to arise here within a local approach to electronic
correlations and disorder. 

More recently the Hubbard model with binary alloy disorder has been
also studied on a Bethe lattice with iterative perturbation theory
as impurity solver \shortcite{semmleretal10}. Particular attention has
been paid to the dependence of the distribution of local densities
of states on the small imaginary part ({}``broadening'') that is
usually added to the frequency in numerical determinations of Green's
functions. The dependence of the distribution of local densities of
states on this parameter can be used to characterize the localized
or extended nature of the electronic states. The main result of that
paper is the determination of the zero-temperature phase diagram of
the model at a particular filling as a function of interaction and
disorder strengths and the identification of regions with metallic,
Mott-Anderson insulating and band insulating behaviors. In particular,
the opening of the Mott gap occurs at values of the interaction and
disorder strengths for which the gapless system is already Anderson
localized. It is difficult to compare this phase diagram with the
one obtained by the infinite-U slave boson mean field theory \shortcite{motand,london}
because the models were solved with different types of disorder and
in different regimes.

\subsection{StatDMFT on realistic lattices}

\label{sub:statDMFTreallattices}

Although the Bethe lattice implementations were very informative,
its exotic connectivity introduces some unwanted features, especially
with regard to the critical behavior of the Anderson transition \shortcite{mirlinfyodorov},
which is believed to correspond to some infinite-dimensional limit.
Therefore, it is important to check what the effects of finite dimensions
are. Motivated by this, the full statDMFT equations have been implemented
in realistic lattices in recent years. Given the successful application
of DMFT to the description of the Mott-Hubbard transition, a natural
candidate is the analysis of the disordered Mott-Hubbard transition.

\subsubsection{The disordered Mott-Hubbard transition}

\label{sub:statDMFTMottHubbardrealistic}

As with any other phase transition, the characterization of the effects
of disorder on the Mott transition poses an important yet difficult
problem. In the particular case of phase transitions incontrovertibly
described by an order parameter, this characterization has seen considerable
advances. In insulating quantum magnets, many examples have been found
in which a whole vicinity of the disordered critical point is described
as a quantum Griffiths phase. This is a phase in which rare regions
of nearly ordered material dominate the physics and the thermodynamic
response becomes divergent \shortcite{Miranda2005,Vojta2006}. The sizes
and energy scales governing the rare regions span several orders of
magnitude and their description requires taking account of very broad
distributions. Furthermore, in many cases, as the critical point itself
is approached, the relative widths of the distributions grow without
limit, a situation generically described as an \emph{infinite randomness
fixed point}. This has been well established in systems with both
Ising and continuous symmetry \shortcite{D.S.Fisher92,D.S.Fisher94,D.S.Fisher95,Hyman1996,yangetal,guo-bhatt-huse-prb96,Hyman1997,Fisher1998,pichetal98,narayanan-vojta-belitz-kirkpatrick-prl99,narayanan-vojta-belitz-kirkpatrick-prb99,motrunichetal01,G.Refael2002}.
More recently, a symmetry-based classification scheme of these Griffiths
phases has been proposed and applied to several different systems
with great success \shortcite{T.Vojta2003,vojta-schmalian-prb05,vojta-schmalian-prl05,Vojta2006,hoyos-vojta-prb06,hoyos-kotabage-vojta-prl07,hoyos-vojta-prl08,vojta-kotabage-hoyos-prb09}.
The Mott transition, however, poses a problem of a different nature,
as it is \emph{not} described by an order parameter in an obvious
way. It is thus not clear how to extend the above insights into its
description. 

The implementation of statDMFT offers a natural way out. In particular,
the clean problem is aptly described by DMFT, as we mentioned at the
end of Section~\ref{sub:cleanDMFT}. The first implementation of
statDMFT for the disordered Hubbard model was performed in \shortcite{songetal08}
using the so-called {}``Hubbard I'' (HI) approximation \shortcite{J.Hubbard1963}
as the impurity solver. This simple impurity solver captures the physics
close to the atomic limit and is therefore convenient for a study
of the Mott insulating phase. However, it suffers from the deficiency
of not giving rise to a quasiparticle peak in the single-impurity
spectral function, a feature known to exist for any finite $U$ when
the hybridization function is that of a good metal. The method was
applied to the cases of half and quarter filling and the density of
states and inverse participation ratio were obtained. Interestingly,
the localization length has a non-monotonic behavior as a function
of the interaction $U$: it increases initially with $U$, showing
a tendency to delocalization, but eventually decreases and becomes
even smaller than the non-interacting value at the Mott transition.
At quarter filling, an Altshuler-Aronov \shortcite{B.L.Altshuler1979a}
density of states anomaly is found, which is however absent at half-filling,
due to an interaction induced-suppression of the charge susceptibility. 

Shortly afterwards, the disordered Hubbard model in a two-dimensional
square lattice at $T=0$ was solved with statDMFT \shortcite{Andrade2009a,andrade09physicsB}
using the Kotliar-Ruckenstein slave boson mean field theory as the
impurity solver \shortcite{kotliarruckenstein}. Very similar results were
also obtained \shortcite{pezzolietal09,pezzolietal10} within an approach
based on a Gutzwiller variational wave function \shortcite{Gutzwiller1963,Gutzwiller1964,Gutzwiller1965}.
This is not too surprising, as the Kotliar-Ruckenstein theory, when
applied to the Mott transition, is known to be equivalent to the Gutzwiller
wave function approach. This kind of approach is known to be able
to capture the low-energy features, such as the disappearance of the
quasiparticle peak as the transition is approached, unlike the HI
approximation. In the impurity problem language, the low-energy sector
is described by two parameters, like in the infinite-$U$ case (see
Section~\ref{sub:statDMFTMottAnderson}): the quasiparticle weight
$Z$, well known from Fermi liquid theory, which determines the width
of the Kondo resonance (essentially the Kondo temperature) and the
resonance position $\tilde{\varepsilon}$, which measures its shift
from the chemical potential. It is important to notice that in the
clean lattice, $Z$ also determines the effective carrier mass ($m/m^{\ast}\sim Z$),
a feature unique to cases in which the self-energy only depends on
the frequency. Indeed, as the interaction strength is tuned to its
critical value $U\to U_{c}$, $Z\sim U_{c}-U\to0$, signaling the
transmutation of the itinerant carriers into localized magnetic moments.
In the statDMFT description, these local quantities vary from site
to site, $Z\to Z_{i}$ and $\tilde{\varepsilon}\to\tilde{\varepsilon}_{i}$,
and their distributions were thoroughly analyzed. Surprisingly, their
critical behaviors were found to be very dissimilar.

\begin{figure}
\begin{centering}
\includegraphics[clip,scale=0.5]{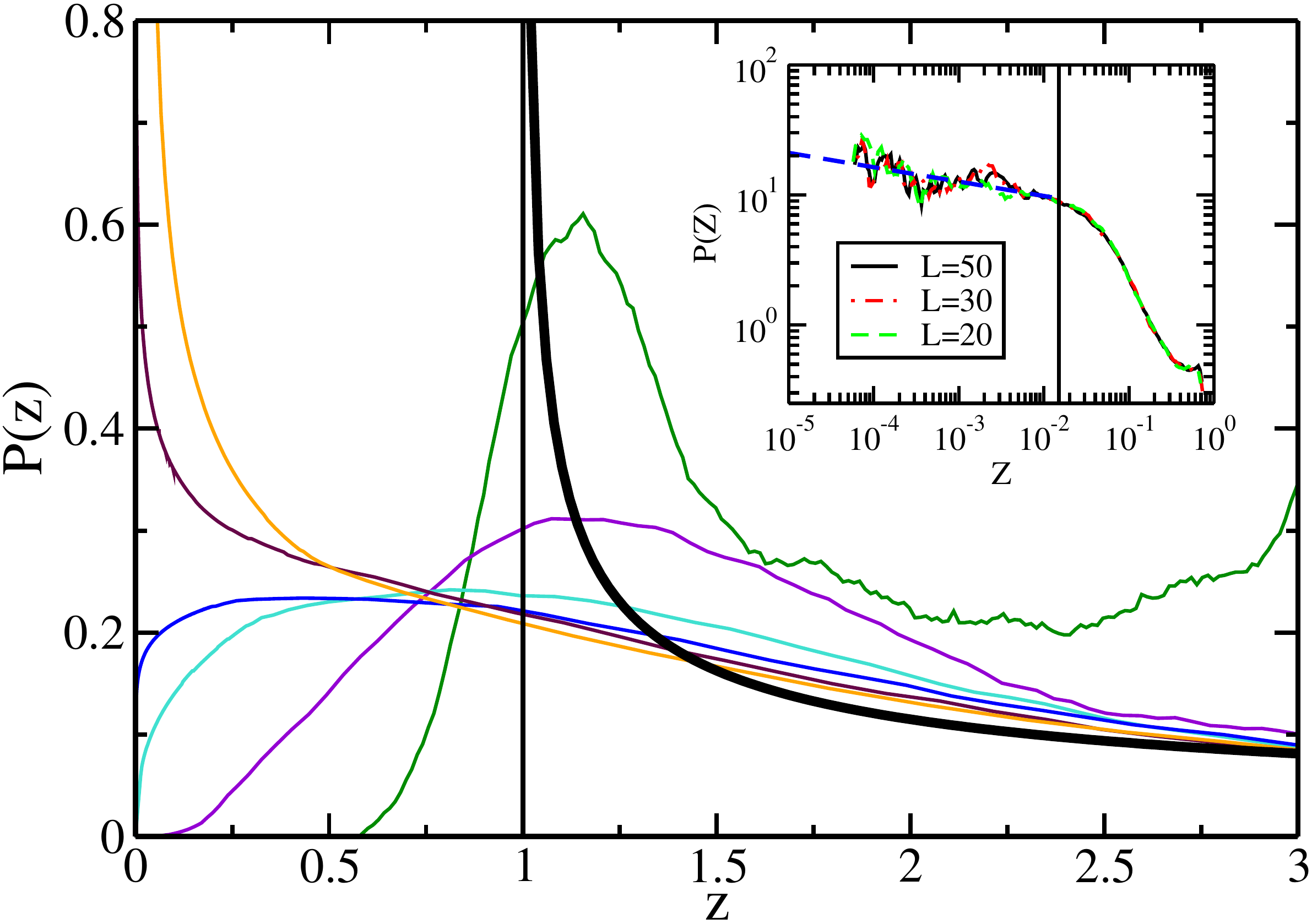}
\par\end{centering}

\caption{\label{fig:powerlawz}Power-law distributions of quasiparticle weights
$Z$ in the half-filled disordered Hubbard model for different interaction
strengths ($U/U_{c}\left(W\right)=0.6,0.8,0.9,0.92,0.94,0.97$) showing
characteritic power-law behavior $P\left(Z\right)\sim Z^{\alpha-1}$:
as $U\to U_{c}$, $\alpha\to0$ and the curves become increasingly
singular at small $Z$. The inset shows the weak dependence on the
lattice size. From reference \protect\shortcite{Andrade2009a}.}

\end{figure}

As the transition is approached, which now happens at a disorder-dependent
critical interaction $U_{c}\left(W\right)$, all $Z_{i}\to0$, just
like in the conventional Gutzwiller-Brinkman-Rice scenario. However,
the quasiparticle weight distribution $P\left(Z\right)$ becomes increasingly
broader as $U\to U_{c}\left(W\right)$. In fact, the typical value
$Z_{typ}=\exp\left(\left\langle \ln Z\right\rangle \right)\to0$,
whereas the mean value remains finite, indicating that although almost
all sites become local moments, some remain empty or doubly occupied
\shortcite{aguiaretal06}. Besides, the $Z$ distribution acquires a generic
power-law shape, as in other Griffiths phases (see Fig.~(\ref{fig:powerlawz}))\begin{equation}
P\left(Z\right)\sim Z^{\alpha-1}.\label{eq:powerlawmott}\end{equation}
This is very similar to the Bethe lattice studies of Section~\ref{sub:statDMFTMottAnderson}
and like in those cases, these power laws generate a singular thermodynamic
response, see Eq.~(\ref{eq:griffsuscandspecheat}). Nevertheless,
whereas before we had a disorder-driven Anderson-type transition,
here this generic behavior is found in the proximity of the \emph{interaction-driven}
Mott transition \emph{for fixed disorder strength}, clearly showing
the amplifying effects of electronic correlations. The similarity
to the quantum Griffiths scenario of magnets is not fortuitous. The
low-$Z$ values which dominate the thermodynamics occur in exponentially
rare regions of suppressed disorder. Finally, we note that just like
in other known quantum Griffiths phases, this electronic Griffiths
phase seems to be tied to a phase transition characterized by an \emph{infinite
randomness fixed point}: we find that, up to the numerical uncertainty,
$\alpha\to0$ as $U\to U_{c}\left(W\right)$.

\begin{figure}
\begin{centering}
\includegraphics[clip,scale=0.5]{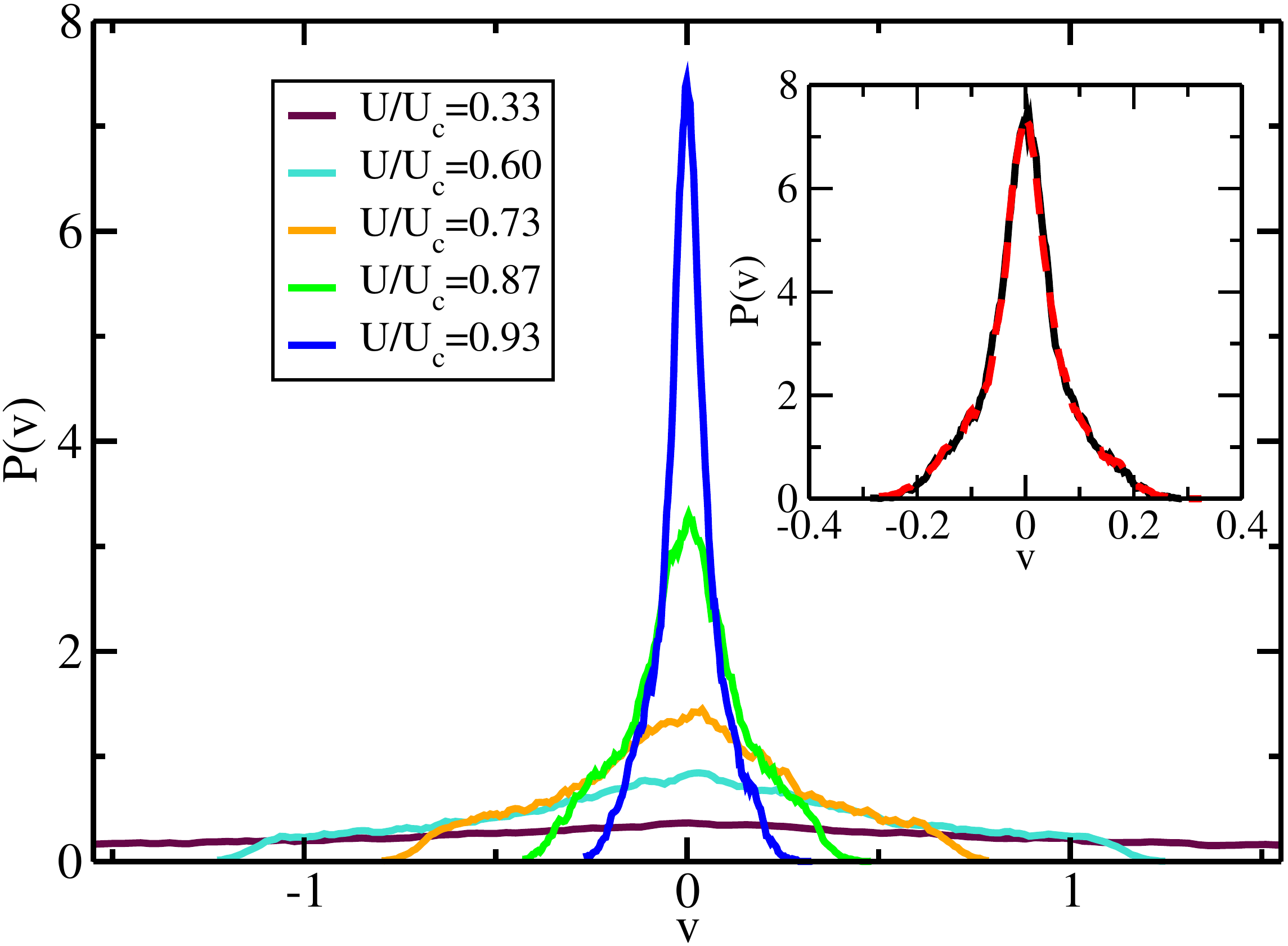}
\par\end{centering}

\caption{\label{fig:distofv} Distributions of the ratio of the local Kondo
resonance position $\tilde{\varepsilon}_{i}$ to the quasiparticle
weight $Z_{i}$, $v_{i}\equiv\tilde{\varepsilon}_{i}/Z_{i}$ in the
half-filled disordered Hubbard model showing a decreasing effective
disorder strength, as measured by $\tilde{\varepsilon}_{i}$, with
increasing interactions. The inset shows the weak dependence on the
lattice size (full line for $L=20$ and dashed line for $L=50$).
From reference \protect\shortcite{andrade09physicsB}.}

\end{figure}

Interestingly, correlations have the opposite effect on the distribution
of resonance positions $\tilde{\varepsilon}_{i}$. Indeed, the width
of the $\tilde{\varepsilon}$ distribution \emph{decreases} as $U\to U_{c}\left(W\right)$
(see Fig.~(\ref{fig:distofv})). This is easily understood from the
pinning of the Kondo resonances to the chemical potential, as already
noted within DMFT \shortcite{Tanaskovi'c2003} (see Section~\ref{sub:elastinelastDisHubModel}).
Just like in DMFT, this leads to a strong \emph{disorder screening
effect}, although, unlike in DMFT, here the screening effect is not
perfect: a small amount of disorder seems to survive as the transition
is approached.

The disordered Hubbard model has also been investigated very recently
in two dimensions with statDMFT and iterative perturbation theory
as impurity solver \shortcite{semmleretal10}. The disorder model used,
however, was not the usual uniform diagonal disorder, but rather a
combination of diagonal and off-diagonal disorder, together with disorder
in the interaction term. This choice was intended to describe the
so-called speckle disorder found in some set-ups of fermionic cold
atoms loaded in optical lattices. The phase diagram was determined
at half filling at zero as well as finite temperatures. A feature
specific to this kind of disorder is the fact that it is unbounded
and arbitrarily large values of on-site potentials occur. As a result,
long (exponential) tails arise flanking the Hubbard bands which contribute
to filling up a possible interaction-induced Hubbard gap. The main consequence
of the presence of these tails is a partial suppression of the Mott
insulating phase, in favor of a disordered strongly correlated metal
phase. As a result, the Mott insulator and Anderson(-Mott) insulator
phases do not share a phase boundary and are separated by a metallic
phase. This is in contrast to the phase diagram obtained within TMT,
see Section~\ref{sub:TMTMottAnderson}.

\subsubsection{A single impurity in a strongly correlated host}

\label{sub:singleimpurityinHubbardmodel}

The smallest element in an analysis of the effects of inhomogeneities
in a crystal is a single isolated point-like impurity in an otherwise
homogeneous host. In the case of a weakly correlated host, the expected
effect is the formation of a potential scattering center, which can
be usually be quite simply described at low energies by a set of phase
shifts with a negligible temperature dependence. If the host is a
weakly correlated metal, the localized impurity generates a radially
oscillatory disturbance of the host charge. The oscillation is characterized
by a wave vector determined by the extremal radii of the Fermi surface
and an envelope which decays as the inverse of the distance to the
impurity raised to the power $d$ in $d$ dimensions\begin{equation}
\delta n\left(\boldsymbol{x}\right)\sim\frac{\cos\left(2k_{F}r\right)}{r^{d}}.\label{eq:Friedeloscillations}\end{equation}
These are known as Friedel oscillations \shortcite{M.A.Ruderman1954,T.Kasuya1956,K.Yosida1957}.
A dilute collection of these impurities can then be treated as independent
and their effects on transport properties is readily computed by conventional
techniques, such as the Boltzmann equation. 

The situation can be very different if the host is a strongly correlated
system. The local inhomogeneity can disrupt the delicate balance of
its host and lead to a spatially dependent pattern which has to be
computed in a fully self-consistent manner. For example, local moments
or antiferromagnetic ordering can be induced in the vicinity of the
impurity \shortcite{allouletal09}. Superconducting correlations, specially
of the unconventional type (e. g., d-wave), also lead to characteristic
spatial patterns in the vicinity of impurities, which can be probed
by scanning tunneling microscopy (STM) techniques \shortcite{balatskyetal06}.
Besides, strong correlations typically generate very low energy scales.
Therefore, one should expect a non-negligible energy and temperature
dependence with possibly observable effects. Indeed, STM studies have
revealed a non-trivial interplay between spatial fluctuations and
the energy dependence of the local density of states in the cuprate
superconductors \shortcite{McElroyetal05}. In this case, the spectral
function, probed by STM, is hardly affected by disorder up to the
superconducting gap energy, whereas much stronger spatial fluctuations
are observed at higher energies.

In general, the localized imperfection can be viewed as a source probe
coupled to all wave vectors of the host charge. Therefore, if this
coupling can be treated within linear response theory, the relevant
quantity to keep track of is the host wave-vector-dependent charge
susceptibility. The vicinity to the Mott metal-insulator transition
causes the charge susceptibility to be strongly suppressed as the
system becomes increasingly more localized. Therefore, we would expect
considerable changes in the spatial pattern of the Friedel oscillations
in a strongly correlated metal. The statDMFT is the tool of choice
to study this situation. A single non-magnetic impurity in a half-filled
Hubbard model has been studied using just such a tool \shortcite{andradeetal10}.
A great deal of analytical insight can then be obtained through an
expansion in the \emph{disorder strength} to first non-trivial order,
using the Kotliar-Ruckenstein slave-boson impurity solver \shortcite{kotliarruckenstein}.
Note that the procedure is fully non-perturbative in the interaction
strength. In the weak coupling limit, $U\ll D$, the approach has
been shown to be fully equivalent to the Hartree-Fock approximation,
which predicts a static self-energy\begin{equation}
\Sigma_{j}\left(i\omega_{n}\right)=Un_{j}^{\left(0\right)},\label{eq:HFselfenergy}\end{equation}
where $n_{j}^{\left(0\right)}$ is the non-interacting charge density\begin{equation}
n_{i}^{\left(0\right)}=1+2\Pi_{i0}^{\left(0\right)}V,\label{eq:FriedelsmallU}\end{equation}
where $\Pi_{ij}^{\left(0\right)}$is the real-space static Lindhard
polarization function and $V$ is the potential of the single impurity
at site 0. This contains the usual Friedel oscillation pattern of
Eq.~(\ref{eq:Friedeloscillations}). 

\begin{figure}
\begin{centering}
\includegraphics[clip,scale=0.5]{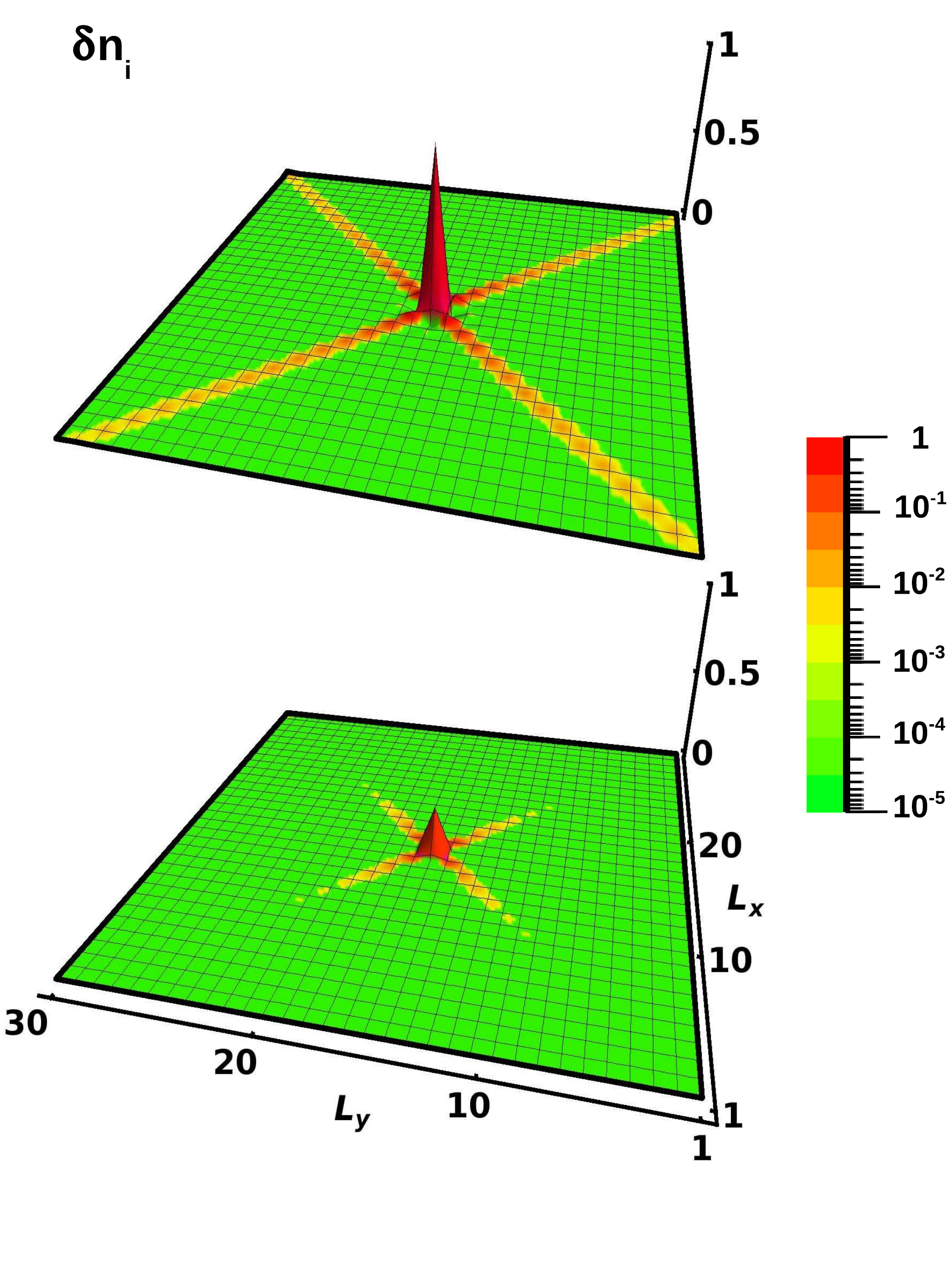}
\par\end{centering}

\caption{\label{fig:Fridel-oscillation-pattern}Friedel oscillation patterns
of charge oscillations generated by a localized non-magnetic impurity
in a half-filled Hubbard model. The long-ranged oscillations of $\delta n_{i}=n_{i}-1$
in the non-interacting case (top figure) are strongly suppressed in
the strongly correlated regime (bottom figure, $m/m^{\ast}=0.3$).
From reference \protect\shortcite{andradeetal10}. }

\end{figure}

However, as the interaction strength is tuned to be close to the Mott
localization limit $U_{c}$, a very different situation arises. In
this case, the charge density pattern decays as \shortcite{andradeetal10}
\begin{equation}
n_{0}\approx-4\frac{\left(U_{c}-U\right)^{3}}{U_{c}^{5}}\left[\Pi^{\left(0\right)}\right]_{i0}^{-1}V\qquad\left(U\to U_{c}\right).\label{eq:FriedellargeU}\end{equation}
Two features of this result stand out. First, the spatial pattern
is determined by the \emph{inverse} of the static Lindhard function.
Surprisingly, however, up to logarithmic corrections, the inverse
static Lindhard function behaves very much the same as the usual Lindhard
function. Second, the \emph{amplitude} of the strongly-correlated
Friedel oscillations is significantly suppressed as $U\to U_{c}$
down by the third power of the parametric distance to the transition.
As a result, the charge perturbation dies out significantly as one
recedes from the impurity, implying a short {}``healing length''
(see Fig.~\ref{fig:Fridel-oscillation-pattern}). Comparison with
numerical solutions of the statDMFT equations shows that the analytical
expressions are quite accurate even for $\left|V\right|\lesssim D$.

The scattering T-matrix, which governs the transport properties, can
also be computed with the same technique. We find its real space expression
to be\begin{equation}
T_{i}=\left[\delta_{i,0}+U\Pi_{i0}^{\left(0\right)}\right]V\qquad\left(U\ll U_{c}\right),\label{eq:TmatrixsmallU}\end{equation}
at small $U$ and \begin{equation}
T_{i}=-\frac{U_{c}-U}{U_{c}^{2}}\left[\Pi^{\left(0\right)}\right]_{i0}^{-1}V\qquad\left(U\to U_{c}\right),\label{eq:TmatrixlargeU}\end{equation}
as one approaches the Mott transition. Here, the point to note is
once again the strong suppression of scattering in the critical region,
a situation by now already familiar ({}``disorder screening'').

The temperature dependence of the scattering is also very special.
Indeed, it is known that the resistivity in the ballistic regime in
two dimensions exhibits an anomalous, non-Fermi liquid linear in
temperature dependence \shortcite{Zala2001}, due to the coherent scattering
off the Friedel oscillations created by the dilute impurities. Since
we find these oscillations to be strongly suppressed by correlations
it is important to study if the anomalous behavior survives. Besides,
strong interactions also give rise to large inelastic scattering effects,
and it is unclear if these are strong enough to mask the anomalous
temperature dependence. The full analysis shows that \shortcite{andradeetal10}:
\begin{itemize}
\item The interaction suppression of scattering (disorder screening), Eq.~(\ref{eq:TmatrixlargeU}),
acts to weaken both elastic and inelastic contributions from impurities,
but \emph{does not destroy} the non-Fermi liquid, anomalous behavior.
\item However, inelastic scattering limits considerably the \emph{temperature
range} in which the linear in $T$ behavior of the resistivity is
observed. These effects come both from the Friedel-oscillation regions,
close to the impurities, as well as from the bulk, impurity-free regions
of the system. It is found that, although both contributions are deleterious
to the anomalous behavior, it is the bulk contribution which always
dominates. In practice, even for quite modest mass renormalizations,
the linear in $T$ regime is probably already unobservable: for $m/m^{\ast}\sim0.6$,
for example, it is restricted to $T\lesssim10^{-4}T_{F}$, where $T_{F}\approx D$.
\end{itemize}
Both effects above seem to make the observation of the anomalous resistivity
in two dimensions very unlikely in the strongly correlated regime. 

In more general terms, the strategy of expansion in weak disorder
strength outlined above, which is quite general, seems to be a good
avenue for exploration of strongly correlated spatially inhomogeneous
systems.

\subsection{Inhomogeneous systems with slab geometries}

\label{sub:slabgeometries}

Although the local approach outlined in Sections~\ref{sub:statDMFTformulation},
\ref{sub:statDMFTBethe} and \ref{sub:statDMFTreallattices} has been
described in the context of random systems, its microscopic formulation
\shortcite{motand,london} is clearly applicable
in any spatially non-uniform situation, whether random or not. Indeed,
in the past several years this method has been applied to a number
of non-random, yet spatially non-uniform systems. Because of this
conceptual similarity, we chose to describe it here under the statDMFT
{}``umbrella'', although the systems have nothing {}``statistical''
in them. A particular focus has been put on systems composed of different
materials arranged in a slab geometry. Motivation for these investigations
has come both from advances in growth techniques of epitaxial heterostructures
of complex materials, particularly transition metal oxides \shortcite{ohtomoetal02},
and also from the use of surface sensitive probes of strongly correlated
materials, such as photoemission experiments \shortcite{sekiyamaetal00,moetal03,rodolakisetal09}.
A number of different spatial arrangements of complex, strongly correlated
materials have been considered: semi-infinite strongly correlated
systems with a free-standing planar boundary ({}``strongly correlated
surface physics'') \shortcite{potthoffnolting99,liebsch03,schwiegeretal03,borghietal09,borghietal10},
planar interfaces between two semi-infinite materials \shortcite{Helmes2008},
heterostructures \shortcite{okamotomillis04,okamotomillis04b}, semi-infinite
metallic leads sandwiching barriers of strongly correlated metals
or insulators \shortcite{freericks04,chenfreericks07,zeniaetal09,borghietal10}.
Studies of other types of inhomogeneities have also been performed
\shortcite{snoeketal08}.

The abrupt change of chemical properties at the planes of surfaces
or interfaces can lead to novel phenomena, some of which have been
identified and characterized by the local approach we are describing
here. This a fast moving area of research and we will not attempt
to cover the richness of behavior that has been and continues to be
uncovered. We will confine ourselves to a few important findings which
serve to convey the flavor of these phenomena.

\subsubsection{The interface between a metal and a Mott insulator}

\label{sub:mottmetalinterface}

Metal-insulator and metal-semiconductor interfaces have been produced
and studied for many years. Phenomena such as band bending due to
charge rearrangement are well known. However, the question of what
happens when a metal is grown epitaxially onto a Mott insulator has
received much less attention. A first attempt to elucidate this question
has been made by considering an ideal interface between particle-hole
symmetric metallic and Mott insulating systems \shortcite{Helmes2008,borghietal09,borghietal10}.
In the absence of antiferromagnetism and within the DMFT picture,
the spins of a bulk Mott insulator are not quenched because there
are no conduction electrons available at the Fermi level with which
they can form a singlet. When a Mott insulator is brought into contact
with a metal, however, the layers which are closest to the interface
can have access to the metallic carriers (through tunneling) and be
quenched by means of the Kondo effect. This has been dubbed the {}``Kondo
proximity effect''. Evidently, the metallization and Kondo quenching
processes are established in a spatially smooth fashion as one enters
the insulator, thus creating a strongly correlated metallic surface
layer. The metallic character can be characterized by means of the
by now familiar quasiparticle weight $Z\left(x\right)$, which is
here a spatially varying quantity that depends on the distance $x$
from the interface. It is found that $Z\left(x\right)$ decays exponentially
as a function of $x$\begin{equation}
Z\left(x\right)\sim\exp\left(-x/\xi\right),\label{eq:zofxmetalMottinsul}\end{equation}
with a characteristic length scale $\xi$ which sets the size of the
metallic surface layer and is determined by the parametric distance
to the Mott transition, $\xi\sim\left(U_{c}-U\right)^{-1/2}$. The
1/2 value of the critical exponent is expected from the mean-field
character of the approach. At Mott criticality, this characteristic
length diverges and \begin{equation}
Z\left(x\right)=\frac{A}{x^{2}}.\label{eq:zofxmetalcrititalMott}\end{equation}
The numerical value of the constant $A$ above is found to be extremely
small ($\sim0.008$), showing the Kondo induced penetration of the
metal into the Mott insulator to be extremely ineffective. The small
values of the quasiparticle residues, which govern the local Fermi
energy scale, also lead to a strong dependence on temperature, energy
or applied voltage. Very small values of these perturbations are able
to completely destroy the metallic behavior, leading to the concept
of a {}``fragile'' Fermi liquid \shortcite{zeniaetal09}.

\subsubsection{Surface dead layer}

\label{sub:deadlayer}

Another important phenomenon occurs at the free planar surface bounding
a strongly correlated metal \shortcite{borghietal09,borghietal10}. Here
again the proximity to the interface with the vacuum induces variations
of the quasiparticle residue $Z$. Interestingly, since the outermost
layer electrons cannot tunnel further out of the material, they have
less of a chance to hybridize and stabilize the metallic behavior.
As a consequence, there is a strong suppression of the quasiparticle
weight close to the open surface. In effect, the outermost electrons
are a very {}``fragile'' Fermi liquid if not completely Mott localized,
thus creating a surface layer of almost Mott insulating character,
the so-called {}``dead layer''. Once again the characteristic width
of this layer also depends on the proximity to criticality as $\xi\sim\left(U-U_{c}\right)^{-1/2}$. 

This has important consequences for the interpretation of photoemission
experiments. For a long time, a quasiparticle peak had been sought
in the photoemission spectra of strongly correlated materials with
little success. While there have been many attempts to explain this
mystery, it is clear now that one expects on general grounds that
the quasiparticle peak width (which is proportional to $Z$) becomes
extremely narrow close to the surface. Since conventional photoemission
spectra only probe the outer surface layers of the material, it is
not too surprising that it has been difficult to detect a well-formed
quasiparticle peak in the past. More recently, however, higher-energy
photons have been used, which are able to penetrate deeper into the
compound and thus probe the behavior more typical of the bulk. As
expected from the theoretical results outlined above, the quasiparticle
becomes much better defined with increasing incident photon energy
\shortcite{sekiyamaetal00,moetal03,rodolakisetal09}.

\section{Glassy behavior of correlated electrons}

\label{sub:Itinerant-quantum-glass}

We discussed so far the nature of strongly inhomogeneous metallic  
phases, resulting from the interplay of strong correlations and disorder. 
Such `electronic Griffiths phases", which can be viewed as precursors to the 
metal-insulator transition, reflect the formation of rare regions with anomalously 
slow dynamics. The resulting non-Fermi-liquid behavior is generically characterized by power law anomalies, with
non-universal, rapidly varying exponents. In contrast, many experimental data, 
especially in Kondo alloys, seem to show reasonably weak anomalies, 
close to marginal Fermi liquid behavior \shortcite{stewartNFL}.

\subsection{Instability of the electronic Griffiths phases to
spin-glass ordering}

\label{sub:instabilitygriff}

Physically, it is clear what is missing from the theory. Similarly as magnetic
Griffiths phases \shortcite{Miranda2005,Vojta2006}, the electronic Griffiths phase is characterized
\shortcite{Miranda2001,tanaskovicetal04} by a broad distribution $P\left(
T_{K}\right)  \sim\left(  T_{K}\right)  ^{\alpha-1}$ of local energy scales
(Kondo temperatures), with the exponent $\alpha\sim W^{-2}$ rapidly decreasing
with disorder $W$. At any given temperature, the local moments with
$T_{K}(i)<T$ remain unscreened. As disorder increases, the number of such
unscreened spins rapidly proliferates. Within the existing theory
\shortcite{mirandavladgabi1,mirandavladgabi2,Miranda2001,tanaskovicetal04} these
unscreened spins act essentially as free local moments and provide a very
large contribution to the thermodynamic response. In a more realistic
description, however, even the Kondo-unscreened spins are \emph{not}
completely free, since the metallic host generates long-ranged
Ruderman-Kittel-Kasuya-Yosida (RKKY) interactions even between relatively
distant spins.

In a disordered metal, impurity scattering introduces random phase
fluctuations in the usual periodic oscillations of the RKKY interaction,
which, however, retains its power law form (although its \textit{average}
value decays exponentially \shortcite{jagannathanetal88,narozhnyetal00}). Hence,
such an interaction acquires a random amplitude $J_{ij}$ of zero mean but
finite variance \shortcite{jagannathanetal88,narozhnyetal00}
\begin{equation}\langle J_{ij}^{2}(R)\rangle \sim \frac{1}{R^{2d}}.\end{equation}
 As a result, in a disordered metallic
host, a given spin is effectively coupled with random but long range
interactions to many other spins, often leading to spin-glass freezing at the
lowest temperatures. How this effect is particularly important in Griffiths
phases can also be seen from the mean-field stability criterion
\shortcite{braymoore80} for spin glass ordering, which takes the form
\begin{equation}
J\chi_{loc}(T)=1. \label{eq:sg}
\end{equation}
Here, $J$ is a characteristic interaction scale for the RKKY interactions, and
$\chi_{loc}(T)$ is the disorder average of the local spin susceptibility. As
we generally expect $\chi_{loc}(T)$ to diverge within a Griffiths phase, this
arguments strongly suggests that in the presence of RKKY interactions such
systems should have an inherent instability to finite (even if very low)
temperature spin glass ordering.

Similarly as other forms of magnetic order, the spin glass ordering is
typically reduced by quantum fluctuations (e.g. the Kondo effect) which are
enhanced by coupling of the local moments to itinerant electrons. Sufficiently
strong quantum fluctuations can completely suppress spin-glass ordering even
at $T=0$, leading to a quantum critical point separating a metallic spin glass
from the conventional Fermi liquid ground state. As in other QCP's, one
expects the precursors to magnetic ordering to emerge even before the
transition is reached and produce non-Fermi liquid behavior within the
corresponding quantum critical region. Since many systems where
disorder-driven non-Fermi-liquid behavior is observed are not too far from incipient
spin-glass ordering, it is likely that these effects play an important role
and should be theoretically examined in detail.

From the theoretical point of view, a number of recent works have examined the
general role of quantum fluctuations in glassy systems and the associated
quantum critical behavior. Most of the results obtained so far have
concentrated on the behavior within the mean-field picture (i.e., in the limit
of large coordination), where a consistent description of the quantum critical point (QCP) behavior has
been obtained for several models. In a few cases \shortcite{re:Read95}, corrections
to mean-field theory have been examined, but the results appear inconclusive
and controversial at this time. In the following, we briefly review the most
important results obtained within the mean-field approaches.

\subsubsection{Quantum critical behavior in insulating and metallic spin glasses}

\label{sub:qcp-msg}

\begin{itemize}
\item \emph{Ising spin glass in a transverse field}
\end{itemize}

\begin{figure}[h]
\begin{center}
\includegraphics[width=3.4in]{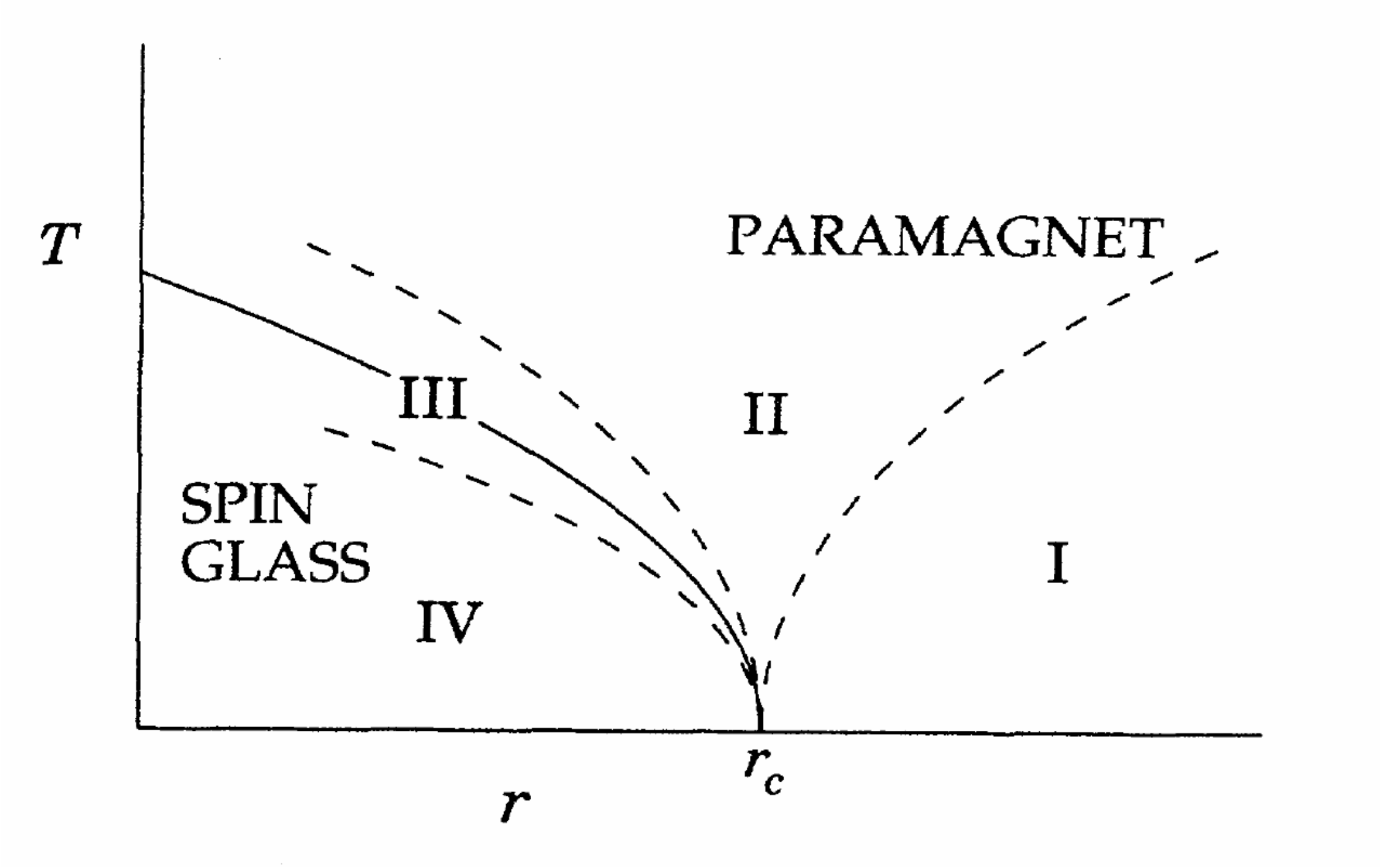}
\end{center}
\caption{Generic phase diagram (following \protect\shortcite{re:Read95}) of the quantum
critical behavior for spin glasses. The parameter $r$, which measures the
quantum fluctuations, can represent the transverse field for localized spin
models or the Fermi energy in metallic spin glasses.}%
\label{cap:fig14}%
\end{figure}

The simplest framework to study the quantum critical behavior of spin glasses
is provided by localized spin models such as the infinite-range Ising model in
a transverse field (TFIM) with random exchange interactions $J_{ij}$ of zero
mean and variance $J^{2}/N$ ($N\longrightarrow\infty$ is the number of lattice
sites).%
\begin{equation}
H_{TFIM}=-\sum_{ij}J_{ij}\sigma_{i}^{z}\sigma_{j}^{z}-\Gamma\sum_{i}\sigma
_{i}^{x}. \label{eqisingsg}%
\end{equation}
In the classical limit ($\Gamma=0$), this model reduces to the well-studied
Sherrington-Kirkpatrick model \shortcite{re:Mezard86}, where spins freeze with
random orientations below a critical temperature $T_{SG}(\Gamma=0)=J$. Quantum
fluctuations are introduced by turning on the transverse field, which induces
up-down spin flips with tunneling rate $\sim\Gamma$. As $\Gamma$ grows, the
critical temperature $T_{SG}(\Gamma)$ decreases, until the quantum critical
point is reached at $\Gamma=\Gamma_{c}\approx0.731J$, signaling the $T=0$
transition from a spin-glass to a quantum-disordered paramagnetic state
(Fig.~\ref{cap:fig14}).

Similarly as in DMFT theories for electronic systems, such infinite range
models can be formally reduced to a self-consistent solution of an appropriate
quantum impurity problem, as first discussed in the context of quantum spin
glasses by Bray and Moore \shortcite{braymoore80}. Early work quickly established
the phase diagram \shortcite{dobrosavljevic87} of this model, but the dynamics near
the quantum critical point proved more difficult to unravel, even when the
critical point is approached form the quantum-disordered side. Here, the
problem reduces to solving for the dynamics of a single Ising spin in a
transverse field, described by an effective Hamiltonian \shortcite{re:Miller93} of
the form%
\[
H=-\frac{1}{2}J^{2}\int\int d\tau d\tau^{\prime}\sigma^{z}(\tau)\chi(\tau
-\tau^{\prime})\sigma^{z}(\tau^{\prime})+\Gamma\int d\tau\sigma^{x}(\tau).
\]
Physically, the interaction of the considered spin with the spin fluctuations
of its environment generates the retarded interaction described by the
{}``memory kernel'' $\chi(\tau-\tau^{\prime})$. An appropriate
self-consistency condition relates the memory kernel to the disorder-averaged
local dynamical susceptibility of the quantum spin
\[
\chi(\tau-\tau^{\prime})=\overline{\left\langle T\sigma^{z}(\tau)\sigma
^{z}(\tau^{\prime})\right\rangle }.
\]
A complete solution of the quantum critical behavior can be obtained, as first
established in a pioneering work by Miller and Huse \shortcite{re:Miller93}. These
authors have set up a diagrammatic perturbation theory for the dynamic
susceptibility, showing that the leading loop approximation already captures
the exact quantum critical behavior, as the higher order corrections provide
only quantitative renormalizations. The dynamical susceptibility takes the
general form%
\[
\chi(\omega_{n})=\chi_{o}+\left(  \omega_{n}^{2}+\Delta^{2}\right)  ^{1/2},
\]
where the local static susceptibility $\chi_{o}$ remains finite throughout the
critical regime, and the spin excitations exist above a gap
\[
\Delta\sim\left(  r/\left|  \ln r\right|  \right)  ^{1/2}%
\]
which vanishes as the transition is approached from the paramagnetic side
(here $r=\left(  \Gamma-\Gamma_{c}\right)  /\Gamma_{c}$ measures the distance
from the critical point).

\begin{itemize}
\item \emph{The quantum spin-glass phase and the replicon mode }
\end{itemize}

The validity of this solution was confirmed by a generalization \shortcite{ye93} to
the $M$-component rotor model (the Ising model belongs to the same
universality class as the $M=2$ rotor model), which can be solved in closed
form in the large $M$ limit. This result, which proves to be exact to all
orders in the $1/M$ expansion, could be extended even to the spin glass phase,
where a full replica symmetric solution was obtained. Most remarkably, the
spin excitation spectrum remains gapless ($\Delta=0$) throughout the ordered
phase. Such gapless excitations commonly occur in ordered states with broken
continuous symmetry, but are generally not expected in classical or quantum
models with a discrete symmetry of the order parameter. In glassy phases (at
least within mean-field solutions), however, gapless excitations generically
arise for both classical and quantum models. Here, they reflect the marginal
stability \shortcite{re:Mezard86} found in the presence of replica symmetry
breaking, a phenomenon which reflects the high degree of frustration in these
systems. The role of the Goldstone mode in this case is played by the
so-called {}``replicon'' mode, which describes the collective low energy
excitations characterizing the glassy state.

A proper treatment of the low energy excitations in this regime requires
special attention to the role of replica symmetry breaking (RSB) in the
$T\rightarrow0$ limit. The original work \shortcite{ye93} suggested that RSB is
suppressed at $T=0$, so that the simpler replica symmetric solution can be
used at low temperatures. Later work \shortcite{georgesetal01}, however,
established that the full RSB solution must be considered before taking the
$T\rightarrow0$ limit, and only then can the correct form of the leading low
temperature corrections (e.g., the linear $T$-dependence of the specific heat)
be obtained.

\begin{itemize}
\item \emph{Physical content of the mean-field solution }
\end{itemize}

In appropriate path-integral language
\shortcite{dobrosavljevic87,ye93}, the problem can be shown to reduce
to solving a one dimensional classical Ising model with long-range
interactions, the form of which must be self-consistently
determined. Such classical spin chains with long range
interactions in general can be highly nontrivial. Some important
examples are the Kondo problem
\shortcite{Yuval-Anderson1,Yuval-Anderson2,Yuval-Anderson3}, and the
dissipative two-level system \shortcite{leggettetal87}, both of which
map to an Ising chain with $1/\tau^{2}$ interactions. Quantum
phase transitions in these problems correspond to the
Kosterlitz-Thouless transition found in the Ising chain
\shortcite{kosterlitz76prl}, the description of which required a
sophisticated renormalization-group analysis. Why then is the
solution of the quantum Ising spin glass model so simple? The
answer was provided in the paper by Ye, Sachdev, and Read
\shortcite{ye93}, which emphasized that the critical state (and the RSB
spin-glass state) does not correspond to the critical point, but
rather to the high-temperature phase of the equivalent Ising
chain, where a perturbative solution is sufficient. In Kondo
language, this state corresponds to the Fermi-liquid solution
characterized by a finite Kondo scale, as demonstrated by a
quantum Monte-Carlo calculation of Rozenberg and Grempel
\shortcite{rozenberg98prl}, which also confirmed other predictions of
the analytical theory.

From a more general perspective, the possibility of obtaining a simple
analytical solution for quantum critical dynamics has a simple origin. It
follows from the fact that all corrections to Gaussian (i.e. Landau) theory
are irrelevant above the upper critical dimension, as first established by the
Hertz-Millis theory \shortcite{hertz,millis} for conventional quantum criticality.
The mean-field models become exact in the limit of infinite dimensions, hence
the Gaussian solution of Refs. \shortcite{re:Miller93,ye93} becomes exact. Leading
corrections to mean-field theory for rotor models were examined by an
$\varepsilon$-expansion below the upper critical dimension $d_{c}=8$ for the
rotor models by Read, Sachdev, and Ye, but these studies found run-away flows,
presumably indicating non-perturbative effects that require more sophisticated
theoretical tools. Most likely these include Griffiths phase phenomena
controlled by the infinite randomness fixed point, as already discussed in
Section~\ref{sub:statDMFT}.

\begin{itemize}
\item \emph{Metallic spin glasses }
\end{itemize}


A particularly interesting role of the low-lying excitations associated with
the spin-glass phase is found in metallic spin glasses. Here the quantum
fluctuations are provided by the Kondo coupling between the conduction
electrons and local moments, and therefore can be tuned by controlling the
Fermi energy in the system. The situation is again the simplest for Ising
spins where an itinerant version of the rotor model of Sengupta and Georges
\shortcite{senguptageorges} can be considered, and similar results have obtained
for the {}``spin-density glass'' model of Sachdev, Read, and Oppermann
\shortcite{sachdevreadopper}. The essential new feature in these models is the
presence of itinerant electrons which, as in the Hertz-Millis approach
\shortcite{hertz,millis}, have to be formally integrated out before an effective
order-parameter theory can be obtained. This is justified \textit{provided}
that the quasiparticles remain well defined at the quantum critical point,
i.e. that the quasiparticle weight $Z\sim T_{K}$ remains finite and the Kondo
effect remains operative. The validity of these assumptions is by no means
obvious, and led to considerable controversy before a detailed quantum Monte
Carlo solution of the model became available \shortcite{rozenberg99prb}, confirming
the proposed scenario.

Under these assumptions, the theory can again be solved in closed form, and we
only quote the principal results. Physically, the essential modification is
that the presence of itinerant electrons now induces Landau damping, which
creates dissipation for the collective mode. As a result, the dynamics is
modified, and the local dynamic susceptibility now takes the following form%
\begin{equation}
\chi(\omega_{n})=\chi_{o}+\left(  \left|  \omega_{n}\right|  +\omega^{\ast
}\right)  ^{1/2}.
\end{equation}
The dynamics is characterized by the crossover scale $\omega^{\ast}\sim r$
($r$ measures the distance to the transition) which defines a crossover
temperature $T^{\ast}\sim\omega^{\ast}$ separating the Fermi liquid regime (at
$T\ll T^{\ast}$) from the quantum critical regime (at $T\gg T^{\ast}$). At the
critical point $\chi(\omega_{n})=\chi_{o}+\left|  \omega_{n}\right|  ^{1/2}$,
leading to non-Fermi liquid behavior of all physical quantities, which acquire
a leading low-temperature correction of the $T^{3/2}$ form. This is a rather
mild violation of Fermi liquid theory, since both the static spin
susceptibility and the specific heat coefficient remain finite at the QCP. A
more interesting feature, which is specific to glassy systems, is the
persistence of such quantum critical non-Fermi-liquid behavior \textit{throughout} the
metallic glass phase, reflecting the role of the replicon mode.

\subsubsection{Spin-liquid behavior, destruction of the Kondo effect by
bosonic dissipation, and fractionalization}

\label{sub:spinliquid}

\begin{itemize}
\item \emph{Quantum Heisenberg spin glass and the spin-liquid solution }
\end{itemize}

Quantum spin glass behavior proves to be much more interesting in the case of
Heisenberg spins, where the Berry phase term \shortcite{Fradkin} plays a highly
nontrivial role, completely changing the dynamics even within the paramagnetic
phase. While the existence of a finite temperature spin-glass transition was
established even in early work \shortcite{braymoore80}, solving for the details of
the dynamics proved difficult until the remarkable work of Sachdev and Ye
\shortcite{sachdevye}. By a clever use of large-$N$ methods, these authors
identified a striking \textit{spin-liquid} solution within the paramagnetic
phase. In contrast to the nonsingular behavior of Ising or rotor quantum spin
glasses, the dynamical susceptibility now displays a logarithmic singularity
at low frequency. On the real axis it takes the form
\[
\chi(\omega)\sim\ln(1/\left|  \omega\right|  )+i\frac{\pi}{2}\mathrm{sgn}(\omega).
\]
A notable feature of this solution is that it is precisely of the form
postulated for {}``marginal Fermi liquid'' phenomenology \shortcite{mfl} of doped
cuprates. The specific heat is also found to assume a singular form
$C\sim\sqrt{T}$, which was shown \shortcite{georgesetal01} to reflect a nonzero
extensive entropy if the spin liquid solution is extrapolated to $T=0$. Of
course, the spin-liquid solution becomes unstable at a finite ordering
temperature, and the broken symmetry state has to be examined to discuss the
low temperature properties of the model.

Subsequent work \shortcite{georgesetal01} demonstrated that this mean-field
solution remains valid for all finite $N$ and generalized the solution to the
spin-glass (ordered) phase. A closed set of equations describing the low
temperature thermodynamics in the spin glass phase was obtained, which was
very recently re-examined in detail \shortcite{rozenberg04prl} to reveal fairly
complicated behavior.

\begin{itemize}
\item \emph{Metallic Heisenberg spin glasses and fractionalization }
\end{itemize}

Even more interesting is the fate of this spin liquid solution in itinerant
systems, where an additional Kondo coupling is added between the local moments
and the conduction electrons. The mean-field approach can be extended to this
interesting situation by examining a Kondo-Heisenberg spin glass model
\shortcite{burdinetal} with the Hamiltonian
\begin{equation}
H_{KH}=-t\sum_{\langle ij\rangle\sigma}(c_{i\sigma}^{\dagger}c_{j\sigma
}^{\phantom{\dagger}}+\mbox{H. c.})+J_{K}\sum_{i}\mathbf{S}_{i}\cdot
\mathbf{s}_{i}+\sum_{\langle ij\rangle}J_{ij}\mathbf{S}_{i}\cdot\mathbf{S}%
_{j}. \label{eq:spinliqham}%
\end{equation}

In the regime where the scale of the RKKY interaction $J=\left\langle
J_{ij}^{2}\right\rangle ^{1/2}$ is small compared to the Kondo coupling
$J_{K}$, one expects Kondo screening to result in standard Fermi liquid
behavior. In the opposite limit, however, the spin fluctuations associated
with the retarded RKKY interactions may be able to adversely affect the Kondo
screening, and novel metallic behavior could emerge. This intriguing
possibility can be precisely investigated in the mean-field (infinite range)
limit, where the problem reduces to a single-impurity action of the form
\shortcite{burdinetal}
\begin{align}
S_{eff}^{KH}  &  =\sum_{\sigma}\int_{0}^{\beta}d\tau c_{\sigma}^{\dagger
}\left(  \tau\right)  \left(  \partial_{\tau}-\mu+v_{j}\right)  c_{\sigma
}^{\phantom{\dagger}}\left(  \tau\right) \nonumber\\
&  -t^{2}\sum_{\sigma}\int_{0}^{\beta}d\tau\int_{0}^{\beta}d\tau^{\prime
}c_{\sigma}^{\dagger}(\tau)G_{c}(\tau-\tau^{\prime})c_{\sigma}%
^{\phantom{\dagger}}(\tau^{\prime})\nonumber\\
&  +J_{K}\int_{0}^{\beta}d\tau\mathbf{S}(\tau)\cdot\mathbf{s}(\tau)\nonumber\\
&  -\frac{J^{2}}{2}\int_{0}^{\beta}d\tau\int_{0}^{\beta}d\tau^{\prime}%
\chi(\tau-\tau^{\prime})\mathbf{S}(\tau)\cdot\mathbf{S}(\tau^{\prime}).
\label{eq:spinliqaction}
\end{align}

Such a single-impurity action (\ref{eq:spinliqaction}) describes the so-called
Bose-Fermi Kondo (BFK) impurity model
\shortcite{qimiao96prl,sengupta,zhusi02,zaranddemler02} where, in addition to the
coupling to the fermionic bath of conduction electrons, the Kondo spin also
interacts with a bosonic bath of spin fluctuations, with local spectral
density $\chi\left(  \omega_{n}\right)  $. Because the same BFK model also
appears in {}``extended'' DMFT theories \shortcite{qimiao00prb} of quantum
criticality in clean systems
\shortcite{qimiao96prl,sengupta,sietal,zhusi02,zaranddemler02}, its properties have
been studied in detail and are by now well understood.

\begin{figure}[ptb]
\begin{center}
\includegraphics[  width=3.4in,
keepaspectratio]{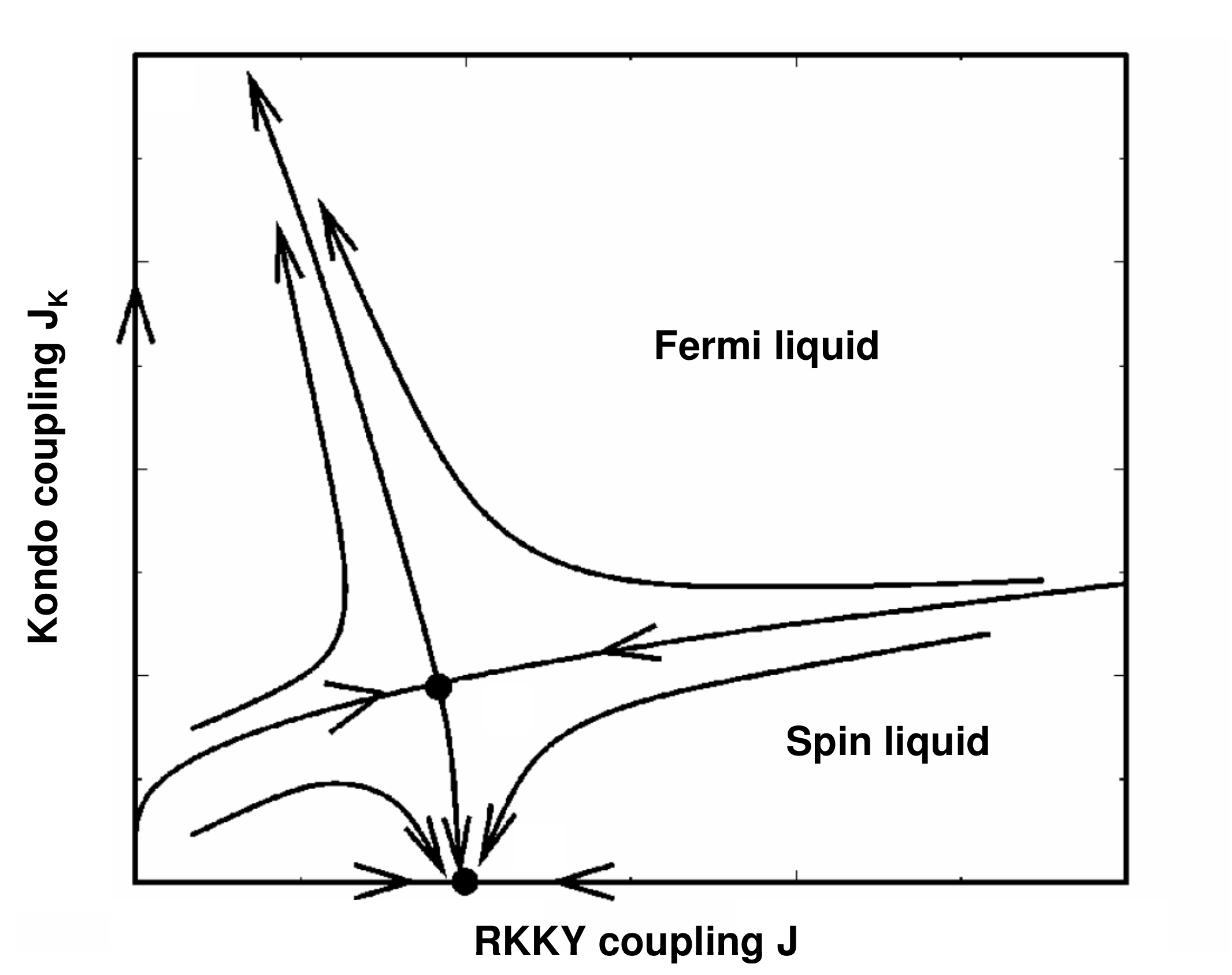}
\end{center}
\caption{Phase diagram of the Bose-Fermi Kondo model in the presence of a
sub-Ohmic bosonic bath \protect\shortcite{qimiao96prl,sengupta,zhusi02,zaranddemler02}.
Kondo screening is destroyed for sufficiently large dissipation (RKKY coupling
to spin fluctuations). }
\label{cap:fig15}
\end{figure}

In the absence of the RKKY coupling ($J=0$), the ground state of the impurity
is a Kondo singlet for any value of $J_{K}\neq0$. By contrast, when $J>0$, the
dissipation induced by the bosonic bath tends to destabilize the Kondo effect.
For a bosonic bath of {}``Ohmic'' form ($\chi\left(  \omega_{n}\right)
=\chi_{o}-C\left|  \omega_{n}\right|  $), this effect only leads to a finite
decrease of the Kondo temperature, but the Fermi liquid behavior persists. In
contrast, for {}``sub-Ohmic'' dissipation ($\chi\left(  \omega_{n}\right)
=\chi_{o}-C\left|  \omega_{n}\right|  ^{1-\epsilon}$ with $\epsilon>0)$ two
different phases exist, and for sufficiently large RKKY coupling the Kondo
effect is destroyed. The two regimes are separated by a quantum phase
transition (see Fig.~\ref{cap:fig15}).

Of course, in the considered Kondo lattice model with additional RKKY
interactions, the form of the bosonic bath $\chi\left(  \omega_{n}\right)  $
is self-consistently determined and can take different forms as the RKKY
coupling $J$ is increased. The model was analytically solved within a large
$N$ approach by Burdin \emph{et al.} \shortcite{burdinetal}, who calculated the
evolution of the Fermi liquid coherence scale $T^{\ast}$ and the
corresponding quasiparticle weight $Z$ in the presence of RKKY interactions.
Within the paramagnetic phase both $T^{\ast}(J)$ and $Z(J)$ are found to
decrease with $J$ until the Kondo effect (and thus the Fermi liquid) is
destroyed at $J=J_{c}\approx10T_{K}^{o}$ (here $T_{K}^{o}$ is the $J=0$ Kondo
temperature), where both scales vanish \shortcite{burdinetal,tanaskovic-2005-95}. At
$T>T^{\ast}(J)$ (and of course at any temperature for $J>J_{c}$) the spins
effectively decouple from conduction electrons and spin liquid behavior,
essentially identical to that of the insulating model, is established. Thus,
sufficiently strong and frustrating RKKY interactions are able to suppress
Fermi liquid behavior, and marginal Fermi liquid behavior emerges in a
metallic system. This phenomenon, corresponding to spin-charge separation
resulting from the destruction of the Kondo effect, is sometimes called
{}``fractionalization''
\shortcite{colemanandrei,kaganetal92,demleretal02,senthiletal,senthiletal2}. Such
behavior has often been advocated as an appealing scenario for exotic phases
of strongly correlated electrons, but with the exception of the described
model, there are very few well established results and model calculations to
support its validity. Finally, we should mention related work
\shortcite{parcolletgeorges} on doped Mott insulators with random exchanges, with
many similarities with the above picture.

We should note, however, that this exotic solution is valid only within the
paramagnetic phase, which is generally expected to become unstable to magnetic
(spin glass) ordering at sufficiently low temperatures. Since
fractionalization emerges only for sufficiently large RKKY coupling (in the
large $N$ model $J_{c}\approx10T_{K}^{o}$), while in general one expects
magnetic ordering to take place already at $J\sim T_{K}^{o}$ (according to the
famous Doniach criterion \shortcite{Doniach}), one expects \shortcite{burdinetal} the
system to magnetically order much before the Kondo temperature vanishes. If
this is true, then one expects the quantum critical behavior to be very
similar to metallic Ising spin glasses, i.e., to assume the conventional
Hertz-Millis form, at least for the mean-field spin glass models we discussed
here. The precise relevance of this paramagnetic spin liquid solution thus
remains unclear, at least for systems with weak or no disorder in the
conduction band.

\begin{itemize}
\item \emph{Fractionalized two-fluid behavior of electronic Griffiths phases }
\end{itemize}

The situation seems more promising in the presence of sufficient amounts of
disorder, where the electronic Griffiths phase forms. Here the disorder generates a very broad distribution
of local Kondo temperatures, making the system much more sensitive to RKKY
interactions. This mechanism has recently been studied within an extended DMFT
approach \shortcite{tanaskovic-2005-95}, which is able to incorporate both the
formation of the Griffiths phase and the effects of frustrating RKKY
interactions leading to spin-glass dynamics. At the local impurity level, the
problem is still reduced to the Bose-Fermi Kondo model, but the presence of
conduction electron disorder qualitatively modifies the self-consistency
conditions determining the form of $\chi\left(  \omega_{n}\right)  $.

To obtain a sufficient condition for decoupling, we examine the stability of
the Fermi liquid solution, by considering the limit of infinitesimal RKKY
interactions. To leading order we replace 
\[\chi(\omega_{n})\longrightarrow
\chi_{o}\left(  \omega_{n}\right)  \equiv\chi(\omega_{n} ;J=0),\] 
and the calculation reduces to the {}\textquotedblleft bare
model\textquotedblright\ of Ref.~\shortcite{tanaskovic-2005-95}. In this case, $P\left(  T_{K}\right)  \sim T_{K}^{\alpha-1}$, where
$\alpha\sim1/W^{2}$, and%
\begin{equation}
\chi_{0}\left(  \omega_{n}\right)  \sim\int dT_{K}P\left(  T_{K}\right)
\chi\left(  \omega_{n},T_{K}\right)  \sim\chi_{0}\left(  0\right)
-C_{0}\left|  \omega_{n}\right|  ^{1-\epsilon}, \label{eq:chizero}%
\end{equation}
where $\epsilon=2-\alpha$. Thus, for sufficiently strong disorder (i.e. within
the electronic Griffiths phase), even the {}``bare'' bosonic bath is
sufficiently singular to generate decoupling. The critical value of $W$ will
be modified by self-consistency, but it is clear that decoupling will occur
for sufficiently large disorder.

Once decoupling is present, the system is best viewed as composed of two
fluids, one made up of a fraction $n$ of decoupled spins, and the other of a
fraction $(1-n)$ of Kondo screened spins. The self-consistent $\chi\left(
\omega_{n}\right)  $ acquires contributions from both fluids
\begin{equation}
\chi\left(  \omega_{n}\right)  =n\chi_{dc}\left(  \omega_{n}\right)  +\left(
1-n\right)  \chi_{s}\left(  \omega_{n}\right)  . \label{eq:twofluidchi}
\end{equation}
A careful analysis \shortcite{tanaskovic-2005-95} shows that, for a bath
characterized by an exponent $\epsilon$
\begin{align}
\chi_{dc}\left(  \omega_{n}\right)   &  \sim\chi_{dc}\left(  0\right)
-C\left|  \omega_{n}\right|  ^{1-\left(  2-\epsilon\right)  }%
;\label{eq:decoupledchi}\\
\chi_{s}\left(  \omega_{n}\right)   &  \sim\chi_{s}\left(  0\right)
-C^{\prime}\left|  \omega_{n}\right|  ^{1-\left(  2-\epsilon-1/\nu\right)  },
\label{eq:screenedchi}
\end{align}
where $\nu=\nu\left(  \epsilon\right)  $ is a critical exponent governing how
the Kondo scale vanishes at the quantum critical point of the Bose-Fermi
model. Since $\nu>0$, the contribution of the decoupled fluid is more singular
and dominates at lower frequencies. Self-consistency then yields $\epsilon=1$,
as in the familiar spin liquid state of Sachdev and Ye \shortcite{sachdevye}. For
$\epsilon=1$, the local susceptibility is logarithmically divergent (both in
$\omega_{n}$ and $T$). This does not necessarily mean that the bulk
susceptibility, which is the experimentally relevant quantity, behaves in the
same manner \shortcite{parcolletgeorges}. More work remains to be done to determine
the precise low temperature form of this and other physical quantities and to
assert the relevance of this mechanism for specific materials.

\begin{figure}[ptb]
\begin{center}
\includegraphics[  width=3.4in,keepaspectratio]{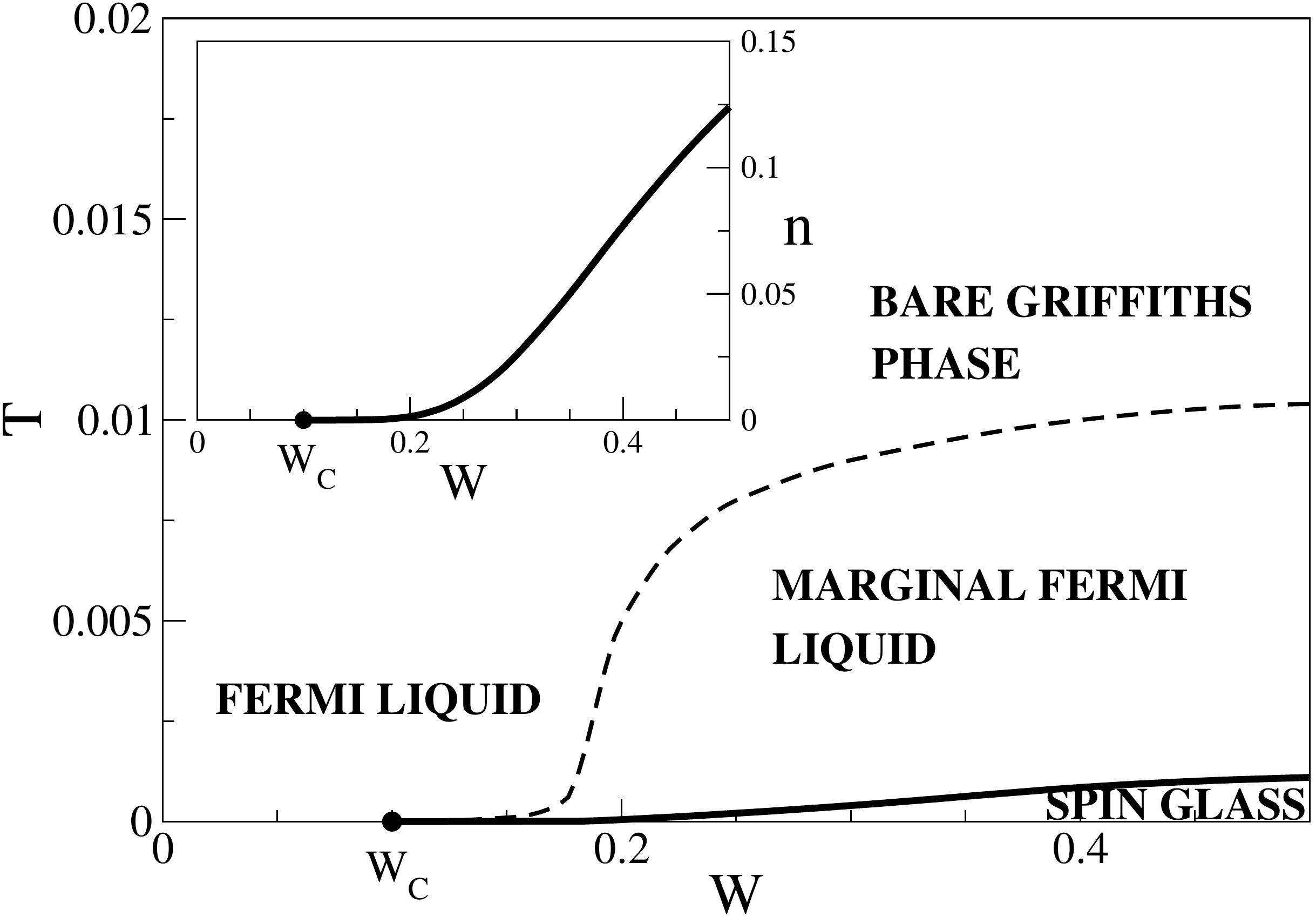}
\end{center}
\caption{Phase diagram of the electronic Griffiths phase model with RKKY
interactions \protect\shortcite{tanaskovic-2005-95}. The inset shows the fraction of
decoupled spins as a function of the disorder strength $W$.}
\label{cap:fig16}
\end{figure}

As in the case where conduction electron disorder is absent, the spin liquid
state is unstable towards spin-glass ordering at sufficiently low
temperatures. However, numerical estimates for the Griffiths phase model
\shortcite{tanaskovic-2005-95} suggest a surprisingly wide temperature window where
the marginal behavior should persist above the ordering temperature.
Fig.~\ref{cap:fig16} represents the predicted phase diagram of this model. For
weak disorder the system is in the Fermi liquid phase, while for $W>W_{c}$ the
marginal Fermi liquid phase emerges. The crossover temperature (dashed line)
delimiting this regime can be estimated from the frequency up to which the
logarithmic behavior of the local dynamical susceptibility $\chi(i\omega) $ is
observed. The spin glass phase, obtained from Eq.~(\ref{eq:sg}), appears only
at the lowest temperatures, well below the marginal Fermi liquid boundary.
Interestingly, recent experiments \shortcite{dougetal} have indeed found evidence
of dynamical spin freezing in the milliKelvin temperature range for the same
Kondo alloys that display normal phase non-Fermi-liquid behavior in a much broader
temperature window. 

The two-fluid phenomenology of the disordered Kondo lattice we have described
above is very reminiscent of earlier work on the clean Kondo lattice, where
the conduction electrons effectively decouple from the local moments, the
latter forming a spin liquid state
\shortcite{colemanandrei,kaganetal92,demleretal02,senthiletal,senthiletal2}. The
major difference between the results presented in this Section and these other
cases is that here local spatial disorder fluctuations lead to an
\emph{inhomogeneous} coexistence of the two fluids, as each site decouples or
not from the conduction electrons depending on its local properties. The
discussed mean-field models should be considered as merely the first examples
of this fascinating physics. The specific features of the spin liquid behavior
that was obtained from these models may very well prove to be too restrictive
and perhaps even inaccurate. For example, the specific heat enhancements may
well be overestimated, reflecting the residual $T=0$ entropy of the mean-field
models. Nevertheless, the physics of Kondo screening being destroyed by the
interplay of disorder and RKKY interactions will almost certainly play a
central role in determining the properties of many non-Fermi-liquid systems, and clearly
needs to be investigated in more detail in the future.

\subsection{Electron glass}

Another aspect of disordered interacting electrons poses a
fundamental problem. Very generally, Coulomb repulsion favors a
uniform electronic density, while disorder favors local density
fluctuations. When these two effects are comparable in magnitude,
one can expect many different low energy electronic
configurations, i.e. the emergence of many {\em metastable
states}. Similarly as in other ``frustrated'' systems with
disorder, such as spin glasses, these processes can be expected to
lead to {\em glassy} behavior of the electrons, and the associated
anomalously slow relaxational dynamics. Indeed, both theoretical
\shortcite{eglass1,eglass2} and experimental
\shortcite{films22,films23,bogdanovich-prl02,JJPRL02} work has found
evidence of such behavior deep on the insulating side of the
transition. However, at present very little is known as to the
precise role of such processes in the critical region.
Nevertheless, it is plausible that the glassy freezing of the
electrons must be important, since the associated slow relaxation
clearly will reduce the mobility of the electrons.  From this
point of view, the glassy freezing of electrons may be considered,
in addition to the Anderson and the Mott mechanism, as a third
fundamental process associated with electron localization.
Interest in understanding the glassy aspects of electron dynamics
has experienced a genuine renaissance in the last few years,
primarily due to experimental advances. Emergence of many
metastable states, slow relaxation and incoherent transport have
been observed in a number of strongly correlated electronic
systems. These included transition metal oxides such as high T$_c$
materials, manganites, and ruthenates. Similar features have
recently been reported in two-dimensional electron gases and even
three-dimensional doped semiconductors such as Si:P.

\begin{itemize}
\item Infinite-dimensional model of an electron glass
\end{itemize}

The interplay of the electron-electron interactions and disorder
is particularly evident deep on the insulating side of the
metal-insulator transition (MIT). Here, both experimental
\shortcite{re:Massey96} and theoretical studies \shortcite{re:Efros75} have
demonstrated that they can lead to the formation of a soft
\textquotedblleft Coulomb gap\textquotedblright, a phenomenon that
is believed to be related to the glassy behavior
\shortcite{films22,films23,bogdanovich-prl02,JJPRL02} of
the electrons. Such glassy freezing has long been suspected
\shortcite{re:Belitz95} to be of importance, but more recent theoretical work
\shortcite{gang4me,Sudip} has suggested that it may even dominate the
MIT behavior in certain low carrier density systems. The classic
work of Efros and Shklovskii \shortcite{re:Efros75} has clarified some
basic aspects of this behavior, but a number of key questions have
remain unanswered.

 As a simplest
example \shortcite{pastor-prl99}  displaying glassy behavior of electrons, we
focus on a simple lattice model of spinless electrons with nearest
neighbor repulsion $V$ in presence of random site energies
$\varepsilon_i$ and inter-site hopping $t$, as given by the
Hamiltonian

\begin{equation}
H=\sum_{<ij>} ( -t_{ij} + \varepsilon_i \delta_{ij})
c^{\dagger}_{i}c_{j} + V\sum_{<ij>}c^{\dagger}_{i}c_{i}
c^{\dagger}_{j}c_{j}.\label{eq:hamil1}
\end{equation}

This model can be solved  in a properly defined limit
of large coordination number \shortcite{georgesrmp}, where an
extended dynamical mean-field theory (EDMFT) formulation becomes exact. We
concentrate on the situation where the disorder (or more generally
frustration) is large enough to suppress any uniform ordering. We
then rescale both the hopping elements and the interaction
amplitudes as $t_{ij} \rightarrow t_{ij}/\sqrt{z}$; \hspace{6pt}
$V_{ij} \rightarrow V_{ij}/\sqrt{z}$. As we will see shortly, the
required fluctuations then survive even in the
$z\rightarrow\infty$ limit, allowing for the existence of the
glassy phase. Within this model:

\begin{description}

\item [$\star$]The universal form  of the Coulomb gap \protect\shortcite{re:Efros75}
proves to be a direct consequence of glassy freezing.

\item  [$\star$] The glass phase is identified through the emergence of an
extensive number of metastable states, which in our formulation is
manifested as a replica symmetry breaking instability
\protect \shortcite{re:Mezard86}.

\item  [$\star$] As a consequence of this ergodicity breaking
\protect \shortcite{re:Mezard86}, the zero-field cooled compressibility is found
to vanish at T=0, suggesting the absence of screening
\protect\shortcite{re:Efros75} in disordered insulators.

\item  [$\star$] The quantum fluctuations can melt this glass even at $T=0$,
but the relevant energy scale is set by the electronic mobility 
and is therefore a nontrivial function of disorder.

\end{description}

We should stress that although this model allows us to
examine the interplay of glassy ordering and quantum fluctuations
due to itinerant electrons, it is too simple to describe the
effects of Anderson localization. These effects require extensions
to lattices with finite coordination and will be discussed in
the next section.

For simplicity, we focus on a Bethe lattice at half filling and
examine the $z\rightarrow\infty$ limit.  This strategy
automatically introduces the correct order parameters and after
standard manipulations \shortcite{dk-prb94} the problem
reduces to a self-consistently defined single site problem, as
defined by an the effective action of the form

\begin{eqnarray}
&&S_{eff} (i) =\sum_{a}\int_o^{\beta}\int_o^{\beta }d\tau d\tau
'\;[ c^{\dagger a}_i (\tau )( \delta (\tau -\tau ')
\partial_{\tau} +\varepsilon_i \nonumber
+ t^2 G (\tau,\tau ') )c^{a}_i (\tau ')\\
&& +\frac{1}{2}V^2\delta n_i^{a}(\tau )\chi (\tau ,\tau ') \delta
n_i^{a} (\tau ')] +\frac{1}{2}V^2\sum_{a\neq b}\int_o^{\beta
}\int_o^{\beta }d\tau d\tau ' \; \delta n_i^{a}(\tau )\; q_{ab}\;
\delta n_i^{b} (\tau ').\label{eq:ch2_seff}
\end{eqnarray}
Here, we have used functional integration over replicated
Grassmann fields \shortcite{dk-prb94} $c_{i}^{a} (\tau )$
that represent electrons on site $i$ and replica index $a$, and
the random site energies $\varepsilon_i$ are distributed according
to a given probability distribution $P(\varepsilon_i )$.  The
operators $\delta n_{i}^{a}(\tau )=( c^{\dagger a}_{i} (\tau
)c_{i}^{a} (\tau ) - 1/2)$ represent the {\em density
fluctuations} from half filling. The order parameters $G (\tau
-\tau ')$, $\chi(\tau -\tau ')$ and $q_{ab}$ satisfy the following
set of self-consistency conditions

\begin{eqnarray}
&&G (\tau -\tau ')=\int d\varepsilon_i P(\varepsilon_i )
< c^{\dagger a}_{i} (\tau )c_{i}^{a} (\tau ')>_{eff},\label{self1}\\
&&\chi (\tau -\tau ')=\int d\varepsilon_i P(\varepsilon_i )
< \delta n^{\dagger a}_{i} (\tau )\delta n_{i}^{a} (\tau ')>_{eff},\label{self2}\\
&&q_{ab}=\int d\varepsilon_i P(\varepsilon_i ) < \delta n^{\dagger
a}_{i} (\tau )\delta n_{i}^{b} (\tau ')>_{eff}\label{self3}.
\end{eqnarray}

\begin{itemize}
\item Order parameters
\end{itemize}

In these equations, the averages are taken with respect to the
effective action of Eq. (\ref{eq:ch2_seff}). Physically, the
``hybridization function'' $t^2 G (\tau -\tau ')$ represents the
single-particle electronic spectrum of the environment, as seen by
an electron on site $i$.  In particular, its imaginary part at
zero frequency can be interpreted \shortcite{dk-prb94} as
the inverse lifetime of the local electron and as such remains
finite as long as the system is metallic.  We recall
\shortcite{dk-prb94} that for $V=0$ these equations reduce
to the familiar CPA description of disordered electrons, which is
exact for $z=\infty$.  The second quantity $\chi(\tau -\tau ')$
represents an (interaction-induced) {\em mode-coupling} term that
reflects the {\em retarded} response of the density fluctuations
of the environment. Note that very similar objects appear in the
well-known mode-coupling theories of the glass transition in dense
liquids \shortcite{re:Cummins94}.  Finally the quantity $q_{ab}$
$(a\neq b)$ is nothing but the familiar Edwards-Anderson order
parameter $q_{EA}$. Its nonzero value indicates that the time
averaged electronic density is spatially non-uniform.

\begin{itemize}
\item Equivalent Infinite Range model
\end{itemize}

From a technical point of view, a RSB analysis is typically
carried out by focusing on a free energy expressed as a functional
of the order parameters. In our Bethe lattice approach, one
directly obtains the self-consistency conditions from appropriate
recursion relations \shortcite{dk-prb94}, without invoking a
free energy functional. However, we have found it useful to map
our $z=\infty$ model to another {\em infinite range} model, which
has {\em exactly} the same set of order parameters and
self-consistency conditions, but for which an appropriate free
energy functional can easily be determined. The relevant model is
still given Eq. (\ref{eq:hamil1}), but this time with {\em random} hopping
elements $t_{ij}$ and {\em random} nearest-neighbor interaction
$V_{ij}$, having zero mean and variances $t^2$, and $V^2$,
respectively. For this model, standard manipulations
\shortcite{dk-prb94} result in the following free energy
functional

\begin{eqnarray}
F [G , \chi , q_{ab}] =
&-&\frac{1}{2}\sum_{a}\int_o^{\beta}\int_o^{\beta }d\tau d\tau'
\;[ t^2 G^2 (\tau ,\tau ' ) + V^2 \chi^2 (\tau ,\tau ' )]
-\frac{1}{2}\sum_{a\neq b}(\beta V)^2 q^2_{ab}\nonumber\\ &-& \ln
\left[ \int d\varepsilon_i P(\varepsilon_i ) \int Dc^{\dagger a}_i
Dc^a_i \exp\left\{ -S_{eff} (i) \right\}\right],\label{free1}
\end{eqnarray}
with $S_{eff} (i)$ given by Eq. (\ref{eq:ch2_seff}). The self-consistency
conditions, Eqs. (\ref{self1}-\ref{self3}) then follow from

\begin{equation}
0=\delta F/\delta G (\tau ,\tau ' ); \; 0=\delta F/\delta \chi
(\tau ,\tau ' ); \; 0=\delta F/\delta q_{ab}.
\end{equation}

We stress that Eqs. (\ref{self1}-\ref{self3}) have been derived for the model with
{\em uniform} hopping elements $t_{ij}$ and interaction amplitudes
$V_{ij}$, in the $z\rightarrow\infty$ limit, but the {\em same}
equations hold for an infinite range model where these parameters
are random variables.

\begin{itemize}
\item The glass transition
\end{itemize}

In our electronic model, the random site energies $\varepsilon_i$
play a role of static random fields. As a result, in the presence of
disorder, the Edwards-Anderson  parameter $q_{EA}$ remains nonzero
for any temperature, and thus cannot serve as an order parameter.
To identify  the glass transition, we search for a replica
symmetry breaking (RSB) instability, following standard methods
\shortcite{re:AT,re:Mezard92}. We define $\delta q_{ab} = q_{ab} -q$,
and expand the free energy functional of Eq. (\ref{free1}) around the replica symmetric
solution. The resulting quadratic form (Hessian matrix) has the
matrix elements given by
\begin{eqnarray}
\frac{\partial^2 F}{\partial q_{ab} \partial q_{cd}}  &=& (\beta
V)^2 \delta_{ac}\delta_{bd} + [< \delta n _a (\tau_1 ) \delta n _b (\tau_2 )>_{RS} < \delta n_c
(\tau_3 )  \delta n _d (\tau_4 )>_{RS}\\
&-&V^4\int_{0}^{\beta}\int_{0}^{\beta}
\int_{0}^{\beta}\int_{0}^{\beta} d\tau_1 d\tau_2 d\tau_3 d\tau_4
[< \delta n _a (\tau_1 ) \delta n _b (\tau_2 ) \delta n _c (\tau_3
) \delta n _d (\tau_4 )>_{RS}]\nonumber,
\label{eq:ch2_hessian}
\end{eqnarray}
where the expectation values are calculated in the RS solution.
Using standard manipulations \shortcite{re:AT}, and after lengthy
algebra, we finally arrive at the desired RSB stability criterion
that takes the form

\begin{equation}
1 = V^2 \left[\left(\chi_{loc} (\varepsilon_i
)\right]^2\right]_{dis}.\label{rsb}
\end{equation}
Here, $[...]_{dis}$ indicates the average over disorder and
$\chi_{loc} (\varepsilon_i )$ is the {\em local compressibility},
that can be expressed as
\begin{equation}
\chi_{loc} (\varepsilon_i )= \frac{\partial}{\partial\varepsilon_i
} \frac{1}{\beta}\int_{o}^{\beta}d\tau <\delta n_i (\tau )>,\label{eq:13}
\end{equation}
and which is evaluated by carrying out quantum averages for a
fixed realization of disorder. The relevant expectation values
have to be carried with respect to the full local effective action
$S_{eff} (i)$ of Eq. (2), evaluated in the RS theory. In general,
the required computations cannot be carried out in closed form,
primarily due to the unknown ``memory kernel'' $\chi(\tau -\tau
')$. However, as we will see, the algebra simplifies in several
limits, where explicit expressions can be obtained.

\begin{itemize}
\item Classical electron glass
\end{itemize}

In the classical $(t=0)$ limit, the problem can be easily solved
in closed form. We first focus on the replica symmetric (RS)
solution and set $q_{ab}=q$ for all replica pairs. The
corresponding equation reads

\begin{equation}
q = \frac{1}{4}\int_{-\infty}^{+\infty}\frac{dx}{\sqrt{\pi}}
e^{-x^2/2}\tanh^2\left[ \frac{1}{2}x \left( (\beta V)^2 q + (\beta
W)^2\right)^{1/2}\right],
\end{equation}
where we have considered a Gaussian distribution of random site
energies of variance $W^2$. Note that the interactions introduce
an effective, {\em enhanced} disorder strength
\begin{equation}
W_{eff}=\sqrt{W^2 +V^2q}\label{eq:rescaled_disorder},
\end{equation}
since the frozen-in density fluctuations introduce an added
component to the random potential seen by the electrons. As
expected, $q\neq 0$ for any temperature when $W\neq 0$. If the
interaction strength is appreciable as compared to disorder, we
thus expect the resistivity to display an appreciable {\em
increase} at low temperatures. We emphasize that this mechanism is
{\em different} from Anderson localization, which is going to be
discussed in the next section, but which also gives rise to a
resistivity increase at low temperatures.

Next, we examine the instability to glassy ordering. In the
classical ($t=0$) limit Eq. (\ref{rsb}) reduces to

\begin{equation}
1=\frac{1}{16}(\beta V)^2
\int_{-\infty}^{+\infty}\frac{dx}{\sqrt{\pi}}
e^{-x^2/2}\cosh^{-4}\left[ \frac{1}{2}x \beta W_{eff}(q)\right],
\end{equation}
with $W_{eff}(q)$ given by Eq.(\ref{eq:rescaled_disorder}).  The
resulting RSB instability line separates a low temperature glassy
phase from a high temperature ``bad metal'' phase. At large
disorder, these expressions simplify and we find

\begin{equation}
T_G \approx \frac{1}{6\sqrt{2\pi}}\frac{V^2}{W},\;
W\rightarrow\infty .\label{eq:14}
\end{equation}
We conclude that $T_G$ decreases at large disorder. This is to be
expected, since in this limit the electrons drop into the lowest
potential minima of the random potential. This defines a unique
ground state, suppressing the {\em frustration} associated with
the glassy ordering and thus reducing the glassy phase. It is
important to note that for the well known de Almeida-Thouless (AT)
line $T_{RSB}$ decreases {\em exponentially} in the strong field
limit. In contrast, we find that in our case, $T_G\sim 1/W$
decreases only slowly in the strong disorder limit. This is
important, since the glassy phase is expected to be most relevant
for disorder strengths sufficient to suppress uniform ordering. At
the same time, glassy behavior will only be observable if the
associated glass transition temperature remains appreciable.

\begin{itemize}
\item The glassy phase
\end{itemize}

\begin{figure}
\begin{center}
\includegraphics[width=4in]{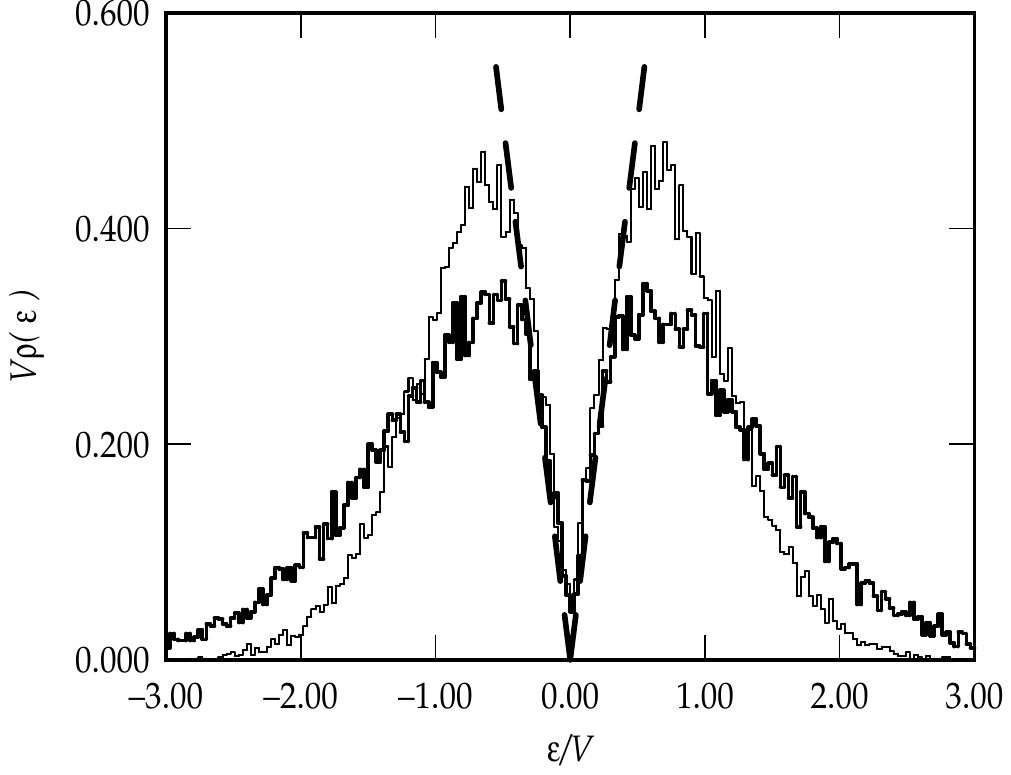}
\caption{Single particle density of states in the classical
($t=0$) limit at $T=0$, as a function of disorder strength.
Results are shown from a simulation on $N=200$ site system, for
$W/V = 0.5$ (thin line) and $W/V = 1.0$ (full  line). Note that
the low energy form of the gap takes a {\em universal} form,
independent of the disorder strength $W$. The dashed line follows
Eq. (\ref{eq:16}).}
\label{fig:dosclassical}
\end{center}
\end{figure}

To understand this behavior, we investigate the structure of the
low-temperature glass phase. Consider the single-particle density
of states at T=0, which in the classical limit can be expressed as
\begin{equation}
\overline{\rho} (\varepsilon , t=0)= \frac{1}{N} \sum_i \delta
(\varepsilon -\varepsilon^{R}_i ),
\end{equation}
where $\varepsilon_i^{R} \equiv \varepsilon_i + \sum V_{ij} n_j$
are the renormalized site energies.  In the thermodynamic limit,
this quantity is nothing but the probability distribution $P_R
(\varepsilon^{R}_i )$.  It is analogous to the ``local field
distribution'' in the spin-glass models and can be easily shown
to reduce to a simple Gaussian distribution in the RS theory,
establishing the {\em absence} of any gap for $T>T_G$. Obtaining
explicit results from a replica calculation in the glass phase is
more difficult, but useful insight can be achieved by using
standard simulation methods \shortcite{re:Palmer79,re:Pazmandi99} on
our equivalent infinite-range model; some typical results are
shown in Fig. \ref{fig:dosclassical}.
We find that as a result of glassy freezing, a pseudo-gap emerges
in the single-particle density of states, reminiscent of the
Coulomb gap of Efros and Shklovskii (ES) \shortcite{re:Efros75}. The
low energy form of this gap appears {\em universal},
\begin{equation}
\rho (\varepsilon)  \approx C\varepsilon^{\alpha}/V^2,\; \mathrm{for}\;\varepsilon < E_g; \;\;\;
C=\alpha = 1,\label{eq:16}
\end{equation}
independent of the disorder strength $W$, again in striking
analogy with the predictions of ES. To establish this result, we
have used stability arguments very similar to those developed for
spin-glass (SG) models \shortcite{re:Pazmandi99}, demonstrating that
the form of Eq. (\ref{eq:16}) represents an exact {\em upper bound} for
$\rho(\varepsilon )$. For {\em infinite-ranged} SG models, as in
our case, this bound appears to be {\em saturated}, leading to
universal behavior. Such universality is often associated with a
critical, self-organized state of the system. Recent work
\shortcite{re:Pazmandi99} finds strong numerical evidence of such
criticality for SG models; we believe that the universal gap form
in our case has the same origin. Furthermore, assuming that the
universal form of Eq. (\ref{eq:16}) is obeyed immediately allows for an
estimate of $T_G (W)$.  Using Eq. (\ref{rsb}) to estimate the gap size
for large disorder gives $T_G \sim E_g \sim V^2 /W$, in agreement
with Eq. (\ref{eq:14}).

The ergodicity breaking associated with the glassy freezing has
important consequences for our model. Again, using the close
similarity of our classical infinite range model to standard SG
models \shortcite{re:Mezard86}, it is not difficult to see that the
{\em zero-field cooled} (ZFC) compressibility vanishes at $T=0$,
in contrast to the field-cooled one, which remains finite.
Essentially, if the chemical potential is modified {\em after} the
system is cooled to $T=0$, the system immediately falls out of
equilibrium and displays hysteretic behavior \shortcite{re:Pazmandi99}
with vanishing {\em typical} compressibility. If this behavior
persists in finite dimensions and for more realistic Coulomb
interactions, it could explain the absence of screening in
disordered insulators.

\begin{itemize}
\item Arbitrary lattices and finite coordination: mean-field glassy
phase of the random-field Ising model
\end{itemize}

The simplest theories of glassy freezing \shortcite{re:Mezard86} are
obtained by examining models with random inter-site interactions.
In the case of disordered electronic systems, the interactions are
not random, but glassiness still emerges due to frustration
introduced by the competition of the interactions and disorder. As
we have seen for the Bethe lattice \shortcite{pastor-prl99}, random
interactions are generated by renormalization effects, so that
standard DMFT approaches can still be used. However, one would
like to develop systematic approaches for arbitrary lattices and
in finite coordination. These issues already appear on the
classical level, where our model reduces to the random-field Ising
model (RFIM) \shortcite{re:Nattermann97}. To investigate the glassy
behavior of the RFIM \shortcite{horbach02}, a systematic
approach that can incorporate short-range fluctuation corrections
to the standard Bragg-Williams theory is the method of
Plefka \shortcite{re:Plefka} and Georges et al. \shortcite{Georges90}.
This work has shown that: (i) corrections to even the lowest nontrivial order immediately
result in the appearance of a glassy phase for sufficiently strong
randomness; (ii) this low-order treatment is sufficient in the joint limit
of large coordination and strong disorder; (iii) the structure of the resulting glassy phase is characterized
by universal hysteresis and avalanche behavior emerging from the
self-organized criticality of the ordered state.

\begin{itemize}
\item Long-range Coulomb interactions and the Efros-Shklovskii gap
\end{itemize}

So far we focused on the limit of large coordination, where the EDMFT approximation 
is essentially exact and the resulting electron glass phase shows many similarities to 
the standard Parisi theory of spin glasses \shortcite{re:Mezard86}. While it produced many appealing results,
this mean-field approach has remained very controversial for short-range spin glasses
in finite dimensions; a competing ``droplet" theory \shortcite{fisherandhuse86} approach suggested that many of Parisi's
predictions may not survive in physical dimensions $d=2,3$. The situation is more promising 
in the case of the physically relevant long-ranged Coulomb interaction, where the prediction of the 
mean-field approach may possibly persist even in low dimensions. 

The most interesting test of these ideas relates to the possibility of describing the emergence of the ``Coulomb gap" in $d=2,3$ for localized electrons interacting via long-range Coulomb interactions, 
as first predicted, based on heuristic arguments, a long time ago \shortcite{re:Efros75}. Here, the 
single particle density of states is predicted to assume a powerlaw form
$g(\varepsilon) \sim\varepsilon^{\gamma}$, notably with a dimensionality-dependent exponent $\gamma = d-1$. 

Conventional mean-field theories typically produce universal dimensionality independent 
exponents and thus cannot be expected to explain this unfamiliar situation. On the other hand, 
the EDMFT \shortcite{chitra00prl}  approach does include the effects of spatial correlations, as the 
bosonic collective modes describing the ``cavity field'' are treated at a Gaussian level, similar 
to the familiar Hertz-Millis theories of quantum criticality  \shortcite{hertz,millis}. 
When applied to the case of long-range Coulomb interactions, the corresponding plasmon 
propagator does reflect \shortcite{chitra00prl} both the specific 
form of the long-range interactions and the dimensionality of the system. When applied 
to disordered electrons, the replica version of this method
produces self-consistency conditions which assume a very similar form as in ordinary 
Parisi theory or the simple infinite range electron glass model
we examined in previous sections. Indeed, recent work has generalized our EDMFT method 
to the long-range models, where the glassy phase of the 
generalized Coulomb interactions \shortcite{pankov05prl} of the form $V(R) \sim 1/R^{\alpha}$ 
has been examined in finite dimensions. We will not elaborate of the details of these theories 
here, but we emphasize that most qualitative features of the resulting glassy phase have 
been found to share very similar behavior \shortcite{muller04prl} as for the infinite-range model 
of electron glasses, including the corresponding pattern of replica symmetry breaking 
\shortcite{pankov06prl,pankov07prb}. Most remarkably, however, this theory produced \shortcite{pankov05prl} 
an interaction-range and dimensionality dependent Coulomb gap exponent
\begin{equation} 
\gamma=(d-\alpha)/\alpha,
\end{equation}
in precise agreement with an appropriate generalization of the Efros-Shklovskii argument \shortcite{re:Efros75}. 
This early result was later confirmed \shortcite{pankov07prb} by a more detailed low-temperature solution 
of the same equations deep in the replica-symmetry broken phase, based on a $T=0$ new solution 
of the Parisi equation due to Pankov \shortcite{pankov06prl}. 

\begin{figure}[h]
\begin{center}
\includegraphics[width=3in]{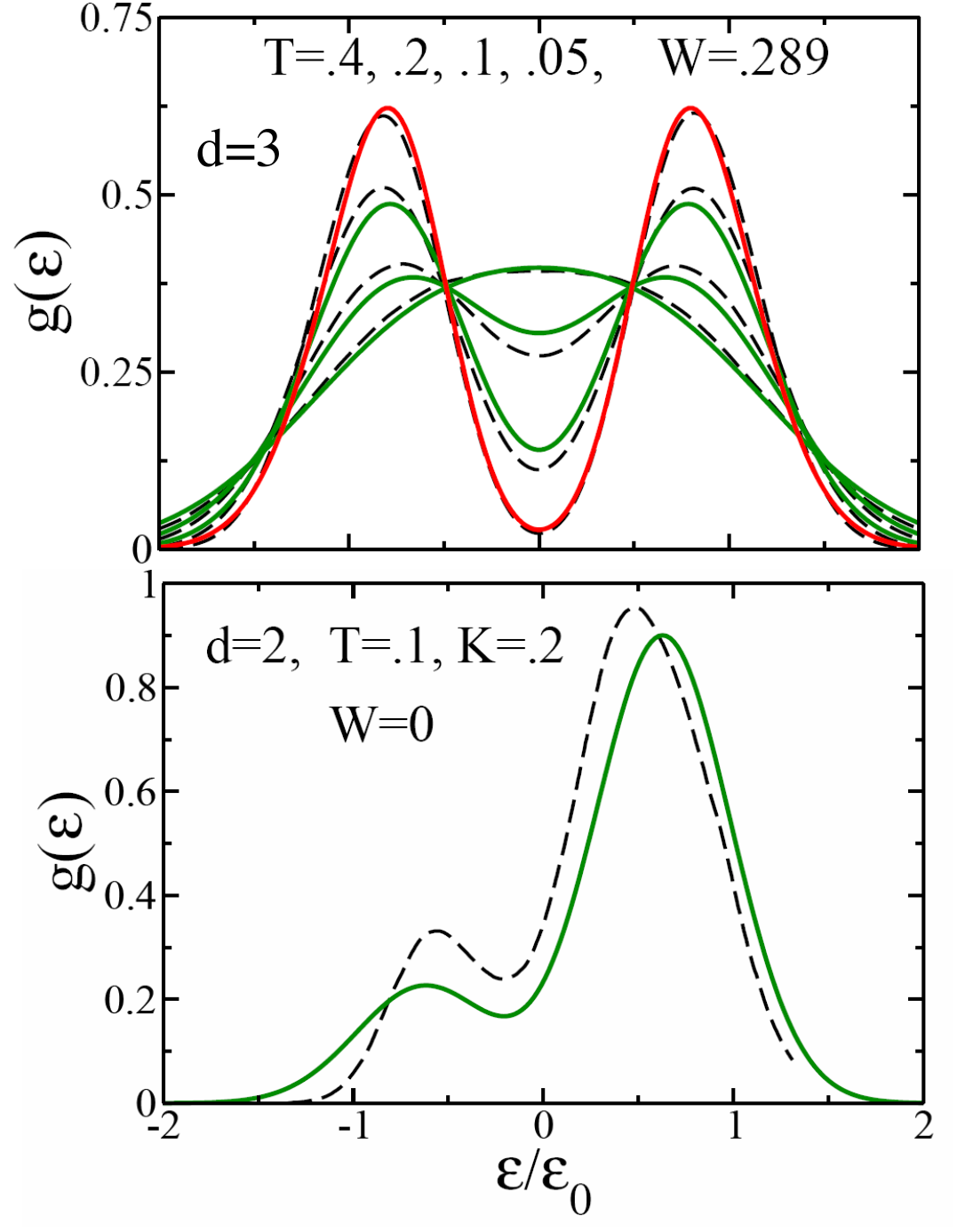}\end{center}
\caption{The analytical EDMFT predictions \protect\shortcite{pankov05prl}  for the single-particle
density of states (full lines) are found to be in excellent
quantitative agreement with simulation results (dashed lines),
with no adjustable parameters. Shown are results for the three
dimensional case studied in Ref. \protect\shortcite{re:Sarvestani95},
corresponding to $W=1/(2\sqrt{3})$, and temperatures $T =0.4, 0.2,
0.1, 0.05$ (top panel), and the two dimensional model of Ref.
\protect\shortcite{efros92prl}, corresponding to $W=0$, $T=0.1$, $K=0.2$. Green
lines correspond to the fluid pseudogap phase,
while the red curve corresponds to $T=0.05$, close to the glass transition temperature. }%
\label{coulomb_gap}%
\end{figure}

Another striking result of these theories should be noted. Namely, one finds \shortcite{pankov05prl} 
that the Coulomb pseudogap starts to form around a crossover temperatures 
$T^* \sim T_C$ (the Coulomb energy), much above the glass transition 
temperature $T_G \approx 0.05 T_C$, and in perfect quantitative agreement 
(Fig. \ref{coulomb_gap} with earlier numerical work \shortcite{clareyu93prl}. 
The EDMFT theory, in fact, demonstrated the following: (1) the universal 
value of the Efros-Shklovskii exponent $\gamma$ is a direct consequence 
of the marginal stability of the glassy phase; (2) in contrast, the emergence 
of a non-universal Coulomb pseudogap in the high temperature regime 
$T_G < T < T^*$ is not related to glassy freezing and the breakdown of 
screening \shortcite{muller04prl,pankov07prb}, and is a robust effect found 
\shortcite{efros92prl} even in the absence of disorder. The two mechanisms for 
pseudogap formation have often been confused  \shortcite{clareyu93prl} in 
previous work, leading to incorrect and even misleading interpretations
of what determines its form in a given temperature range.

\begin{itemize}
\item Quantum melting of the electron glass
\end{itemize}

Next, we investigate how the glass transition temperature can be
depressed by quantum fluctuations introduced by inter-site
electron tunneling. For simplicity, we again focus on the simplest 
infinite-range model of the electron glass. As in other quantum glass problems, quantum
fluctuations introduce dynamics in the problem, and the relevant 
self-consistency equations cannot be solved in closed form for
general values of the parameters. In the following, we will see
that in the limit of large randomness an exact solution is
possible. The main source of difficulty in general quantum glass problems
relates to the existence of a self-consistently determined
``memory kernel'' $\chi (\tau -\tau ' )$ in the local effective
action. By the same reasoning as in the classical case, one can
also ignore this term since this quantity is also bounded.

\begin{figure}[h]
\begin{center}
\includegraphics[width=4in]{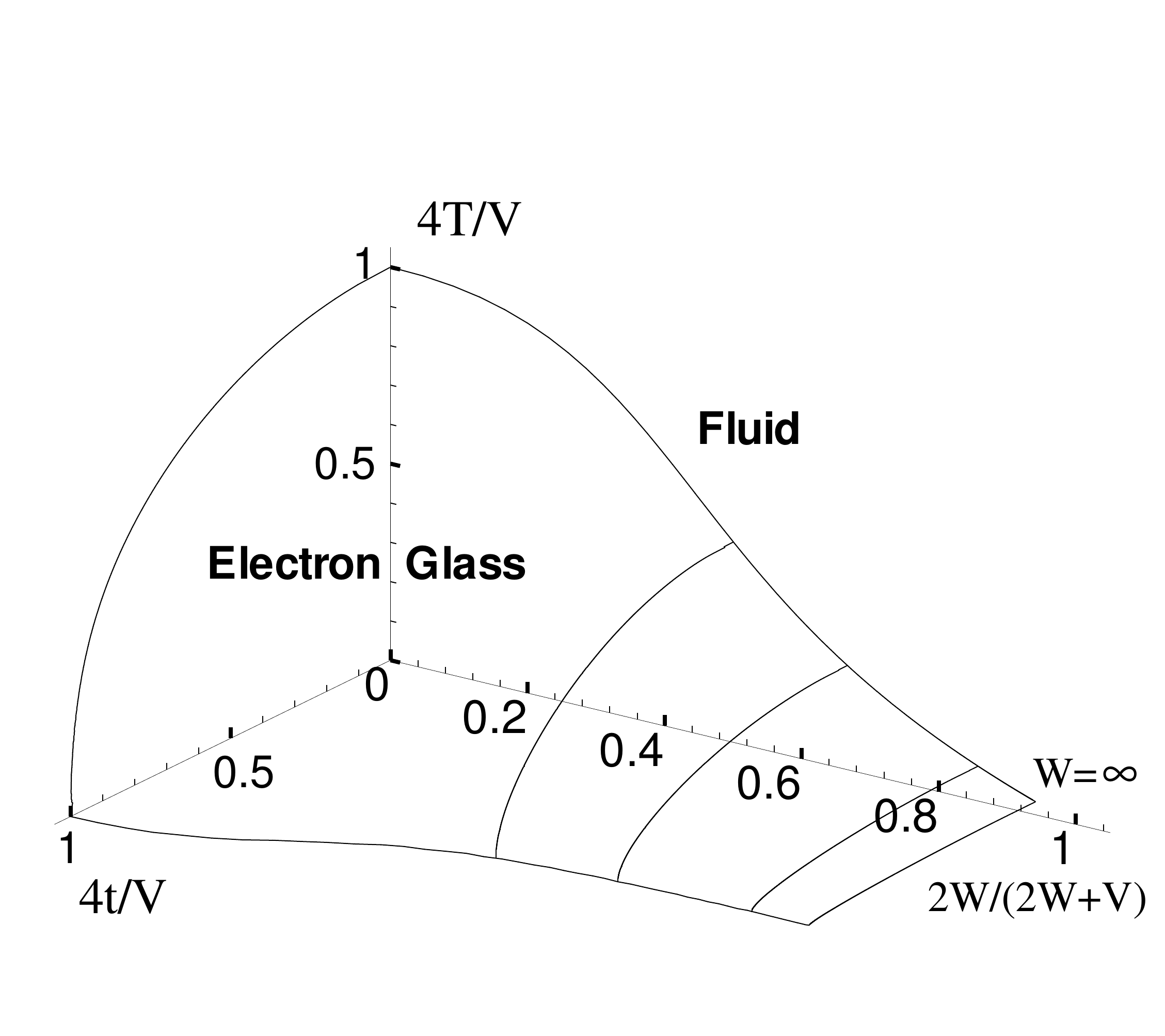}
\caption{Phase diagram as a function of quantum hopping $t$,
temperature $T$ and disorder strength $W$. The glass transition
temperature $T_G$ decreases only slowly (as $\sim1/W$) in the strong
disorder limit. In contrast, the critical value of the hoping
element $t_G$ remains finite as $W\rightarrow\infty$}
\label{fig:glassDMFT}
\end{center}
\end{figure}

The remaining action is that of \emph{noninteracting} electrons in the presence of a strong random potential. The resulting {\em  local}
compressibility then takes the form

\begin{equation}
\chi_{loc} (\varepsilon ) =\frac{\beta}{4}\int_{-\infty}^{+\infty}
d\omega \rho_{\varepsilon}(\omega ) \cosh^{-2}
(\frac{1}{2}\beta\omega ).
\end{equation}
Here,  $\rho_{\varepsilon}(\omega )$ is the local density of
states, which in the considered large $z$ limit is determined by
the solution of the CPA equation
\begin{equation}
\rho_{\varepsilon}(\omega )=-\frac{1}{\pi}{\rm Im} G(\omega);\;\;
G(\omega )= \int \frac{d\varepsilon P(\varepsilon )}{ \omega+i\eta
-\varepsilon -t^2 G(\omega )},
\end{equation}
In the limit $W/t >> 1$, it reduces to a narrow resonance of width
$\Delta =\pi t^2 P(0)\sim t^2/W$
\begin{equation}
\rho_{\varepsilon}(\omega ) \approx \frac{1}{\pi } \frac{\Delta
}{(\omega -\varepsilon )^2 +\Delta^2}.
\end{equation}

The resulting expression for the quantum critical line in the
large disorder limit takes the form
\begin{equation}
t_G (T=0, W\rightarrow\infty )= V/\sqrt{\pi }.\label{eq:24}
\end{equation}

At first glance, this result is surprising, since it means that a
{\em finite} value of the Fermi energy is required to melt the
electron glass at $T=0$, {\em even} in the $W\rightarrow\infty$
limit! This is to be contrasted with the behavior of $T_G$ in the
classical limit, which according to Eq. (\ref{eq:14}) was found to decrease
as $1/W$ for strong disorder. At first puzzling, the above result
in fact has a simple physical meaning. Namely, the small resonance
width (or ``hybridization energy'') $\Delta\sim t^2 /W$ can be
interpreted \shortcite{anderson58,motand} as the
characteristic energy scale for the electronic motion. As first
pointed out by Anderson \shortcite{anderson58}, according to Fermi's
golden rule, the transition rate to a neighboring site is
proportional to $\Delta$ and not $t$, and thus becomes extremely
small at large disorder. Thus the ``size'' of quantum
fluctuations, which replace the thermal fluctuations at $T=0$, is
proportional to $\Delta\sim 1/W$, and thus becomes very small in
the large $W$ limit. We can now easily understand the qualitative
behavior shown in Eq. (\ref{eq:24}) by replacing $T_G\rightarrow \Delta\sim
t_G^2 /W$ in Eq. (\ref{eq:14}). The leading $W$ dependence {\em cancels out},
and we find a {\em finite} value for $t_G$ in the
$W\rightarrow\infty$ limit.

More generally, we can write an expression for the glass
transition critical line in the large disorder limit, as a
function of $\beta = 1/T$ and $t$ in the scaling form

\begin{equation}
1=(V/t)^2 \phi (\beta t^2 /W),
\end{equation}
with
\begin{equation}
\phi(z) = \frac{1}{4} z^2 \int_{-\infty}^{+\infty} dx \left[
\int_{-\infty}^{+\infty} dy \frac{1}{\pi} \frac{1}{1+ (x-y)^2}
\cosh^{-2} (\frac{1}{2} zy )\right]^2 .
\end{equation}

At finite disorder an exact solution is not possible, but we can
make analytical progress  motivated by our discussion of the large
$W$ limit. Namely, one can imagine evaluating the required local
compressibilities in Eq. (\ref{eq:13}) by a ``weak coupling'' expansion in
powers of the interaction $V$. To leading order, this means
evaluating the compressibilities at $V=0$, an approximation which
becomes exact for $W$ large. Such an approximation can be tested
for other spin glass problems. We have carried out the
corresponding computations for the infinite range Ising spin glass
model in a transverse field, where the exact critical transverse
field is known from numerical studies. We can expect the leading
approximation to {\em underestimate} the size of the glassy
region, i. e. the critical field, since the omitted ``memory
kernel'' introduces long range correlations in time, which make
the system more ``classical''. Indeed, we find that the leading
approximation underestimates the critical field by only about
30\%, whereas the next order correction gives an error of less
than 5\%. Encouraged by these arguments, we use this
``weak-coupling'' approximation for arbitrary disorder strength
$W$. Again, the computation of the compressibility reduces to that
of noninteracting electrons in a CPA formulation; the resulting
phase diagram is shown in Fig. \ref{fig:glassDMFT}.

\begin{itemize}
\item Quantum critical behavior of the electron glass
\end{itemize}

So far, we have seen how our extended DMFT equations can be
simplified for large disorder, allowing an exact computation of
the phase boundary in this limit.  In our case, this quantum
critical line separates a (non-glassy) Fermi liquid phase from a
metallic glass phase which, as we will see, features non-Fermi
liquid behavior. If one is interested in the details of the {\em dynamics}
of the electrons near the quantum critical line, the above
simplifications do not apply and one is forced to
self-consistently calculate the form of the ``memory kernel" (local
dynamic compressibility) $\chi (\tau -\tau ')$. Fortunately, this
task can be carried out using methods very similar to those
developed for DMFT models for metallic spin
glasses \shortcite{re:Read95}. Formulating such a theory is technically
possible because the exact quantum critical behavior is captured
when the relevant field theory is examined at the Gaussian
level \shortcite{re:Miller93}, in the considered limit of large
dimensions.

Because of the technical complexity of this calculation, we only
report the main results, while the details can be found in 
\shortcite{Denis}. In this paper, the full replica-symmetry broken
(RSB) solution was found, both around the quantum critical line and
in the glassy phase. In the Fermi liquid phase, the memory kernel
was found to take the form
\[
V^2 \chi (\omega_n )=D(\omega_n)+\beta q_{{\rm
EA}}\delta_{\omega_n,0}, \]
with
 \[
D(\omega_n)= -y q_{{\rm EA}}^2 /V^4 -\sqrt{|\omega_n|+\Delta}.\]
Here, $\Delta$ is a characteristic energy scale that vanishes on
the critical line, which also determines the crossover temperature
scale separating the Fermi liquid from the quantum-critical
regime. In contrast to conventional quantum critical phenomena,
but similarly to metallic spin glasses, the ``gap" scale $\Delta$ 
is equal to zero not only on the critical line, but remains zero
\emph{throughout the entire glassy phase}. As a result, the
excitations in this region assume a non-Fermi liquid form
\[ D(\omega_n)= -y q_{{\rm EA}}^2 /V^4 -\sqrt{|\omega_n|}.\]
This behavior reflects the emergence of soft ``replicon"
modes \shortcite{re:Mezard86} describing in our case low
energy charge rearrangements inside the glassy phase. At finite
temperatures, electrons undergo inelastic scattering from such
collective excitations, leading to a temperature dependence of
the resistivity that takes the following non-Fermi liquid form
\[\rho (T) =\rho(o) +AT^{3/2}.\]
Interestingly, very recent experiments \shortcite{bogdanovich-prl02} on 
two-dimensional electron gases in silicon have revealed precisely such
a temperature dependence of the resistivity. This behavior has been
observed in what appears to be an intermediate metallic glass
phase separating a conventional (Fermi liquid) metal at high
carrier density from an insulator at the lowest densities.

Another interesting feature of the predicted quantum critical
behavior relates to the disorder dependence of the crossover exponent
$\phi$ describing how the gap scale $\Delta\sim\delta r^{\phi}$
vanishes as a function of the distance $\delta r$ from the
critical line. Calculations \shortcite{arrachea}  show that $\phi =2$ in the
presence of site energy disorder, which for our model plays the role
of a random symmetry breaking field, and $\phi =1$ in its absence.
This indicates that site disorder, which is common in disordered
electronic systems, produces a particularly large quantum critical
region, which could be the origin of the large dephasing observed in
many materials near the metal-insulator transition.

\begin{itemize}
\item Glassy behavior near the Mott-Anderson transition
\end{itemize}

As we have seen, the stability of the glassy phase is crucially
determined by the electronic mobility at $T=0$.  More precisely,
we have shown that the relevant energy scale that determines the
size of quantum fluctuations introduced by the electrons is given
by the local ``resonance width'' $\Delta$. It is important to
recall that precisely this quantity may be considered
\shortcite{anderson58} as an order parameter for Anderson
localization of noninteracting electrons. Recent work
\shortcite{motand,london} demonstrated that
the {\em typical} value of this quantity plays the same role even
at a Mott-Anderson transition. We thus expect $\Delta$ to
generally vanish in the insulating state. As a result, we expect
the stability of the glassy phase to be strongly affected by
Anderson localization effects, as we will explicitly demonstrate
in the next section.

On physical grounds, one expects the quantum fluctuations
\shortcite{pastor-prl99} associated with mobile electrons \ to suppress
glassy ordering, but their precise effects remain to be
elucidated. Note that even the \emph{amplitude }of such quantum
fluctuations must be a singular function of the distance to the
MIT, since they are dynamically determined by processes that
control the electronic mobility.

To clarify the situation, the following basic questions need to be
addressed: (1) Does the MIT\ coincide with the onset of glassy
behavior? (2) How do different physical processes that can
localize electrons \ affect the stability of the glass phase? In
the following, we provide simple and physically transparent \
answers to both questions. We find that: (a) Glassy behavior
generally emerges before the electrons localize; (b) Anderson
localization \shortcite{anderson58} enhances the stability of the
glassy phase, while Mott localization \shortcite{mott-book90} tends to
suppress it.

In order to be able to examine both the effects of Anderson and
Mott localization, we concentrate on extended Hubbard models given
by the Hamiltonian

\[
H=\sum_{ij\sigma}(-t_{ij}+\varepsilon_{i}\delta_{ij})c_{i,\sigma}^{\dagger
}c_{j,\sigma}+U\sum_{i}n_{i\uparrow}n_{i\downarrow}+\sum_{ij}V_{ij}\delta
n_{i}\delta n_{j}.
\]
Here, $\delta n_{i}=$ $n_{i}-$ $\left\langle n_{i}\right\rangle $
represent local density fluctuations ( $\left\langle
n_{i}\right\rangle $ is the site-averaged electron density), $U$
is the on-site interaction, and $\varepsilon_{i\ }$are$_{\
}$Gaussian distributed random site energies of variance $W^{2}$.
In order to allow for glassy freezing of electrons in the charge
sector, we introduce weak inter-site density-density interactions
$V_{ij}$, which we also choose to be Gaussian distributed
random variables of variance $V^{2}$ /$z$ ($z$ is the coordination
number). We emphasize that, in contrast to previous work
\shortcite{tmt}, we shall now keep the coordination number $z$
finite, in order to allow for the possibility of Anderson
localization. To investigate the emergence of glassy ordering, we
formally average over disorder by using standard replica methods
\shortcite{mitglass-prl03} and introduce collective $Q$-fields to decouple the
inter-site $V$-term \shortcite{mitglass-prl03}. A mean-field is then obtained by
evaluating the $Q$-fields at the saddle-point level. The resulting
stability criterion takes a form similar to the one previously discussed (Eq. \ref{rsb})

\begin{equation}
1-V^{2}%
{\displaystyle\sum\limits_{j}}
[\chi_{ij}^{2}]_{dis}=0. \label{instab}%
\end{equation}
Here, the non-local static compressibilities are defined (for a
fixed realization of disorder) as
\begin{equation}
\chi_{ij} =-\partial n_{i}/\partial\varepsilon_{j},%
\end{equation}
where $n_{i}$ is the local quantum expectation value of the
electron density, and $[\cdots]_{dis}$ represents the average over
disorder. Obviously, the stability of the glass phase is
determined by the behavior of the fourth-order correlation function
$\chi^{(2)} =\sum\limits_{j} [\chi_{ij}^{2}]_{dis}$ in the
vicinity of the metal-insulator transition. We emphasize that this
quantity is to be calculated in a disordered Hubbard model with
finite range hopping, i.e. in the vicinity of the Mott-Anderson
transition. The critical behavior of $\chi^{(2)}$ is very
difficult to calculate in general, but we will see that simple
results can be obtained in the limits of weak and strong disorder,
as follows.

\subsection{Large disorder}

As the disorder grows, the system approaches the Anderson
transition at $t=t_{c}(W)\sim W$. The first hint of singular
behavior of $\chi^{(2)}$ in an Anderson insulator is seen by
examining the deeply insulating, i. e., atomic limit $W\gg t,$
where to leading order we set $t=0$ and obtain
$\chi_{ij}=\delta(\varepsilon_{i}-\mu )\delta_{ij}$, i.e.
$\chi^{(2)}=[\delta^{2}(\varepsilon_{i}-\mu)]_{dis}=+\infty$! 
Since we expect all quantities to behave in
qualitatively the same fashion throughout the insulating phase, we
anticipate $\chi^{(2)}$ to diverge already at the Anderson
transition. Note that, since the instability of the glassy phase
occurs already at $\chi^{(2)}=V^{-2}$, the glass transition must
\emph{precede} the localization transition. Thus, for any finite
inter-site interaction $V$, we predict the emergence of an
intermediate \emph{metallic glass phase }separating the Fermi
liquid from the Anderson insulator. Assuming
that near the transition%

\begin{equation}
\chi^{(2)}\simeq\frac{A}{W^{2}}((t/W)-B)^{-\alpha}%
\end{equation}
($A$ and $B=t_{c}/W$ are constants of order unity), from Eq.
$\left( \ref{instab}\right)  $ we can estimate the form of the
glass transition line and we get
\begin{equation}
\delta t(W)=t_{G}(W)-t_{c}(W)\sim
V^{2/\alpha}W^{1-2/\alpha};\;W\rightarrow
\infty. \label{anderson}%
\end{equation}
The glass transition and the Anderson transition lines are
predicted to converge at large disorder for $\alpha<2$ and
diverge for $\alpha>2$. Since all the known exponents
characterizing the localization transition seem to grow with
dimensionality, we may expect a particularly large metallic glass
phase in large dimensions.

\begin{figure}[ptb]
\begin{center}
\includegraphics[width=4in ]{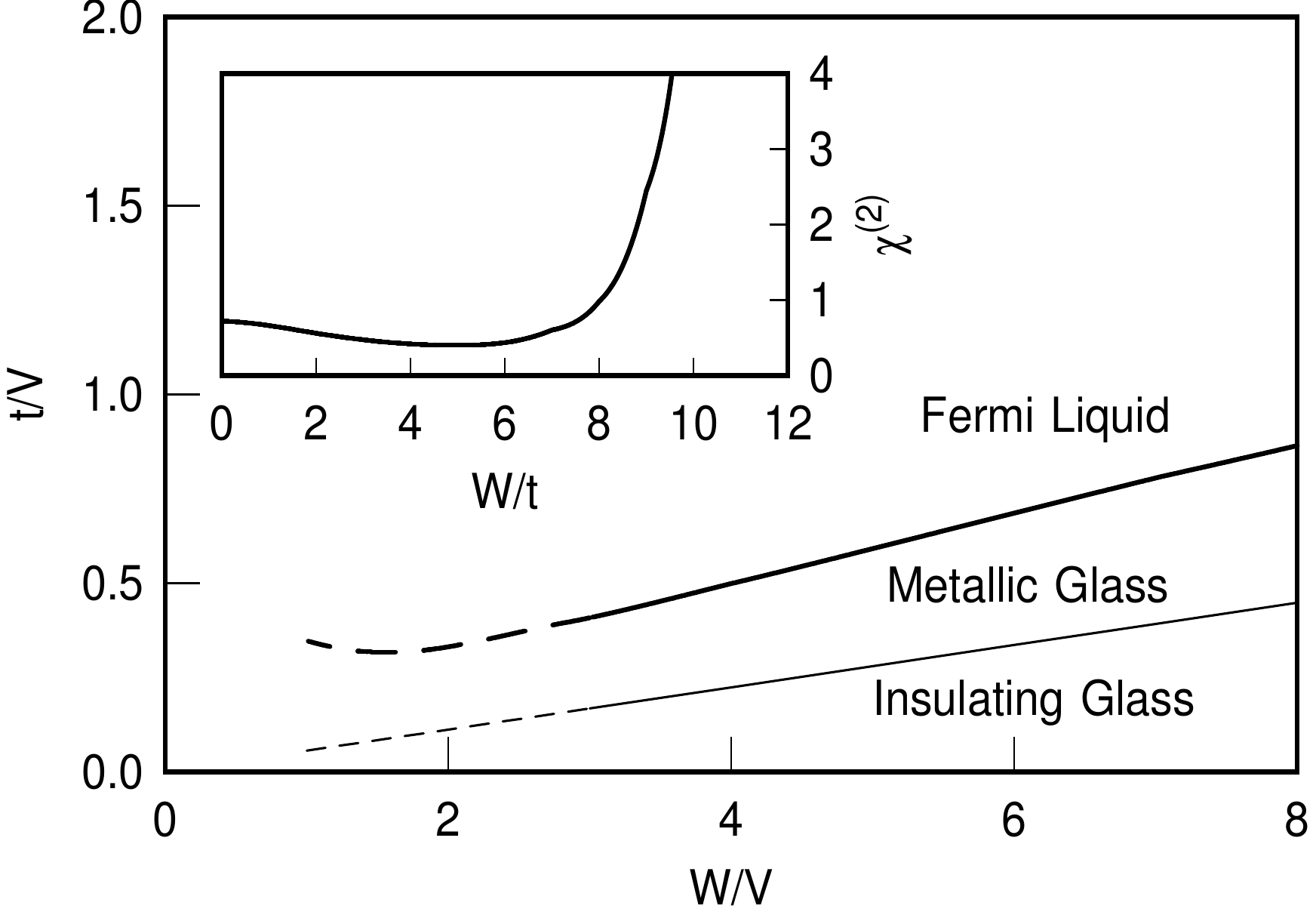}
\end{center}
\caption{Phase diagram for the $z=3$ Bethe lattice, valid in the
large
disorder limit. The inset shows $\chi^{(2)}$ as a function of disorder $W$.}%
\label{fig1}%
\end{figure}

\textit{Bethe lattices}. In order to confirm this scenario by explicit
calculations, we compute the behavior of $\chi^{(2)}$ at the
Anderson transition of a half-filled Bethe lattice of coordination
$z=3.$ We use an essentially exact numerical approach
\shortcite{motand} based on the recursive structure of
the Bethe lattice \shortcite{abouetal}. In this approach, local
and non-local Green's functions on a Bethe lattice can be sampled
from a large ensemble and the compressibilities $\chi_{ij}$ can
then be calculated by examining how a local charge density $n_{i}$
is modified by an infinitesimal variation of the local site energy
$\varepsilon_{j}$ on another site. To do this, we have taken
special care in evaluating the local charge densities $n_{i}$ by
numerically computing the required frequency summations over the
Matsubara axis, where the numerical difficulties are minimized.
Using this method, we have calculated $\chi^{(2)}$ as a function of
$W/t$ (for this lattice at half-filling $E_{F}=$ 2$\sqrt{2}t$),
and find that it decreases exponentially \shortcite{mirlinfyodorov} as the
Anderson transition is approached. We emphasize that only a finite
enhancement of $\chi^{(2)}$ is required to trigger the instability
to glassy ordering, which therefore occurs well before the
Anderson transition is reached. The resulting $T=0$ phase diagram,
valid in the limit of large disorder, is presented in Fig.~\ref{fig1}. Note
that the glass transition line in this case has the form
$t_{G}(W)\sim W$, in agreement with the fact that exponential
critical behavior of $\chi^{(2)}$ corresponds to $\alpha
\rightarrow\infty$ in the above general scenario. These results
are strikingly different from those obtained in a theory which
ignores localization \shortcite{pastor-prl99}, where $t_{G}(W)$ was found to
be weakly dependent on disorder and remain\emph{ finite} as
$W\longrightarrow\infty$. Anderson localization effects thus
strongly enhance the stability of the glass phase at sufficiently
large disorder. Nevertheless, since the Fermi liquid to metallic
glass (FMG) transition occurs at a finite distance \emph{before}
the localization transition, we do not expect the leading quantum
critical behavior \shortcite{Denis} at the FMG transition to be
qualitatively modified by localization effects.

\begin{figure}
\begin{center}
\includegraphics[width=4in]{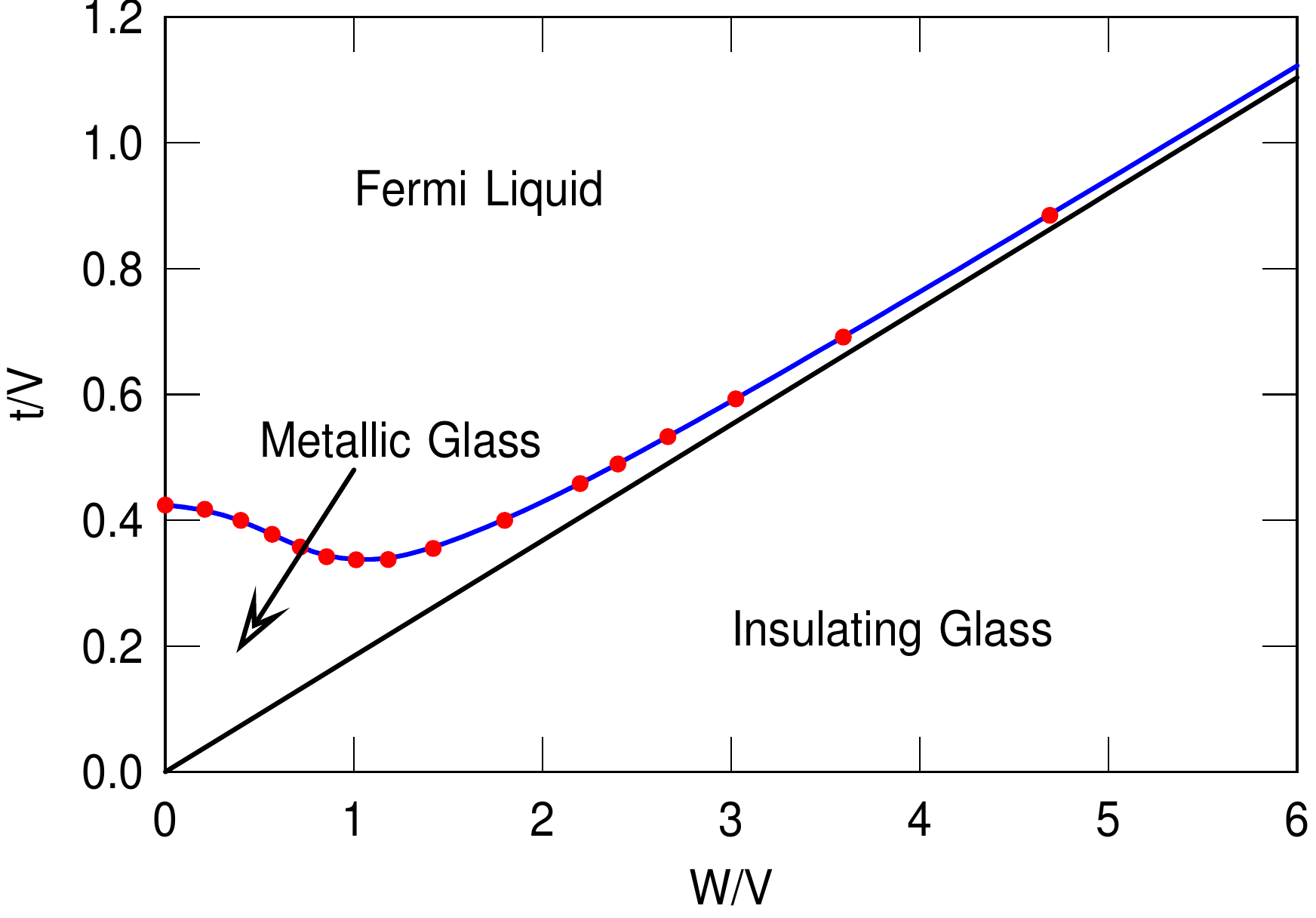}\caption{Phase
diagram from Typical Medium Theory of Anderson
localization \protect\shortcite{tmt}, giving $\alpha  =1$. The intermediate metallic
glass phase shrinks as disorder $W$ grows. Compare
this to the Bethe lattice case (Fig. \ref{fig1}), where $\alpha =\infty$.
\label{fig6}}\end{center}
\end{figure}

\emph{Typical medium treatment}. As an alternative approach to the Bethe lattice calculation, in
this section we introduce Anderson localization to the problem by
using the formalism of  ``Typical Medium Theory" (TMT) \shortcite{tmt}, which was explained in detail in Section \ref{sec:beyondDMFT}. We
calculate the cavity field $\Delta_{TYP}(\omega)$ by solving the
relevant self-consistency condition \shortcite{tmt}, which in turn
allows us to find local compressibilities: \begin{eqnarray}
 \chi_{ii}&=&-{\partial n\over\partial \varepsilon_i}={1\over\pi}{\partial\over\partial\varepsilon_i}\int_{-\infty}^0
 d\omega Im G(\varepsilon_i,\omega,W),\\
  G(\varepsilon_i,\omega,W)&=&{1\over\omega-\varepsilon_i-\Delta_{TYP}(\omega)},
\end{eqnarray} needed to determine the critical line of the glass
transition. These calculations were performed using a model
semicircular bare density of states $\rho_0(\omega)$  and a box distribution of
disorder $P(\varepsilon_i)$. The resulting phase diagram is shown in Fig. \ref{fig6}.
The intermediate  metallic glassy phase still exists, but shrinks
as $W\rightarrow\infty$, reflecting the small value of the
critical exponent $\alpha=1$, which can be shown analytically
within TMT. A more realistic value for this exponent, corresponding
to $d=3$, requires more detailed numerical calculations, which
remains a challenge for future work.

\emph{Low disorder - Mott transition}. In the limit of weak disorder $W\ll U,V$, interactions drive
the metal-insulator transition. Concentrating on the model at
half-filling, the system will undergo a Mott transition
\shortcite{mott-book90} as the hopping $t$ is sufficiently reduced. Since
for the Mott transition $t_{Mott}(U)\sim U$, near the transition
$W\ll t$ and to leading order we can ignore localization
effects. In addition, we assume that $V\ll U$ and to leading
order the compressibilities have to be calculated with respect to
the action $S_{el}$ of a disordered Hubbard model. The simplest
formulation that can describe the effects of weak disorder on such
a Mott transition is obtained from the dynamical mean-field theory
(DMFT) \shortcite{georgesrmp}. This formulation, which ignores
localization effects, is obtained by rescaling the hopping
elements $t\rightarrow t/\sqrt{z}$ and then formally taking the
limit of large coordination $z\rightarrow\infty$ (see Section \ref{sub:disDMFT}). To obtain
qualitatively correct analytical results describing the vicinity
of the disordered Mott transition at $T=0,$ we have solved the
DMFT equations using a 4-boson method \shortcite{mitglass-prl03}. At weak
disorder, these equations can be easily solved in closed form and
we simply report the relevant results. The critical value of
hopping for the Mott transition is
found to decrease with disorder as%
\begin{equation}
t_{c}(W)\approx t_{c}^{o}\,(1-4(W/U)^{2}+\cdots),
\end{equation}
where for a simple semi-circular density of states
\shortcite{georgesrmp} $t_{c}^{o}=3\pi U/64$ (in this model, the
bandwidth $B=4t$). Physically, the disorder tends to suppress the
Mott insulating state, since it broadens the Hubbard bands and
narrows the Mott-Hubbard gap. At sufficiently strong disorder
$W\geq U$, the Mott insulator is suppressed even in the atomic
limit $t\rightarrow0$. The behavior of the compressibilities can
also be calculated near the Mott
transition and to leading order we find%
\begin{equation}
\chi^{(2)}=\left[  \frac{8}{3\pi
t_{c}^{o}}(1-\frac{t_{c}(W)}{t})\right] ^{2}(1+28(W/U)^{2}).
\end{equation}
Therefore, as with any compressibility, $\chi^{(2)}$ is found to be
very small in the vicinity of the Mott transition, even in the
presence of finite disorder. As a result, the tendency to glassy
ordering is strongly suppressed at weak disorder when one
approaches the Mott insulating state.
%

Finally, having analyzed the limits of weak and strong disorder,
we briefly comment on what may be expected in the intermediate
region $W\sim U$. 
The Mott gap cannot exist for $W>U$, so
in this region and for sufficiently small $t$ (i. e. kinetic
energy), one enters a gapless (compressible) Mott-Anderson
insulator. For $W\sim U,$ the computation of $\chi^{(2)}$ requires
the full solution of the Mott-Anderson problem. The required
calculations can and should be performed using the formulation of
\shortcite{motand}, but that
difficult task is a challenge for the future. However, based on
general arguments presented above, we expect $\chi^{(2)}$ to
\emph{vanish} as one approaches the Mott insulator $(W<U)$, but to
\emph{diverge} as one approaches the Mott-Anderson insulator
($W>U).$ Near the tetracritical point, we may
expect $\chi^{(2)}\sim\delta W^{-\beta}\delta t^{\alpha },$ where
$\delta W=W-W_{Mott}(t)$ is the distance to the Mott transition
line, and $\delta t=t-t_{c}(W)$ is the distance to the
Mott-Anderson line. Using this Ansatz and Eq. (\ref{instab}), we
find the glass transition line to
take the form%

\begin{equation}
\delta t=t_{G}(W)-t_{c}(W)\sim\delta W^{\beta/\alpha};\;W\sim
W_{Mott}.
\label{tetracritical}%
\end{equation}

We thus expect the intermediate metallic glass phase to be
suppressed as the disorder is reduced, and one approaches the Mott
insulating state. Physically, glassy behavior of electrons
corresponds to many low-lying rearrangements of the charge
density; such rearrangements are energetically unfavorable close
to the (incompressible) Mott insulator, since the on-site
repulsion $U$ opposes charge fluctuations. Interestingly, very
recent experiments on low density electrons in silicon MOSFETs
have revealed the existence of exactly such an intermediate
metallic glass phase in low mobility (highly disordered) samples
\shortcite{bogdanovich-prl02}. In contrast, in high mobility (low disorder)
samples \shortcite{JJPRL02}, no intermediate metallic glass phase is
seen and glassy behavior emerges only as one enters the
insulator, consistent with our theory. Similar conclusions have
also been reported in studies of highly disordered InO$_{2}$ films 
\shortcite{films22,films23}, where the glassy
slowing down of the electron dynamics seems to be suppressed as
the disorder is reduced and one crosses over from an Anderson-like
to a Mott-like insulator. In addition, these experiments
\shortcite{bogdanovich-prl02,JJPRL02} provide striking evidence of
scale-invariant dynamical correlations inside the glass phase,
consistent with the hierarchical picture of glassy dynamics, that
generally emerges from mean-field approaches \shortcite{re:Mezard86}
such as the one used in this work.

\section{Beyond DMFT: loop expansion and diffusion modes}

DMFT approaches to correlation and disorder have already provided significant insight in a number of key phenomena and processes around the metal-insulator transition. Most importantly, these methods have made it possible to describe strong correlation effects associated with both ``Mottness" and the glassy behavior of electrons. The information provided is, in many ways, rather complementary to the conventional weak-coupling approaches which focused on the effects of long-wavelength diffusion modes within the Fermi liquid picture. Because of their focus on local correlation effects, the DMFT theories are ill-suited to describe those phenomena and regimes where long-range spatial correlations dominate. One of the biggest challenges for future work is to find ways to combine the strength of both approaches and introduce systematic methods to incorporate the non-local spatial correlations ignored by DMFT. While little has been done so far to implement this ambitious program, some approaches have already been outlined that provide guidance on how the problem should be formulated. We conclude our discussion by a brief outline of a promising approach to include both the local correlation processes and the diffusion-mode effects within a single theory.

\subsection{Gauge Invariant Models of Wegner}
The task of identifying the leading nonlocal corrections to DMFT can be formulated in a particularly elegant and transparent fashion for certain special models of disorder. Here, the DMFT equations can be obtained  \shortcite{dk-prb94} from a functional-integral formulation at the saddle-point level, allowing systematic corrections using a loop expansion. In this class, the inter-site hopping elements are assumed to be random variables of the form
\begin{equation}
t_{ij}=y_{ij}\; g(x_i ,x_j ),
\end{equation}
and in addition, there can be  an arbitrary
distribution of site energies $\varepsilon_i$.
Here, $y_{ij}$-s are independent {\em bond}
 variables with a symmetric distribution, i. e. 
$\overline{y_{ij}^{2n+1}}=0$, and $g(x_i ,x_j )$ is an arbitrary
function of local {\em site} variables $x_i$. 

The special class of models which have a symmetric distribution of
hopping elements has a very simple physical interpretation. As first
observed by Wegner \shortcite{wegner79,wegner80}, in these 
``gauge invariant'' models, the phases of
the electrons undergo random shifts at every lattice hop and so the
mean free path $\ell$ reduces to one lattice spacing. 
On general grounds, on
length scales longer than $\ell$ the details of the lattice structure
are washed out by disorder, so that for gauge invariant models in the 
large dimensionality limit, the details of the lattice structure 
become irrelevant. We contrast this with the models with arbitrary
disorder discussed previously, which have a well defined pure
limit and accordingly can also have an arbitrarily large mean free
path. The presence of this intermediate length scale ($\ell$ can be
much larger than the lattice spacing $a$, but much smaller than the
localization length $\xi$) is often irrelevant to both
long-wavelength phenomena such as localization and local phenomena
such as the Mott transition. The gauge invariant models avoid these
unnecessary complications without disrupting any of the qualitative
properties on either very short or very long length scales. 

The general properties are the
same for all the models in this class, but for the simplicity of our
presentation, we will restrict our attention to the separable case \shortcite{shiba71}
where $g(x_i ,x_j )=x_i x_j$, with an arbitrary
distribution $P_X (x_i)$ for the site variables $x_i$. For the trivial
choice of 
$P_X (x_i)=\delta (x_i -1)$, the models reduce to the 
gauge invariant models of Wegner \shortcite{wegner79,wegner80}. Nontrivial distributions 
$P_X (x_i)$ which extend to small values of the variable $x_i$ are
useful for the study of disorder-induced local moment formation \shortcite{milovanovic89}.
Sites with small $x_i $ represent sites with weak
hybridization.
At intermediate correlations, we expect sites
with small $x_i$ to behave as local moments and give large
contributions to thermodynamic quantities such as the specific
heat coefficient $\gamma =C/T$, while other sites remain in the
itinerant regime. 

Also, we take the $y_{ij}$-s to be  Gaussian random variables 
with zero mean and variance
\begin{equation}
\overline{y_{ij}^2}=\frac{1}{z}f_{ij}\;  t^2.
\end{equation}
Here, the (uniform) matrix $f_{ij}$ specifies the lattice structure
\begin{equation}
f_{ij}\; =\; \left\{ \begin{array}{l}
			1,\ \mbox{for connected sites,} \\
			0,\ \mbox{for disconnected sites,}
		      \end{array} \right. 
\end{equation}
and we have scaled the (square of the) hopping elements by the
coordination number \mbox{$z=\sum_j f_{ij}$}, in order to obtain finite
results in the \mbox{$z\rightarrow\infty$} limit. 

\subsection{Functional Integral Formulation}

At this point, it is convenient to explicitly perform the averaging
over the Gaussian random (bond) variables $y_{ij}$, using the standard replica formulation   \shortcite{wegner79,wegner80}. The
hopping part of the action then assumes the form \shortcite{dk-prb94}
\begin{equation}
  S_{hop} = \frac{1}{2}t^2\; \sum_{ij}\; \frac{1}{z} 
	    f_{ij}\;x_i^2\; x_j^2\; 
	    \left[
            \sum_{\alpha , s}\int_o^{\beta}\, d\tau\, 
            [c^{\dagger\;\alpha}_{s,i} (\tau )
            c_{s,j}^{\alpha} (\tau ) + h. c. ]\right]^2 .
\end{equation}
As we can see from this expression, the averaging over disorder has
generated a {\em quartic} term in the action that is nonlocal in
(imaginary) time, spin and replica indices. We are now in a position
to introduce  collective $Q$-fields \shortcite{wegner79,wegner80,fink-jetp83,fink-jetp84} of the form (in terms of
Matsubara frequencies $\omega =2n\pi T$; the indices ``n'' are omitted for
brevity) 
\begin{equation}
Q_{\omega_1\omega_2}^{\alpha_1\alpha_2 ,s_1 s_2}(i)=
\frac{1}{z} \sum_j f_{ij}\; x_j^2\; 
c^{\dagger\;\alpha_1}_{j,s_1} (\omega_1 )
c^{\alpha_2}_{j,s_2} (\omega_2 ),
\end{equation}
by decoupling the (quartic) hopping term using a Hubbard-Stratonovich
transformation. For simplicity, as before, we will ignore the
superconducting phases, as well as the fluctuations in the
particle-particle (Cooper) channel, so that the $Q$-field does not
have anomalous components. The procedure can be straightforwardly
generalized to include the omitted terms \shortcite{efetov80}. 

It is now possible to formally integrate out the electron (Grassmann)
fields and the resulting action for the $Q$-fields can be written as
\begin{equation}
S[Q]=S_{hop} [Q]\; +\; S_{loc} [Q].
\end{equation}
The nonlocal part of the action $S_{hop} [Q]$ takes a simple quadratic
form in terms of the $Q$ fields
\begin{equation}
S_{hop} [Q]= -\frac{1}{2}\; t^2\;
\sum_{ij}\sum_{\alpha_1\alpha_2}\sum_{s_1 s_2}\sum_{\omega_1\omega_2}K_{ij}\; 
Q_{\omega_1\omega_2}^{\alpha_1\alpha_2 ,s_1 s_2}(i)
Q_{\omega_2\omega_1}^{\alpha_2\alpha_1 ,s_2 s_1}(j),
\end{equation}
where, $K_{ij}=\frac{1}{z} f^{-1}_{ij}$ 
is the inverse lattice matrix, scaled by coordination number $z$.  In
contrast, all the nonlinearities are contained in the {\em local} part
of the action
\begin{equation}
S_{loc} [Q]= -\sum_i\;\ln
\int dx_i P_X (x_i )\int d\varepsilon_i P_S (\varepsilon_i )
\int Dc^{\dagger}_i Dc_i
\exp\left\{-S_{eff} [c^{\dagger}_i ,c_i ,Q_i , x_i ,\varepsilon_i ]\right\},
\end{equation}
where the local effective action takes the form
\begin{eqnarray}
& & S_{eff} [c^{\dagger}_i ,c_i ,Q_i , x_i ,\varepsilon_i ] =\nonumber\\
& & -\sum_{\alpha_1\alpha_2}\sum_{s_1 s_2}\sum_{\omega_1\omega_2}
c^{\dagger\;\alpha_1}_{i,s_1} (\omega_1 )[
\left( i\omega_1 +\mu -\varepsilon_i\right)
\delta_{\alpha_1\alpha_2}\delta_{s_1 s_2}\delta_{\omega_1\omega_2} 
-x_i^2\;t^2\, Q_{\omega_1\omega_2}^{\alpha_1\alpha_2 ,s_1 s_2} (i)
] c^{\alpha_2}_{i,s_2} (\omega_2 ) \nonumber \\
& & + U\sum_{\alpha}\;\sum_{\omega_1 +\omega_3 =\omega_2 +\omega_4 }\;
c^{\dagger\;\alpha}_{i ,\uparrow} (\omega_1 )
c^{\alpha}_{i ,\uparrow} (\omega_2 )
c^{\dagger\;\alpha}_{i,\downarrow} (\omega_3 ) 
c^{\alpha}_{i,\downarrow} (\omega_4 ).
\end{eqnarray}
The local effective action $S_{eff} [c^{\dagger}_i ,c_i ,Q_i , x_i
,\varepsilon_i ]$ is identical to the action of a (generalized)
Anderson impurity model embedded in an electronic bath characterized by
a hybridization function $x_i^2\;t^2\, Q_{\omega_1\omega_2}^
{\alpha_1\alpha_2 ,s_1 s_2}(i)$. We can thus interpret our system as a 
{\em collection} of Anderson impurity models \shortcite{Anderson1961} that are ``connected''
through the existence of collective $Q$-fields. Here we  note that,
in contrast to an ordinary Anderson model, the hybridization function 
is now {\em non-diagonal} in frequency, spin and replica indices.
Physically, this reflects the fact that in general dimensions a given
site can be regarded as an Anderson impurity model in a {\em
fluctuating} bath which breaks translational invariance in 
time, space and spin. 

\subsection{Saddle-Point Solution}

The action has a general form which is very similar to standard
lattice models investigated in statistical mechanics \shortcite{goldenfeldbook}.
As usual, the
problem simplifies considerably in the limit of large coordination
number, when the spatial fluctuations of the Hubbard-Stratonovich 
field ($Q$ in our case) are suppressed and the mean-field theory
becomes exact. It is worth pointing out that there are two classes of
lattices which can have large coordination:

(a) Lattices with short-range bonds 
but living in a space  of {\em large dimensionality}. For example, on
a hypercubic lattice with nearest neighbor hopping in $d$ dimensions, 
$z=2d$. 

(b) Lattices embedded in a finite dimensional space but having 
{\em long hopping range}. In this case, the lattice matrix $f_{ij}$
takes the form 
\begin{equation}
f_{ij}\; =\; \left\{ \begin{array}{l}
			1, |i-j|<\, L,  \\
			0, \mbox{otherwise,}
		      \end{array} \right. 
\end{equation}
and the coordination number $z\sim L^d$. 

In either case, when $z\rightarrow\infty$, the functional integral
over $Q$-fields, representing the partition function, can be evaluated
(exactly) by a saddle-point method and we obtain a  mean-field theory. 
In order to derive the mean-field equations in our case, we look for
extrema of the action $S[Q]$ with respect to variations of the
$Q$-fields, i. e. 
\begin{equation}
\frac{\delta\, S[Q]}{\delta\, Q_{\omega_1\omega_2}^{\alpha_1\alpha_2 ,s_1 s_2}(i)}=0.
\end{equation}
Since the saddle-point solution is translationally invariant in time and
space and conserves spin, it is {\em diagonal} in all indices
\begin{equation}
\left[ Q_{\omega_1\omega_2}^{\alpha_1\alpha_2 ,s_1 s_2}(i)\right]|_{SP}=
\delta_{\alpha_1\alpha_2}\delta_{s_1 s_2}\delta_{\omega_1\omega_2} Q^{SP}_s (\omega ),
\end{equation}
and the saddle-point equations assume the form
\begin{equation}
Q_s^{SP} (\omega )=
\int d\varepsilon_i P_S (\varepsilon_i )\int dx_{i} P_X (x_{i} ) \,
x_{i}^2\; G_{i,s} (\omega ),
\end{equation}
where
\begin{equation}
  G_{i,s} (\omega )=<c^{\dagger}_{s}(\omega)c_{s}(\omega )>
  _{S_{eff} [c^{\dagger} ,c ,Q^{SP} , x_i ,\varepsilon_i ]}.
\end{equation}
If we identify 
\begin{equation}
W_{s,i} (\omega )\equiv x_i^2\,t^2 Q_s^{SP} (\omega ),
\end{equation}
we see that our saddle-point equations become {\em identical} to the  standard DMFT
($d\rightarrow\infty$) equations, 
when applied to the appropriate model of hopping disorder. 
We emphasize that the present equations are exact at
$z\rightarrow\infty$ for an {\em arbitrary} lattice, due to the
presence of the ``gauge invariant'' form of the hopping disorder. 
Since the saddle point equations determine the local effective action,
this means that all the {\em local} correlation functions will be
insensitive to the lattice structure in this mean-field limit.
However, other properties such as the tendency to the formation of
spin and charge density waves {\em are} very sensitive to the
details of the lattice structure. 

As an example, we can compare the case of simple hopping disorder, 
$t_{ij}=y_{ij}$, in a bipartite lattice such as the Bethe
lattice and the case of a lattice with infinite range hopping (the
limit $L\rightarrow\infty$ of the model (b) above). The
self-consistency (mean-field) equations are {\em identical} in the two
cases and in fact reduce to those of a {\em pure} Hubbard model on a
Bethe lattice with hopping $t$. On the other hand, it is well
established \shortcite{re:Rozenberg92} that in the first case the system is unstable towards the
formation of an antiferromagnetic ground state, even for arbitrarily
small $U/t$, while in the second, the system remains paramagnetic for
any $U/t$, due to the large frustration. 
In many physical systems, such as doped semiconductors \shortcite{paalanen91}, disorder
introduces large amounts of frustration and magnetic ordering does
not occur, even though the system is strongly correlated. In order to
study such situations, it is  useful to have at hand microscopic
models that have a non-magnetic ground state and allow one to study
the approach to the metal-insulator transition which occurs at $T=0$.

\subsection{Loop Expansion}

The present approach is particularly convenient for the study of the 
effects of strong correlations on {\em disorder-driven} transitions
and the interplay of Anderson localization and strong correlations in
general. This is especially true since Anderson localization is not
present in $d=\infty$ (or infinite range) models and so one has to
extend the approach to include spatial fluctuations
missing from the mean-field description. In order to systematically
study the fluctuation effects, we proceed to carry out an expansion in
terms of the deviations of the collective $Q$-fields from their
saddle-point value, i.~e., in powers of $\delta Q(i)=Q(i)-Q^{SP}$. This
procedure, also known as a {\em loop expansion} \shortcite{goldenfeldbook} has been used in other
disordered problems, such as spin-glasses \shortcite{dominicis91}, to generate systematic
corrections to the mean-field theory. The method is particularly
convenient when applied to long-range models \shortcite{dominicis91}
(class (b) above), since
in that case the loop corrections are ordered by a small parameter
$1/z$. The loop expansion can be applied also to large dimensionality
models, but in that case a given order in a loop
expansion can be considered to be an infinite resummation of the
simple $1/d$ expansion \shortcite{Georges90}, 
since each term contains all powers of $1/d$. 

When the expansion of the effective action in terms of $\delta Q$-s is
carried to lowest, quadratic order, we obtain a theory describing
gaussian fluctuations around the saddle point, which represent weakly
interacting collective modes \shortcite{wegner79,wegner80,fink-jetp83,fink-jetp84}. 
Higher order terms in the expansion then
generate effective interactions of these modes, which under
appropriate conditions can lead to fluctuation-driven phase
transitions. In practice, if all the components of the collective
$Q$-fields are retained in this analysis, the theory becomes
prohibitively complicated and cumbersome. However, in order to analyze
the {\em critical behavior}, it is not necessary to keep track of all
the degrees of freedom, but it suffices to limit the analysis to 
{\em soft modes}, i.~e.,
those that represent low energy excitations. In disordered metallic 
phases, charge and spin conservation laws lead to the existence of 
{\em diffusion modes}, which are the hydrodynamic modes describing
charge- and spin-density relaxation. In the Fermi liquid regime
\shortcite{ckl}, 
all the other collective excitations require higher energy and can be
neglected in a hydrodynamic description of the system. One is then
led to construct a theory of interacting diffusion modes, as a theory
of critical phenomena for disordered interacting systems.

This line of reasoning was used in field-theoretical approaches to the
localization problem of noninteracting electrons, 
as first developed by Wegner \shortcite{wegner79,wegner80}. In this theory,
collective $Q$-fields, similar to the ones presented in this paper,
are introduced. At the saddle-point level, no phase transition occurs
and all the states are extended. An analysis of the fluctuations of
the $Q$-fields is then performed and a {\em subset} of those
fluctuations identified, which represent the hydrodynamic (diffusion)
modes. Only the fluctuations of these modes are retained and an
effective hydrodynamic theory is constructed - the {\em non-linear sigma
model} \shortcite{wegner79,wegner80} . The interactions of these modes lead to the metal-insulator
(localization) transition, which was analyzed using
renormalization-group techniques and $2+\varepsilon$ expansions. In subsequent
work, Finkelshtein \shortcite{fink-jetp83,fink-jetp84} was able to apply a similar procedure to {\em
interacting} disordered electrons. However, this theory is based on
a number of implicit assumptions that restrict its validity to {\em
Fermi-liquid} regimes. In the language of $Q$-fields, this theory again
expands around a noninteracting saddle-point and the interaction
effects  appear only at the level of the fluctuation corrections. 

When strongly correlated electronic systems are considered, much of
the physics relates precisely to the {\em destruction of coherent
quasiparticles} by inelastic processes, so that one needs a
description that is not limited to Fermi-liquid regimes. In the
language of hydrodynamics, new soft modes appear, which indicate the
tendency towards  interaction-driven instabilities. In particular, strong
correlations can lead to local moment formation and the Mott
transition. In both cases,  charge fluctuations are suppressed and
low-energy spin fluctuations dominate the physics. 

In the present approach, in contrast to the work of  Wegner and 
Finkelshtein, the strong correlations
are treated in a non-perturbative fashion already at the saddle-point 
(mean-field) level, so that {\em all} the soft modes can be
systematically included. In particular, even at the one-loop level, we
can address the question of how  the disorder-induced local moment
formation and the approach to the Mott transition affect the weak
localization (diffusion) corrections. Ultimately, our approach
indicates how a more general low energy theory can be constructed that
extends the $\sigma$-model description so as to include strong correlation
effects. 

In the present discussion, we will limit our attention to the form of the
Gaussian fluctuations of the $Q$-fields, which allow one in principle to compute the
leading corrections to mean-field theory. 
The Gaussian part of the action takes the form
\begin{eqnarray}
S^{(2)}[Q] & = & -\frac{1}{2}\,t^2\,\sum_{l_1\cdots l_4}
\int\frac{dk}{(2\pi )^d}\,\delta Q_{l_1 l_2} (k)\left[
\left(L^2 k^2 +1\right)\delta_{l_1 l_4}\delta_{l_2 l_3}\right.
\nonumber\\
& & \left. -t^2 W(l_1 )W(l_3 )\delta_{l_1 l_2}\delta_{l_3 l_4}
+t^2 \Gamma (l_1\cdots l_4 )\right]\, \delta Q_{l_3 l_4 } (-k).
\end{eqnarray}
This expression is appropriate for the long-ranged model (b) above, 
in  which case the
inverse lattice matrix in momentum space takes the form
$K(k)\approx 1+L^2 k^2$ and we cut off the momentum integrals at 
$\Lambda=2\pi /L$. 
Note that the coefficient of $k^2$, which can be interpreted as the
{\em stiffness} of the $\delta Q$ modes, is $\sim L^2$, so we see  that
indeed the fluctuations are suppressed at $L\rightarrow\infty$. 
In the above formula, the index $l_m$ is used
to represent the frequency, spin and replica indices. 
The local vertex function  $\Gamma (l_1\cdots l_4 )$ is given by
\begin{equation}
\Gamma (l_1\cdots l_4 )=
\int d\varepsilon_i P_S (\varepsilon_i )\int dx_{i} P_X (x_{i} ) \,
x_i^4 < c^{\dagger}_i  (l_1 ) c_i (l_2 )
c^{\dagger}_i (l_3 ) c_i (l_4 ) >_{S_{eff} [Q^{SP}]} .
\end{equation}
At this level, the dynamics of the collective fluctuations $\delta
Q$ is governed by the form of $S^{(2)} [\delta Q ]$, which is expressed in
terms of the {\em local correlation functions} of the saddle-point
theory, i.~e., of the $d=\infty$ disordered Hubbard model. 
Accordingly, a detailed study of the $d=\infty$ limit does not provide
only a mean-field description of the problem, but also determines the
form of the leading corrections resulting from fluctuations. 
Extensions of the theory to include the effects of these fluctuation corrections
remain to be addressed in more detail in future work. 

\section{Acknowledgments}
This work was supported by FAPESP through grant 07/57630-5 (EM), CNPq
through grant 304311/2010-3 (EM), and by NSF through grants DMR-0542026 and DMR-1005751.
(VD).

\bibliographystyle{OUPnamed_notitle}
\bibliography{all,vlad11}

\end{document}